\documentclass[11pt]{article}
\usepackage{latexsym}
\usepackage{amsmath}
\usepackage{tensor}
\usepackage{multibox}
\usepackage{verbatim}
\usepackage{mathrsfs}

\DeclareMathAlphabet{\mathpzc}{OT1}{pzc}{m}{it}

 \csname
@addtoreset\endcsname{equation}{section}

\def\prd{\pr \cdot}

\def\gz0{\gamma^{0}}

\def\nn{\nonumber}
\newcommand{\eq}[1]{(\ref{#1})}

\def\ft#1#2{{\textstyle{{\scriptstyle #1}
\over {\scriptstyle #2}}}}

\def\ket#1{|#1\rangle}
\def\scs#1{\section{\sc #1}}
\def\scss#1{\subsection{\sc #1}}
\def\scsss#1{\subsubsection{\sc #1}}

\def\a{\alpha}
\def\b{\beta}
\def\g{\gamma}

\def\d{\delta}
\def\D{\Delta}

\def\h{\eta}
\def\th{\theta}

\def\l{\lambda}
\def\L{\Lambda}
\def\m{\mu}
\def\n{\nu}
\def\x{\xi}

\def\P{\Pi}

\def\r{\rho}

\def\t{\tau}

\def\F{\Phi}
\def\vf{\varphi}
\def\c{\chi}

\def\O{\Omega}

\def\cA{{\cal A}}
\def\cB{{\cal B}}
\def\cC{{\cal C}}
\def\cD{{\cal D}}
\def\cE{{\cal E}}
\def\cF{{\cal F}}

\def\cJ{{\cal J}}
\def\cK{{\cal K}}
\def\cL{{\cal L}}

\def\cO{{\cal O}}

\def\be{\begin{equation}}
\def\ee{\end{equation}}
\def\bea{\begin{eqnarray}}
\def\eea{\end{eqnarray}}
\def\ba{\begin{array}}
\def\ea{\end{array}}
\def\bec{\begin{center}}
\def\ec{\end{center}}
\def\ba{\begin{align}}
\def\ena{\end{align}}
\def\ft{\footnote}

\def\hpe{\pe \hat{\phantom{\! \! \pe}}}
\def\pe{\prime}
\def\12{\frac{1}{2}}
\def\fr{\frac}
\def\pr{\partial}
\def\prd{\partial \cdot}
\def\bra{\langle \,}
\def\ket{\, \rangle}
\def\comma{\,,\,}
\def\eq{\equiv}

\def\ra{\rightarrow}

\begin{document}

\begin{flushright}
{\today}
\end{flushright}

\vspace{25pt}

\begin{center}

%%%%%%%%%%%%%%%%%%%%%%%%%%%%%%%%%%%%%%%%%%%%%%%%%%%%%%%%%%%%%%%%%%%%%%%%%%%%%%%%%%%%%%%%%

{\Large\sc Unconstrained Higher Spins of Mixed Symmetry \vskip 12pt I. Bose Fields}\\

%%%%%%%%%%%%%%%%%%%%%%%%%%%%%%%%%%%%%%%%%%%%%%%%%%%%%%%%%%%%%%%%%%%%%%%%%%%%%%%%%%%%%%%%%

\vspace{25pt}
{\sc A.~Campoleoni${}^{\; a,\; b,\; d}$, D.~Francia${}^{\; c,\; b}\ \footnote{Present address.}$, J.~Mourad$^{\; b}$ and A.~Sagnotti$^{\; a,\; b,\; d}$}\\[15pt]

{${}^a$\sl\small
Scuola Normale Superiore and INFN\\
Piazza dei Cavalieri, 7\\I-56126 Pisa \ ITALY \\
e-mail: {\small \it a.campoleoni@sns.it,
sagnotti@sns.it}}\vspace{10pt}

{${}^b$\sl\small AstroParticule et Cosmologie (APC) \footnote{Unit\'e mixte de Recherche du CNRS (UMR 7164.)}\\
Universit\'e Paris VII - Campus Paris Rive Gauche \\
10, rue Alice Domon et Leonie Duquet \\ F-75205 Paris Cedex 13\
FRANCE
 \\ e-mail:
{\small \it francia@apc.univ-paris7.fr, mourad@apc.univ-paris7.fr}}\vspace{10pt}

{${}^c$\sl\small Department of Fundamental Physics\\ Chalmers
University of Technology \\ S-412\ 96 \ G\"oteborg \ SWEDEN
\\ }\vspace{10pt}

{${}^d$\sl\small Centre de Physique Th\'eorique,
\'Ecole Polyt\'echnique\\
F-91128 Palaiseau \ FRANCE}\vspace{10pt}

%%%%%%%%%%%%%%%%%%%%%%%%%%%%%%%%%%%%%%%%%%%%%%%
\vspace{35pt} {\sc\large Abstract}\end{center}
%%%%%%%%%%%%%%%%%%%%%%%%%%%%%%%%%%%%%%%%%%%%%%%

{This is the first of two papers devoted to the local
``metric-like'' unconstrained Lagrangians and field equations for
higher-spin gauge fields of mixed symmetry in flat space. Here we
complete the previous constrained formulation of Labastida for Bose
fields. We thus recover his Lagrangians via the Bianchi identities,
before extending them to their ``minimal'' unconstrained form with
higher derivatives of the compensator fields and to yet another,
non-minimal, form with only two-derivative terms. We also identify
classes of these systems that are invariant under Weyl-like
symmetries.}

\setcounter{page}{1}

\pagebreak

\tableofcontents

\newpage

%%%%%%%%%%%%%%%%%%%%%%%%%%%%%%%%%%%%%%%%%%%%%%%

\scs{Introduction}\label{sec:intro}

%%%%%%%%%%%%%%%%%%%%%%%%%%%%%%%%%%%%%%%%%%%%%%%

The theory of Higher-Spin Gauge Fields is an enticing and still
largely unexplored corner of Field Theory, that has marked
consistently the frontier of our understanding since the 1930's
\cite{solvay}. This difficult subject has attracted over the years
several leading specialists, but despite many efforts a number of
key questions remain unanswered. Surprisingly, perhaps, a few of
these still concern the free theory, and the present work is devoted
to clarifying some of the corresponding issues. Furthermore, our
current understanding of higher-spin interactions is quite
incomplete: classic no-go theorems showed in fact, long ago, that
they are naively inconsistent, at least for massless fields in flat
space
\cite{nogo}, but several important examples of higher-spin vertices
were found nonetheless over the years \cite{otherold,othernew}. Crucial long-term efforts by Vasiliev, initially with Fradkin \cite{vasold,vasnew}, have actually provided paradigmatic examples of
classically consistent interactions between infinitely many totally
symmetric massless higher-spin fields in curved $(A)dS$ backgrounds.
The systematics of such constructions, however, is not fully
understood at this time, while Vasiliev's ``unfolded'' formulation
is not Lagrangian and appears rather remote from ordinary lower-spin
field theories. It is thus natural to try and bridge the apparent
gap between these methods and the ordinary constructions for low
spins, and this paper is meant as a step in that direction. In
addition, and perhaps more importantly, higher spins are a key
element of String Theory \cite{stringtheory}, whose massive spectra
contain infinitely many of them. Hence one might say that
higher-spin interactions have presently two vastly different
realizations. In the first, provided by the Vasiliev setting, their
gauge symmetry is unbroken, while in the second, provided by String
Theory, it is somehow spontaneously broken. Unfortunately, we do not
understand precisely how to bridge the gap between the two, or even
how to attach a precise meaning to the breaking mechanism at work in
String Theory. Still, it is fair to stress that, while the massive
higher-spin modes of String Theory are usually ignored, they are
clearly instrumental in granting some of its most spectacular
properties, such as modular invariance or open-closed duality.

A deeper understanding of these massive modes can not forego a
closer look at higher-spin fields of mixed symmetry, that constitute
the vast majority of them, and this provides at least part of the
motivations behind the present work. This is the first of two papers
devoted to some open issues concerning free higher-spin gauge fields
of mixed symmetry. It deals mostly with multi-symmetric bosonic
fields of the type $\varphi_{\mu_1
\ldots\,
\mu_{s_1};\,\nu_1 \ldots\, \nu_{s_2};\,\ldots}$, while the companion paper \cite{cfms2}
deals mostly with multi-symmetric fermionic fields of the type
$\psi_{\mu_1
\ldots\, \mu_{s_1};\,\nu_1
\ldots\,
\nu_{s_2};\,\ldots}$. Both these types of fields possess several ``families''
of symmetric index sets, are defined in generic $D$-dimensional
space times and are inevitable ingredients for $D>5$. Dealing with
these reducible $gl(D)$ tensors is a necessary complication if one
wishes to establish quantitative links between Higher-Spin Gauge
Theory and String Theory, which, as anticipated, is one of the key
problems today. Indeed these types of reducible $gl(D)$ tensors,
rather than the more familiar Young-projected tensors, accompany in
String Theory products of bosonic oscillators, and therefore it is
both natural and convenient to formulate the theory directly in
terms of them. Let us also remark that the simplest instance of a
reducible field of this type is actually rather familiar. It is
simply a two-tensor, which one would usually split aforehand into a
symmetric tensor and a Kalb-Ramond two-form. We also describe how to
adapt the formalism to reducible multi-antisymmetric fields or
multi-forms, or to more general types of bosonic fields where some
collections of space-time indices are fully symmetric and some are
fully antisymmetric. The relative simplicity of this type of
extension was pointed out in \cite{dvh,bb1,dmh,bb2}, and the result
is clearly of interest in comparisons with superstring models, where
these more general types of fields accompany products of fermionic
and bosonic oscillators.

Allowing mixed symmetry complicates matters to a considerable
extent: for instance, the gauge symmetry is typically reducible, so
that gauge transformations of the gauge transformations emerge, a
feature already displayed by the relatively simple example, which we
just mentioned, of a two-form Kalb-Ramond field. Hence, one can not
forego the need for a proper notation to deal concisely and
effectively with the general case. Our present choice, explained in
some detail in the Appendices, is a natural extension of the
index-free description used in previous work on fully symmetric
(spinor)tensors by some of us. Still, it is far more complicated,
and involves explicit ``family'' indices, needed to identify
particular subsets of the tensor indices to be treated differently
by the various operators entering the field equations or the
Lagrangians. For instance, while the bosonic gauge fields can still
be simply denoted by $\varphi$, as in the fully symmetric case,
their gauge transformations,
\be \delta \, \varphi \ = \ \partial^{\,i} \, \Lambda_{\,i} \ , \ee
now involve explicit family indices. These are needed to identify
the various sets of space-time indices taking part in the gradient,
which are to be fully symmetrized \emph{only} with others of the
same type. At the same time, the gauge parameters $\Lambda_i$ carry
one lower family index and bear one less space-time index of the
$i$-th family than the corresponding gauge fields.

Free higher-spin bosonic gauge fields first took a Lagrangian form
in the late seventies, thanks to the work of Fronsdal
\cite{fronsdal}, who displayed the gauge symmetry emerging for massless totally
symmetric tensors from the massive Singh-Hagen
\cite{singhagen} construction. He thus identified some surprising
algebraic constraints on gauge parameters and gauge fields: in his
formulation the former are in fact to be traceless, while the latter
are to be doubly traceless. In the eighties, the development of free
String Field Theory \cite{sftheory} made it imperative to look more
closely at tensors of mixed symmetry, and important progress was
shortly made by a number of authors
 \cite{mixed} (although some
relevant work on the subject dates back to previous years
\cite{curt}). Most notably, Labastida \footnote{W. Siegel called to our attention
\cite{siegeladd}, where free
actions for Bose and Fermi fields were built within a BRST-type
setting. These works develop some of the ideas that originally led
from light-cone to covariant string fields. The approach, however,
is rather remote from the line of developments that lies at the heart
of this work and that can be largely traced back to the previous
works of Fronsdal and Labastida.}
\cite{labastida1,labastida} arrived at bosonic Lagrangians that generalize
what Fronsdal had previously attained for fully symmetric tensors,
requiring that the complete construction result in self-adjoint
kinetic operators, as was the case for the simpler, symmetric
theory. This condition revealed the need for peculiar
constraints on gauge fields and parameters, that in our present
notation read
\be T_{(\,ij} \, \Lambda_{\,k\,)} \, = \, 0 \ , \qquad T_{(\,ij} \, T_{kl\,)}\,
\varphi \, = \, 0 \ , \label{laba} \ee
where $T_{ij}$ denotes a trace involving a pair of space-time
indices belonging to the $i$-th and $j$-th families. Hence, quite
differently from what Fronsdal's case could naively suggest,
\emph{not all} traces of the gauge parameters and \emph{not all}
double traces of the gauge fields are forced to vanish in
Labastida's formulation. Further developments along these lines may
be found in \cite{mixed2}.

As shown in a number of previous works, the need for algebraic
constraints on gauge fields and parameters can actually be foregone,
arriving at formulations that are more akin, in spirit, to the more
familiar cases of low spins. This can be achieved in two quite
distinct, albeit related, ways. The first is to allow for
\emph{non-local} gauge-dependent additions to the field equations
\cite{fs1,fs2}, that have the virtue of linking the result to the
linearized higher-spin curvatures introduced by de Wit and Freedman
\cite{dewfr}. These results were extended to fields of mixed
symmetry in \cite{bb1, dmh, bb2}, and more recently the setting was shown
to also encompass the important case of massive symmetric fields
\cite{dario07}. Let us stress that, in sharp contrast with the
Singh-Hagen construction, the non-local formulation needs
\emph{no} additional auxiliary fields to this end. The second way is
to allow for additional fields in more conventional \emph{local}
Lagrangian formulations \cite{fs3,fms}. In their ``minimal'' form,
Fronsdal's constraints can then be foregone at the modest price of
introducing only two additional fields and some higher-derivative
terms that are anyway harmless: for a fully symmetric spin-$s$ field
$\varphi_{\mu_1 \ldots\, \mu_s}$, these are a spin-$(s-3)$
compensator $\alpha_{\mu_1 \ldots\, \mu_{s-3}}$, that first emerges
for $s=3$, and a spin-$(s-4)$ Lagrange multiplier $\b_{\mu_1
\ldots\, \mu_{s-4}}$, that first emerges for $s=4$. Or, as recently
shown in
\cite{dario07}, also a further spin-$(s-2)$ compensator and a
further Lagrange multiplier, if one insists on having not more than
two derivatives in the resulting Lagrangians. It should be stressed,
however, that these results, and a similar one obtained in
\cite{buchnew} restricting the ``triplets'' \cite{triplets,fs2,bonelli} of String Field Theory,
were all preceded by what should be regarded as the first instance
of a local unconstrained formulation. This non-minimal form,
obtained by Pashnev, Tsulaia, Buchbinder and others \cite{bpt} with
BRST techniques, makes use of ${\cal O}(s)$ different fields to
describe unconstrained spin-$s$ modes, and was connected to the
minimal on-shell formulation in
\cite{st}. Let us also emphasize that most previous works on
higher-spin fields, and our constructions in particular, proceed
along the lines of the metric formalism for gravity, and this is
also true for the world-line constructions of \cite{worldline}. An
important alternative is the analogue of the vielbein formalism,
first introduced in
\cite{lopvas} and developed further in a number of later works,
including those in \cite{extraframe} that are directly connected to
mixed-symmetry fields. This framework has already led to key
developments, since it is the basis of the Vasiliev construction
\cite{vasold,vasnew}, where the interactions are driven by an
infinite-dimensional extension of the tangent-space Lorentz algebra.
Still, attaining a deeper understanding of the metric-like formalism
appears of interest today, since it has the potential of clarifying
the geometrical origin of higher-spin interactions.

The main purpose of this paper is thus to extend the minimal
``metric-like'' formulation of unconstrained higher spins of
\cite{fs2,st,fs3,fms} to the case of mixed symmetry. As we shall
see, in general the structure of the Labastida constraints for Bose
fields would suggest to introduce compensators $\alpha_{ijk}$,
bearing a triple of family indices and fully symmetric under their
interchange, and Lagrange multipliers $\beta_{ijkl}$, bearing a
quadruple of family indices and fully symmetric under their
interchange. Actually, matters are slightly more complicated, since
this natural choice, motivated also by the analogy with the
symmetric construction of
\cite{fs2,st,fs3,fms}, would lead to the emergence of gauge invariant
combinations of the higher traces of the $\alpha_{ijk}$, that
therefore can not be regarded as independent compensators. This
difficulty reflects an intrinsic problem of the Labastida
constraints \eqref{laba}, that in fact are
\emph{not} independent. A simple way out is then to relate the
$\alpha_{ijk}$ to new fields, here termed $\Phi_i$, whose gauge
transformations are proportional to the parameters $\Lambda_i$, that
enter Lagrangians and field equations only via their symmetrized
traces and are such that
\be \alpha_{\,ijk}\,(\Phi) \, = \, \frac{1}{3}\ T_{(\,ij}
\, \Phi_{\,k\,)}\, . \label{alphaphi}
\ee

While the introduction of these additional fields, compensators and
Lagrange multipliers alike, reflects nicely the nature of the
constraints of eq.~(\ref{laba}), our approach differs somewhat, in
spirit, both from the original work of \cite{labastida} and from the
more recent work of \cite{bb2}, since we are guided throughout by
the Bianchi identities and their traces. This choice proves very
convenient when building the unconstrained theory, and particularly
so for unconstrained Fermi fields, for which in the companion paper
\cite{cfms2} we can present for the first time complete
``metric-like'' local Lagrangians together with their constrained
counterparts. Our ``minimal'' setting to arrive at an unconstrained
gauge symmetry stands out for its relative simplicity, but brings
about higher-derivative terms involving the compensator fields.
While this is not a problem, as can be clearly seen for example in
\cite{fms}, we shall also discuss how to complicate matters a bit in
order to obtain more conventional Lagrangians where all terms
contain at most two derivatives, generalizing the construction of
\cite{dario07}.

As first shown  by Labastida \cite{labastida}, the constrained
formulation of mixed-symmetry fields brings about a surprise with
respect to Fronsdal's case. Once, as in eq.~(\ref{laba}), not all
double traces of the gauge field $\varphi$ are forced to vanish, the
Lagrangian can (and indeed does) include additional terms involving
higher field traces, whose number grows with the number of families.
In a similar fashion, the unconstrained formulation for Bose fields
also involves a number of higher traces of a proper gauge-invariant
tensor built from $\varphi$ and $\alpha_{ijk}(\Phi)$ which grows
with the number of families. For this reason, in Section
\ref{sec:2bose} we begin by constructing Lagrangians and field equations for the
simplest generalization of the fully symmetric case, two-family
fields of the type $\vf_{\mu_1 \ldots\, \mu_{s};\,\nu_1 \ldots
\nu_{s_2}}$, and  study their reduction to the constrained Labastida
formulation. This analysis will also display the emergence of
Weyl-like symmetries for these higher-spin fields in sporadic
low-dimensional cases. We also add some comments on the Lagrangian
field equations of
\cite{labastida}, in a way that sheds some light on a subtlety
originating from the constrained nature of the gauge field $\vf$. We
conclude Section
\ref{sec:2bose} with the explicit examples of reducible rank-$(s,1)$
fields of the type $\vf_{\mu_1
\ldots\, \mu_{s_1};\,\nu}$ and reducible rank-$(4,2)$ fields of the type
$\varphi_{\mu_1 \ldots\,
\mu_4;\,\nu_1\nu_2}$, that are relatively simple but suffice to
illustrate a number of key points of our construction. In Section
\ref{sec:generalb} we extend the discussion to multi-family bosonic
fields, identifying precisely via the Bianchi identities their
unconstrained Lagrangians. When restricted to the case of
constrained fields, our result confirms the Labastida construction,
up to some changes of notation and to some typos that we correct.

In Section \ref{sec:irreducible} we describe how the unconstrained
formulation of bosonic fields can be adapted to tensors transforming
in irreducible representations of the Lorentz group. Whereas the
application to String Theory is more closely related to reducible
tensors, as we have stressed, it is in fact also interesting to
trace how the theory develops along lines closer to what one usually
does for low spins. In Section
\ref{sec:lowerder} we then describe how the minimal
higher-derivative unconstrained Lagrangians can be reduced
systematically to others with only two derivatives. In Section
\ref{sec:multiforms} we show how the present formalism can be simply adapted to
describe multi-forms or even more general types of fields needed to
encompass massive superstring excitations. Finally, Section
\ref{sec:conclusions} contains our conclusions, and the paper closes
with a number of Appendices where our notation is carefully spelled
out.

For the reader's convenience, let us summarize again the main
results contained in this paper:
\begin{itemize}
\item{the formalism previously developed for symmetric fields is extended via the introduction
of ``family indices'' that identify the subsets of space-time
indices acted upon by traces, gradients and other operations;}
\item{the Labastida construction is linked to the Bianchi identities, in strict analogy
with what happens for symmetric fields, and for linearized gravity
in particular. In the multi-symmetric case, the construction rests
on the subclass of their traces that are most antisymmetric in their
family indices or, more precisely, that are subject to two-column
Young projections;}
\item{formulations capable of bypassing the Labastida constraints
on gauge fields and parameters are obtained via a single type of
compensators $\a_{ijk}$, here to be expressed in terms of other more
basic fields $\Phi_i$ in order to account for the linear dependence
of the Labastida constraints, and of a single type of Lagrange
multipliers $\b_{ijkl}$. The resulting Lagrangians display a novel
type of gauge symmetry allowing certain redefinitions of the
$\b_{ijkl}$ and, as in the one-family or symmetric case, contain
higher derivatives of the compensators. However, we also describe
how to systematically modify the construction in order to obtain
Lagrangians with only two derivatives that contain a minimal number
of additional fields;}
\item{a rich pattern of sporadic cases is exhibited where Weyl-like
symmetries emerge, generalizing the well-known property of
two-dimensional gravity;}
\item{the formalism applies, with minor modifications, to both multi-symmetric tensors
and multi-forms, and thus encompasses all types of bosonic fields
that play a role in massive string spectra.}
\end{itemize}
\vskip 24pt

%%%%%%%%%%%%%%%%%%%%%%%%%%%%%%%%%%%%%%%%%%%%%%%%

\scs{Two-family bosonic fields}\label{sec:2bose}

%%%%%%%%%%%%%%%%%%%%%%%%%%%%%%%%%%%%%%%%%%%%%%%%

In this section we derive Lagrangians and field equations for
two-family bosonic fields of the type $\varphi_{\mu_1 \ldots\,
\mu_{s_1} ; \, \nu_1
\ldots\, \nu_{s_2}}$. We aim at a ``minimal'' unconstrained formulation,
and thus initially we allow higher-derivative terms involving the
compensators. We defer to Section \ref{sec:lowerder} a discussion of
how to recast these results in a form involving only two-derivative
compensator terms. While the results for two-family fields are still
relatively handy, this analysis has the virtue of displaying quite
clearly the key differences with respect to the simpler case of
fully symmetric tensors discussed in
\cite{fs3,fms}.

\vskip 24pt

%%%%%%%%%%%%%%%%%%%%%%%%%%%%%%%%%%%%%%%%%%%%%%

\scss{The Lagrangians}\label{sec:lagrangian2b}

%%%%%%%%%%%%%%%%%%%%%%%%%%%%%%%%%%%%%%%%%%%%%%

In the following, all space-time indices will be left implicit as in
the symmetric case of \cite{fs3,fms}, so that a generic two-family
gauge field $\varphi_{\mu_1\ldots\,
\mu_{s_1} ; \, \nu_1 \ldots\, \nu_{s_2}}$ will be simply denoted by
$\varphi$. While our notation is explained in detail in Appendix
\ref{app:MIX}, for the sake of clarity let us begin by summarizing
a few properties that we shall need repeatedly in the following. The
key novelty is the need for family indices specifying the sets of
space-time indices acted upon by gradients, divergences and traces.
Thus, for instance, the divergence and the gradient involving
space-time indices of the first family are here denoted by
\begin{eqnarray}
\pr_{\,1} \, \vf & \equiv & \pr_{\,\l} \, \vf^{\,\l}{}_{\m_1 \ldots\, \m_{s_1-1} ; \, \n_1 \ldots\, \n_{s_2}} \, , \\
\pr^{\,1} \, \vf & \equiv & \pr_{\,(\,\m_1} \, \vf_{\,\m_2 \ldots\,
\m_{s_1+1} \,) ; \, \n_1 \ldots\, \n_{s_2}} \, ,
\end{eqnarray}
where, here as in the rest of the paper, a pair of round parentheses
enclosing a set of indices indicates their total symmetrization with
the \emph{minimal} possible number of terms and normalized with a
\emph{unit} overall coefficient, rather than with unit strength.
Furthermore, we associate \emph{upper} family indices to operators
like the gradient, which add space-time indices, and \emph{lower}
family indices to operators like the divergence, which remove them.
In a similar fashion, we also introduce ``diagonal'' and ``mixed''
traces and metric tensors, so that for example
\begin{eqnarray}
T_{11} \, \vf & \equiv & \vf^{\,\l}{}_{\l \, \m_1 \ldots\, \m_{s_1-2} ; \, \n_1 \ldots\, \n_{s_2}} \, , \nonumber \\
T_{12} \, \vf & \equiv & \vf^{\,\l}{}_{\m_1 \ldots\, \m_{s_1-1} ; \, \l \, \n_1 \ldots\, \n_{s_2-1}} \, , \\
\h^{11} \, \vf & \equiv & \h_{\,(\,\m_1 \m_2} \, \vf_{\, \m_3 \ldots\, \m_{s_1+2} \,) ; \, \n_1 \ldots\, \n_{s_2}} \, , \nonumber \\
\h^{12} \, \vf & \equiv & \12 \, \Big(\, \h_{\,\m_1 (\, \n_1 \, |} \, \vf_{\, \m_2 \ldots\, \m_{s_1+1} ; \, | \, \n_2 \ldots\, \n_{s_2+1}\, )} \, + \,
\h_{\,\m_2 (\, \n_1 \, |} \, \vf_{\, \m_1\m_3 \ldots\, \m_{s_1+1} ; \, | \, \n_2 \ldots\, \n_{s_2+1}\, )} \, + \, \ldots \,\Big) \, , \nonumber
\end{eqnarray}
where the peculiar factor $\frac{1}{2}$ in the last expression
allows a more convenient presentation of a number of results,
including the Lagrangians. Further, in order to simplify the
combinatorics of partial integrations we shall also introduce a suitably
normalized scalar product, which is described in detail in Appendix
\ref{app:MIX}. As we have stressed, in these expressions and
elsewhere in this paper, round parentheses (brackets) enclose fully
(anti)symmetric space-time or family indices. In addition, vertical
bars separate, whenever this is needed for clarity, indices
belonging to different sets.

The starting point to derive a gauge invariant Lagrangian is the
gauge transformation
\be \label{gauge} \d \, \vf \, = \, \pr^{\,i} \, \L_{\,i} \, ,\ee
where, as anticipated in the Introduction, the gauge parameters bear
a lower family index, consistently with the fact that $\L_{\,i}$ carries
only $(s_i-1)$ space-time indices belonging to the $i$-th family.
This concise notation displays directly the reducible nature of the
gauge symmetry of mixed-symmetry fields: eq.~\eqref{gauge} clearly
allows the ``gauge-for-gauge'' transformations
\be
\delta \, \L_{\,i} = \pr^{\,j} \, \L_{\,[\,ij\,]}\,
,\label{gaugexgauge}
\ee
where the new parameters are antisymmetric under the interchange of
their indices. At two families, the process stops here, while in the
general multi-family case to be discussed in Section
\ref{sec:generalb} the chain of gauge-for-gauge transformations
continues further.

One can now introduce the Fronsdal-Labastida operator
\cite{labastida1,labastida}
\be \label{fronsdal}
\cF \, = \, \Box \, \vf \, - \, \pr^{\,i} \pr_{\,i} \, \vf + \12 \, \pr^{\,i} \pr^{\,j} \, T_{ij} \, \vf \, ,
\ee
that, as in the symmetric or one-family case, is not fully invariant
under the transformation of eq.~\eqref{gauge}. One indeed finds
\be \label{gauge_fronsdal} \d \, \cF \, =  \, \frac{1}{6} \, \pr^{\,i} \pr^{\,j} \pr^{\,k} \,
T_{(\,ij} \, \L_{\,k\,)} \, ,\ee
where the remainder on the right-hand side is proportional to the
Labastida constraints on the gauge parameters
\be
T_{(\,ij} \, \L_{\,k\,)} \, = \, 0 \, , \label{labacgaugeb}
\ee
and can be eliminated introducing compensators $\a_{ijk}$, fully
symmetric under the interchange of their three family indices and such
that
\be \label{compensator} \d \, \a_{\,ijk} \, = \, \frac{1}{3} \,
T_{(\,ij} \, \L_{\,k\,)} \, , \ee
so that for an arbitrary number of families they carry
$s_1,\ldots,(s_i-1),\ldots,(s_j-1),\ldots,(s_k-1),\ldots,s_n$
space-time indices. In analogy with the one-family case of
\cite{fs3,fms}, one can then introduce the gauge invariant tensor
\be \label{A} \cA \, = \, \cF - \12 \, \pr^{\,i} \pr^{\,j} \pr^{\,k} \, \a_{\,ijk}\ , \ee
that plays the role of a basic kinetic tensor for the unconstrained
theory. However, as anticipated in the Introduction, allowing
independent compensators $\a_{ijk}$ would lead to some difficulties,
since the very nature of the transformations
\eqref{compensator} would imply the existence of gauge-invariant
constructs built from the $\alpha_{ijk}$ alone. For instance,
\be
\delta  \left[\, T_{i\,(\,j} \, \alpha_{\,klm\,)} \, - \, T_{j\,(\,i} \,
\alpha_{\,klm\,)} \,\right] = 0 \, .
\ee
On the other hand, this gauge invariant combination vanishes identically if
the $\a_{ijk}$ are expressed in terms of other independent
compensators $\Phi_i$ according to
\be \label{newcomp} \a_{\,ijk} \, \equiv \, \a_{\,ijk}(\Phi) \, =  \,\frac{1}{3} \, T_{(\,ij} \, \F_{\,k\,)} \ , \ee
where the $\Phi_i$ transform proportionally to the gauge parameters:
\be \d \, \F_{\,k} \, = \, \L_{\,k} \ . \label{gaugephii} \ee
Indeed, given a pair of traces, the combination
\be T_{i(\,j}\, T_{kl\, )}\, = \, T_{ij}\, T_{kl} \, +\, T_{ik}\,
T_{jl} \, +\, T_{il}\, T_{jk} \ee
is also totally symmetric in $(ijkl)$. This reflects a special
property of products of identical tensors, that can only build Young
diagrams in family-index space with even numbers of boxes in each
row \footnote{Our conventions are spelled out in Appendix
\ref{app:MIX}. Further details on these matters can be found, for
instance, in \cite{branching}.}.

Notice also that the composite compensator $\alpha_{ijk}(\Phi)$ of
eq.~(\ref{newcomp}) would emerge directly if the Stueckelberg-like
shift \be \varphi \ \to \ \varphi \,-\,
\partial^{\,i}\, \Phi_{\,i} \label{stueck} \ee were performed in the
Labastida tensor ${\cal F}$, while the fact that only the
combination (\ref{newcomp}) is present reflects the original,
constrained Labastida gauge symmetry \footnote{M.A. Vasiliev
stressed to us the role of this shift in the symmetric case of
\cite{fs3}. In the following we shall see how to formulate the
procedure in various ways for mixed-symmetry tensors.}. Under the
gauge-for-gauge transformations \eqref{gaugexgauge}, these $\Phi_i$
fields shift like the gauge parameters would, and thus like ordinary
gauge fields. This is as it should be: if they were inert as the
higher-spin field $\vf$, part of the gauge transformations would be
ineffective, in contradiction with the manifest possibility of
removing the $\Phi_i$ undoing the Stueckelberg shift. Hence, the
$\a_{ijk}(\Phi)$, that for brevity from now on will be often simply
called $\a_{ijk}$, can be fully eliminated without affecting the
constrained Labastida gauge symmetry, in complete analogy with the
minimal unconstrained case.

The Bianchi identities satisfied by $\cF$,
\be \label{bianchiF} \pr_{\,i} \, \cF \, - \, \12 \, \pr^{\,j} \, T_{ij} \, \cF \, = \, - \, \frac{1}{12} \, \pr^{\,j} \pr^{\,k} \pr^{\,l} \,
T_{(\,ij}\,T_{kl\,)} \, \vf \, , \ee
are another crucial ingredient of the construction, and as in the
symmetric case the ``classical anomaly'' present in this expression
determines directly the Labastida constraints on the gauge field
$\vf$,
\be
\label{labac}
T_{(\,ij}\,T_{kl\,)} \, \vf \, = \, 0 \, .
\ee
The Bianchi identities for the $\cA$ tensor have the form
\be \label{bianchiA} \pr_{\,i} \, \cA \, - \, \12 \, \pr^{\,j} \, T_{ij} \, \cA \, = \, - \, \frac{1}{4} \, \pr^{\,j}\pr^{\,k}\pr^{\,l} \, \cC_{\,ijkl}\, , \ee
where
\be \label{C} \cC_{\,ijkl} \, = \, \frac{1}{3} \, \left\{ \, T_{(\,ij}\,T_{kl\,)}
\, \vf \, - \, 3 \, \pr_{\,(\,i}\, \a_{\,jkl\,)} \, - \, \frac{3}{2} \, \pr^{\,m}
\left(\, T_{(\,ij} \, \a_{\,kl\,)\,m} \, - \, T_{m\,(\,i} \, \a_{\,jkl\,)} \,\right) \,\right\}
\, ,
\ee
or more simply
\be
\label{C2} \cC_{\,ijkl} \, = \, \frac{1}{3} \, T_{(\,ij}\,T_{kl\,)} \left(\,
\varphi - \partial^{\,i} \, \Phi_{\,i} \,\right) \, ,
\ee
are the proper gauge-invariant extensions of the symmetrized double traces of $\vf$. Notice that arriving at this rather concise form,
that in the one-family case would reduce to the simpler expression
\be \cC_{1111} \, \to \,  \vf^{\,\pe\pe} \, - \, 4 \, \prd \a \, -
\, \pr \, \a^{\, \prime} \, \equiv \, \cC \label{c_one} \ee
of \cite{fs3,fms}, is straightforward if $\Phi_i$ is introduced via
(\ref{stueck}). It is nonetheless instructive to trace the key steps
of the derivation working in terms of $\alpha_{ijk}$, since this
requires both the symmetries of $\cC_{ijkl}$ and those of the
pre-factor and is a prototype of a number of similar calculations
needed to reproduce our results. Naively the Bianchi identity for
$\cA$ would read
\be \label{prebianchi} \pr_{\,i} \, \cA \, - \, \12 \, \pr^{\,j} \, T_{ij}
\, \cA \, = \, - \, \frac{1}{12} \, \pr^{\,j} \pr^{\,k} \pr^{\,l} \left(\,
T_{(\,ij}\,T_{kl\,)} \, \vf \, - \, 3 \, \pr_{\,(\,i}\, \a_{\,jkl\,)}
\, - \, 3 \, \pr^{\,m} \, T_{mi} \, \a_{\,jkl}  \,\right) \, , \ee
or
\be \pr_{\,i} \, \cA \, - \, \12 \, \pr^{\,j} \, T_{ij} \, \cA \, = \, - \,
\frac{1}{4} \, \pr^{\,j} \pr^{\,k} \pr^{\,l} \left(\, \cC_{\,ijkl} \, + \,
\pr^{\,m} \, \cD_{\,i,\,jkl,\,m} \,\right) \, , \ee
a result that would seem to differ from eq.~(\ref{C}) due to the
presence of the remainder
\be \cD_{\,i,\,jkl,\,m} \, = \, \12 \left(\, T_{(\,ij} \, \a_{\,kl\,)\,m} \, - \, T_{m\,(\,i} \,
\a_{\,jkl\,)} \,\right) - \, T_{mi} \, \a_{\,jkl} \, .\ee
Its family indices $jklm$, however, are projected according to a
hooked Young diagram, so that this expression actually contracts to
zero against the four derivatives present in the resulting
expression, and consequently the Bianchi identities take the form
\eqref{bianchiA}.

Notice that, once the $\a_{ijk}$ are expressed in terms of the
$\Phi_i$, the traces of the $\cC_{ijkl}$ satisfy the algebraic
constraints
\be Y_{\{5,1\}}\, T_{mn}\, \cC_{\, ijkl} \, =\, 0 \, , \label{y51tc}
\ee where $Y_{\{5,1\}}$ is a Young projector onto the irreducible
${\{5,1\}}$ representation of the permutation group acting on the
family indices. The reason, as we already stressed, is that
identical $T$ tensors can only build Young diagrams in family-index
space with even numbers of boxes in each row. This is a manifest
property of the first contribution to $\cC_{ijkl}$,
$T_{(\,ij}\,T_{kl\,)}\,\vf$, and becomes a property of all
$\cC_{ijkl}$ once the $\a_{ijk}$ are expressed in terms of
independent compensators $\Phi_i$ as in \eqref{C2}. Some non-trivial
identities follow, as for instance
\be
T_{i\,(\,j} \, \cC_{\,klmn\,)} \, = \, T_{(\,ij} \, \cC_{\,klmn\,)} \,
. \label{extC}
\ee
Let us stress again that these subtleties are by no means a
peculiarity of the unconstrained formalism, but reflect the fact
that the Labastida constraints
\eqref{labac} are nicely covariant but \emph{not} independent, so that
different traces of these equations can indeed turn out to be
proportional. The reason behind the need for relating the $\a_{ijk}$
to the $\Phi_i$ fields can be similarly traced to the lack of
independence of the Labastida constraints \eqref{labacgaugeb} on the
gauge parameters.

In complete analogy with the one-family case, the Bianchi identities
suggest to begin from the trial Lagrangian
\be \label{trial} \cL_0 \, = \, \12 \, \bra \vf \comma \cA \, - \, \12
\, \h^{ij} \, T_{ij} \, \cA \ket \, , \ee
whose gauge variation, up to a total divergence, is
\be \label{var_zero}
\begin{split}
\d \, \cL_0 \, & = \, - \, \12 \, \bra \L_{\,i} \,,\, \pr_{\,i} \, \cA \, -
\, \12 \, \pr^{\,j} \, T_{ij} \, \cA \, - \, \12 \, \h^{\,jk} \, \pr_{\,i} \, T_{jk} \, \cA \ket \\ & = \, - \, \frac{1}{24} \, \bra \pr_{\,(\,i\,}\pr_{\,j\,}\pr_{\,k}\, \L_{\,l\,)} \comma \cC_{\,ijkl}
\ket \, + \, \frac{1}{8} \, \bra T_{jk} \, \L_{\,i} \,,\, \pr_{\,i} \, T_{jk} \, \cA \ket \, .
\end{split}
\ee
It is at this point that the new subtleties of the two-family case
first emerge, since the last term in \eqref{var_zero} is
\emph{not} directly related to the gauge transformation of the $\alpha_{ijk}$
compensators. Rather, it contains a reducible tensor in index space,
$T_{ij} \, \L_{k}$, that however can be Young projected into the
$\{3\}$ and $\{2,1\}$ representations, so that
\be
\begin{split}
\label{first_proj} \d \, \cL_0 \, & = \, - \, \frac{1}{24} \, \bra \pr_{\,(\,i\,}\pr_{\,j\,}\pr_{\,k}\, \L_{\,l\,)} \comma \cC_{\,ijkl}
\ket \, + \, \frac{1}{72} \, \bra T_{(\,ij} \, \L_{\,k\,)}
\comma \pr_{\,(\,i}\, T_{jk\,)} \, \cA \ket \\
& + \, \frac{1}{24} \, \bra
T_{jk} \, \L_{\,i} \comma \left(\, 2 \, \pr_{\,i}\, T_{jk} - \, \pr_{\,(\,j}\, T_{k\,)\,i} \,\right) \cA \ket \, ,
\end{split}
\ee
where in the last term we have actually projected only the right entry of the scalar product. Out of the last two contributions, only the first can be eliminated by a term involving the compensators, but interestingly this complication is accompanied by another novelty. Indeed the trace of the Bianchi identity
now reads
\be \label{first_bianchi} \pr_{\,i} \, T_{jk} \, \cA \, - \, \12 \,
\pr_{\,(\,j}\, T_{k\,)\,i} \, \cA \, = \, \12 \, \pr^{\,l} \, T_{il} \,
T_{jk} \, \cA \, - \, \frac{1}{4} \, T_{jk} \, \pr^{\,l}\pr^{\,m}\pr^{\,n}\, \cC_{\,ilmn}  \, , \ee
and thus, in sharp contrast with the one-family case, contains divergences of single traces of $\cA$, with family
indices nicely projected in the ``hooked'' $\{2,1\}$ representation.
Notice that this expression actually admits two independent Young
projections. The first is the symmetric $\{3\}$,
\be
\pr^{\,l} \, T_{(\,ij} \,
T_{kl\,)} \, \cA \, = \, 9 \, \Box \, \pr^{\,l}\, \cC_{\,ijkl} \, + \, 3 \, \pr^{\,l}\pr^{\,m} \, \pr_{\,(\,i}\,\cC_{\,jk\,)\,lm} \, + \, \12 \, \pr^{\,l}\pr^{\,m}\pr^{\,n} \, T_{(\,ij}\, \cC_{k\,)\,lm}   \, ,
\label{bosebtriv}
\ee
that, as in the one-family case, is not particularly interesting. It
relates in fact the symmetrized double trace of the kinetic tensor
$\cA$ to the gauge invariant constraints
\eqref{C}, consistently with an algebraic identity satisfied by $\cA$
that can be simply extracted from the trace rules collected in
Appendix \ref{app:bose}, according to which
\be
T_{(\,ij}\,T_{kl\,)}\, \cA \, = \, 3 \, \left\{\, 3 \, \Box \, \cC_{\,ijkl} \, + \, \pr^{\,m}
\left(\, \pr_{\,(\,i}\, \cC_{\,jkl\,)\,m} \, - \, \pr_{\,m} \, \cC_{\,ijkl} \,\right) \, + \,
\frac{1}{2} \, \pr^{\,m}\pr^{\,n} \, T_{mn}\,\cC_{\,ijkl} \,\right\} \, .
\label{aconstrdt}
\ee
On the other hand, the $\{2,1\}$ projection is a novel feature of
the mixed-symmetry case, that first emerges at two families and
links in a non-trivial fashion divergences and gradients of the
traces of $\cA$:
\be \label{first_bianchi_proj}
\begin{split}
& \pr_{\,i} \, T_{jk} \, \cA \, - \, \12 \,
\pr_{\,(\,j}\, T_{k\,)\,i} \, \cA \, = \, \frac{1}{6} \, \pr^{\,l}
\left(\, 2 \, T_{il} \, T_{jk} - T_{i\,(\,j} \, T_{k\,)\,l} \,\right) \cA \\
& + \, \frac{1}{4} \, \pr^{\,l}\pr^{\,m} \left(\, 2\, \pr_{\,i}\,\cC_{jklm} \, - \, \pr_{\,(\,j}\,\cC_{\,k\,)\,ilm} \,\right) \, - \, \frac{1}{12}\,  \pr^{\,l}\pr^{\,m}\pr^{\,n} \left(\, 2\, T_{jk}\, \cC_{ilmn} \, - \, T_{i\,(\,j}\, \cC_{\,k\,)\,lmn} \,\right)\, .
\end{split}
\ee
As a result, the gauge variation \eqref{var_zero} can be
turned into
\be \label{var_zeroII}
\begin{split}
\d \, \cL_0 \, & = \, - \, \frac{1}{24} \, \bra \pr_{\,(\,i\,}\pr_{\,j\,}\pr_{\,k}\, \L_{\,l\,)} \, - \,
\frac{1}{9} \, \pr_{\,(\,i\,}\pr_{\,j\,}\pr_{\,k\,|} \, \h^{mn} \left(\, 2\, T_{mn} \, \L_{\,|\,l\,)} \, -
\, T_{|\,l\,)\,(\,m}\, \L_{\,n\,)} \,\right) \comma \cC_{\,ijkl} \ket \\
& + \, \frac{1}{24} \,
\bra \frac{1}{3} \, T_{(\,ij} \, \L_{\,k\,)} \comma \pr_{\,(\,i}\, T_{jk\,)} \, \cA \ket \, + \frac{1}{72}
\, \bra T_{jk} \, \L_{\,i} \,,\, \pr^{\,l} \left( \, 2
\, T_{il} \, T_{jk} \, - \, T_{i\,(\,j} \, T_{k\,)\,l} \, \right) \cA
\ket \, ,
\end{split}
\ee
where the first term only contains the constraints and therefore can
be compensated via Lagrange multipliers. In a similar fashion, the
rest can be also \emph{partly} compensated by a pair of new terms.
Only the first of these, however,
\be \label{l1} \cL_1 \, = \, - \, \frac{1}{24} \, \bra \a_{\,ijk} \comma \pr_{\,(\,i}\, T_{jk\,)} \, \cA \ket \, , \ee
has a direct analogue in the one-family case, while the second,
\be \label{l2} \cL_2 \, = \, \frac{1}{72} \, \bra \vf \,,\, \h^{ij}
\, \h^{kl} \left( \, 2 \, T_{ij} \, T_{kl} \, - \, T_{i\,(\,k} \,
T_{l\,)\,j} \, \right) \cA \ket \, , \ee
is again a genuine novelty of the two-family case, since it involves
double traces and a $\{2,1\}$ Young projection for the family
indices, automatically promoted to a $\{2,2\}$ Young projection due
to the presence of a pair of identical $T$ tensors. It should be
appreciated that this extension of the symmetry is a further
manifestation of the phenomenon mentioned after eq.~(\ref{y51tc}).

The resulting Lagrangian is still \emph{not} gauge invariant, but
its gauge variation,
\be \label{var_due}
\begin{split}
& \d \left(\cL_0 + \cL_1 + \cL_2\right) =   - \, \frac{1}{24} \,
\bra \pr_{\,(\,i\,}\pr_{\,j\,}\pr_{\,k\,|}\!
 \left[\,\L_{\,|\,l\,)} - \frac{1}{9} \, \h^{mn} \left(\, 2\, T_{mn} \, \L_{\,|\,l\,)} - T_{|\,l\,)\,(\,m}\,
 \L_{\,n\,)} \,\right)\,\right] \!,\, \cC_{\,ijkl} \ket \\
& - \, \frac{1}{864} \, \bra \left(\, 2 \, T_{jk} \, T_{lm} \, - \,
T_{j\,(\,l} \, T_{m\,)\,k} \,\right) \L_{\,i} \comma \pr_{\,i}\, \left(\, 2 \, T_{jk} \,
T_{lm} \, - \, T_{j\,(\,l} \, T_{m\,)\,k} \,\right) \cA
\ket
\, ,
\end{split}
\ee
can be compensated following steps similar to the previous ones. The
key observation is, once more, that the combination of two identical
$T$ tensors above is not only $\{2,1\}$, but actually $\{2,2\}$
Young projected. As a result, the left-hand side of the second
scalar product contains
\emph{only} two irreducible Young components, as summarized by the
diagrams
\begin{center}
\begin{picture}(200,30)(30,0)
\multiframe(0,20)(10.5,0){1}(10,10){} \multiframe(10.5,20)(10.5,0){1}(10,10){} \multiframe(0,9.5)(10.5,0){1}(10,10){}
\put(27,21){$\to$}
\multiframe(43,20)(10.5,0){1}(10,10){} \multiframe(53.5,20)(10.5,0){1}(10,10){} \multiframe(43,9.5)(10.5,0){1}(10,10){}
\multiframe(53.5,9.5)(10.5,0){1}(10,10){} \put(76,21){$\Rightarrow$} \multiframe(100,20)(10.5,0){1}(10,10){}
\multiframe(110.5,20)(10.5,0){1}(10,10){} \multiframe(100,9.5)(10.5,0){1}(10,10){} \multiframe(110.5,9.5)(10.5,0){1}(10,10){}
\put(127,21){$\otimes$} \multiframe(143,20)(10.5,0){1}(10,10){} \put(163,20){$\eq$} \multiframe(180,20)(10.5,0){1}(10,10){}
\multiframe(190.5,20)(10.5,0){1}(10,10){} \multiframe(201,20)(10.5,0){1}(10,10){} \multiframe(180,9.5)(10.5,0){1}(10,10){}
\multiframe(190.5,9.5)(10.5,0){1}(10,10){} \put(220,20){$\oplus$} \multiframe(240,20)(10.5,0){1}(10,10){}
\multiframe(250.5,20)(10.5,0){1}(10,10){} \multiframe(240,9.5)(10.5,0){1}(10,10){} \multiframe(250.5,9.5)(10.5,0){1}(10,10){}
\multiframe(240,-0.8)(10.5,0){1}(10,10){}
\put(270,20){.}
\end{picture}
\end{center}
In the general multi-family case, one would connect the $\{3,2\}$
Young projection to compensators and the $\{2,2,1\}$ Young
projection to a double trace of the Bianchi identity
\eqref{bianchiA}. However, with only two families the $\{2,2,1\}$
projection simply does not exist, since it would involve
anti-symmetrizations over \emph{three} distinct family indices. As a
result, in order to complete the two-family construction, one need
only reconstruct the gauge transformations of some compensator
traces in the $\{3,2\}$ projection. After some algebra, one finds
that
\be
\left(\, 2 \, T_{ij}\,T_{kl} \, - \, T_{i\,(\,k}\,T_{l\,)\,j} \,\right) \L_{\,m} \xrightarrow{\{3,2\}} \,
\frac{3}{4} \left(\, 3 \, T_{ij} \, \d \, \a_{\,klm} \, + \, 3 \,
T_{kl} \, \d \, \a_{\,ijm} \, - \, T_{(\,ij} \, \d
\,\a_{\,kl\,)\,m}\,\right) \label{32projection}
\, , \ee
which indeed fixes the remaining compensator terms.

Summarizing, in the two-family case the terms to be compensated with
Lagrange multipliers do not receive further corrections, and as a
result the unconstrained gauge invariant Lagrangian is
\begin{align} \label{lag}
\cL\left(\,\vf,\Phi_i,\beta_{ijkl}\,\right) \, & = \, \12 \, \bra \vf \comma \cA \, - \, \12 \, \h^{ij} \, T_{ij} \, \cA \, +
\, \frac{1}{36} \, \h^{ij} \, \h^{kl} \left(\, 2 \, T_{ij} \, T_{kl} \, - \,
T_{i\,(\,k} \, T_{l\,)\,j} \,\right) \cA \ket \nonumber \\
& - \, \frac{1}{24} \, \bra \a_{\,ijk}(\Phi) \comma \pr_{\,(\,i} \, T_{jk\,)} \,
\cA \, - \, \frac{1}{12} \, \h^{lm} \left(\,
2 \, \pr_{\,(\,i}\, T_{jk\,)}\, T_{lm} - \, \pr_{\,(\,i\,|}\, T_{l\,|\,j} \, T_{k\,)\,m} \,\right) \cA \ket \nonumber \\
& + \, \frac{1}{8} \, \bra \b_{\,ijkl} \comma \cC_{\,ijkl} \ket \, ,
\end{align}
where for the sake of clarity we have stressed once more that here
the $\alpha_{ijk}$ are to be regarded as functions of the $\Phi_i$.
Furthermore, the $\b_{ijkl}$ are Lagrange multipliers, whose gauge
transformations,
\be \label{mult}
\begin{split}
\d \, \b_{\,ijkl} \, = & \, \frac{1}{4} \, \Big\{ \, \pr_{\,(\,i}\, \pr_{\,j}\,
\pr_{\,k}\, \L_{\,l\,)} \, + \, \frac{1}{9} \, \pr^{\,m} \, \pr_{\,(\,i}\, \pr_{\,j\,|} \left(\, 2 \, T_{|\,kl\,)} \, \L_{\,m} \, - \, T_{m\,|\,k} \, \L_{\,l\,)} \,\right) \\
& \, - \, \frac{1}{9} \, \h^{mn} \, \pr_{\,(\,i}\, \pr_{\,j}\,
\pr_{\,k\,|} \left(\, 2\, T_{mn} \, \L_{|\,l\,)} \, - \, T_{|\,l\,)\,(\,m}
\, \L_{\,n\,)} \,\right) \, \Big\} \ ,
\end{split}
\ee
can be obtained from the term of \eqref{var_due} that contains
$\cC_{ijkl}$. Notice, however, that when this transformation is
inserted in the Lagrangian
\eqref{lag}, the $\eta^{mn}$ in the last term gives rise to a
contribution involving the trace $T_{mn}\,
\cC_{ijkl}$. Only the resulting $\{4,2\}$ projection of the last term
is then effective on the $\cC_{ijkl}$, on account of the constraint
\eqref{y51tc}. Hence, eq.~(\ref{mult}) could well be presented in a
different form, restricting the last term directly to its $\{4,2\}$
Young projection. Interestingly, this also implies that, if the
$\a_{ijk}$ are expressed in terms of the independent compensators
$\Phi_i$ via eq.~\eqref{newcomp}, the Lagrangian
\eqref{lag} possesses a further gauge symmetry, related to shifts of
the Lagrange multipliers of the type
\be
\delta \, \b_{\,ijkl} \, = \, \eta^{mn}\, L_{\,ijkl,\,mn} \, , \label{betalsym}
\ee
where $L_{\,ijkl,mn}$ is $\{5,1\}$ projected in its family indices.
Indeed, this shift would produce the contribution
\be
\delta \, \cL = \frac{1}{16} \ \bra L_{\,ijkl,\,mn} \comma T_{mn}\,
\cC_{\,ijkl} \ket \, ,
\ee
an expression that vanishes on account of the constraints
\eqref{y51tc} on the $\cC_{ijkl}$, that hold once the $\a_{ijk}$ are
expressed in terms of the $\Phi_i$.

It is actually possible to recast the Lagrangian \eqref{lag} in an
alternative form that will soon prove convenient to derive its field
equations. To this end, let us consider a field $\phi$ that is
\emph{not} subject to the double trace constraints
\eqref{labac}, but whose gauge parameters $\L_i$ are still
constrained according to \eqref{labacgaugeb}. A convenient
presentation of the corresponding Lagrangian is then
\begin{align}
\cL_{\,C}\left(\,\phi,\gamma_{\,ijkl}\,\right) \, & = \, \12 \, \bra \phi \comma \cF(\phi) \, - \, \12 \, \h^{ij}
\, T_{ij} \, \cF(\phi) \, +
\, \frac{1}{36} \, \h^{ij} \, \h^{kl} \left(\, 2 \, T_{ij} \, T_{kl} \, - \,
T_{i\,(\,k} \, T_{l\,)\,j} \,\right) \cF(\phi) \ket \nonumber \\
& + \, \frac{1}{24} \, \bra \g_{\,ijkl} \comma T_{(\,ij}\,T_{kl\,)}\,
\phi \ket\, \label{stuecklag} \, ,
\end{align}
that differs from the Labastida form of \cite{labastida} simply
because here gauge invariance requires projected traces and Lagrange
multipliers $\gamma_{ijkl}$. These fields enforce the double trace
constraints, and their gauge transformations are actually those
given for the $\beta_{ijkl}$ in eq.~\eqref{mult},
\be \label{multgamma}
\begin{split}
\d \, \g_{\,ijkl} \, = & \, \frac{1}{4} \, \Big\{ \, \pr_{\,(\,i}\, \pr_{\,j}\,
\pr_{\,k}\, \L_{\,l\,)} \, + \, \frac{1}{9} \, \pr^{\,m} \, \pr_{\,(\,i}\, \pr_{\,j\,|} \left(\, 2 \, T_{|\,kl\,)} \, \L_{\,m} \, - \, T_{m\,|\,k} \, \L_{\,l\,)} \,\right) \\
& \, - \, \frac{1}{9} \, \h^{mn} \, \pr_{\,(\,i}\, \pr_{\,j}\,
\pr_{\,k\,|} \left(\, 2\, T_{mn} \, \L_{|\,l\,)} \, - \, T_{|\,l\,)\,(\,m}
\, \L_{\,n\,)} \,\right) \, \Big\} \ .
\end{split}
\ee
The relation between the two Lagrangians of eqs.~\eqref{lag} and
\eqref{stuecklag} is then
\be
\cL\left(\,\vf,\Phi_{i},\beta_{\,ijkl}\,\right)\, = \, \cL_{\,C}\left(\, \vf \,-\, \partial^{\,i}\,
\Phi_{i} \comma \beta_{\,ijkl}\,-\,\Delta_{\,ijkl}(\Phi)\,\right)\, ,
\label{lcunc}
\ee
where
\be
\Delta_{\,ijkl}(\Phi) \, = \, \frac{1}{4} \
\pr_{\,(\,i\,}\pr_{\,j\,}\pr_{\,k\,|} \left[\, \Phi_{\,|\,l\,)} \, -
\, \frac{1}{9} \ \h^{mn} \left(\, 2\, T_{mn} \, \Phi_{\,|\,l\,)}
\, - \, T_{|\,l\,)\,(\,m} \, \Phi_{\,n\,)} \,\right) \,\right] \, .
\label{stueckmult}
\ee

Finally, in the one-family case the Lagrangian (\ref{lag}) reduces
to the result of \cite{fs3,fms}, that in this notation would read
\be {\cal L} \, = \, \frac{1}{2} \, \bra \varphi \comma \cA - \frac{1}{2} \, \eta \, \cA^{\; \prime} \ket \, - \, \frac{1}{8}\, \bra \alpha
\comma
\partial \cdot \cA^{\; \prime} \ket \, + \, \frac{1}{8} \, \bra \beta \comma \cC \ket \ ,  \label{lagone}
\ee
where ``primes'' denote traces and the symbol $\pr\, \cdot$, to be
identified with $\pr_{\, 1}$, is used to denote a divergence.

One could also work with formally independent $\alpha_{ijk}$, adding
to the Lagrangian \eqref{lag} new
\emph{gauge invariant} multipliers $\ell_{\,ijk}$, according to
\be \label{lagextra}
\cL_{4} \, = \, \bra \ell_{\,ijk} \comma \frac{1}{3} \, T_{(\,ij}\,\F_{\,k\,)} \, - \, \a_{\,ijk} \ket \ .
\ee
Let us stress, however, that in this case the $\{5,1 \}$ part of the
last term in the gauge transformation of the $\b_{ijkl}$ would be
also effective, since it would couple to the gauge invariant
combination
\be \label{gaugeinv}
\begin{split}
& Y_{\{5,1\}}\, T_{mn}\, \cC_{\,ijkl} \, \sim \ \pr^{\,p} \Big\{ \,
\left(
\, T_{n(\,i} \, T_{jk} \,
\a_{lm\,)p} \, + \, T_{m(\,i} \, T_{jk} \, \a_{ln\,)p} \, \right) \\
& - \,2 \left( \, T_{(\,ij} \, T_{kl} \, \a_{m\,)np} \,  + \, T_{(\,ij} \, T_{kl} \, \a_{n\,)mp} \, \right) + \, \left( \, T_{p\,(\,i} \, T_{jk} \, \a_{lm\,)n} \, + \,
T_{p\,(\,i} \, T_{jk} \, \a_{ln\,)m} \, \right) \\
& - \,2 \left( \, T_{p\,(\,i\,|} \, T_{n\,|j} \, \a_{klm\,)} \, + \, T_{p\,(\,i\,|} \, T_{m\,|j} \, \a_{kln\,)} \, \right) + \, \left(\, T_{pn} \, T_{(\,ij} \, \a_{klm\,)} \, + \, T_{pm} \,
T_{(\,ij} \, \a_{kln\,)} \,\right)\, \Big\}\, ,
\end{split}
\ee
that would not vanish when working with independent $\a_{ijk}$. On
shell, however, one would return anyway to $\cC_{ijkl}$ tensors
subject to the proper constraints \eqref{y51tc}, that as we stressed
are driven by the double traces of $\varphi$ that they contain.
Notice that in this way of formulating the theory the symmetry of
eq.~\eqref{betalsym} would extend to a simultaneous redefinition of
the $\b_{ijkl}$ and $\ell_{ijk}$,
\begin{alignat}{2}
& \delta \, \b_{\,ijkl} \, &&=\, \h^{\,mn} \, L_{\,ijkl,\,mn} \, , \nonumber \\
& \delta \, \ell_{\,ijk} \, &&= \, \frac{1}{4} \
\h^{\,mn}\,\h^{\,pq} \left(\, \pr_{\,(\,i}\, L_{\,jk\,)\,mn,\,pq} \,
- \, \pr_{\,(\,m}\, L_{\,n\,)\,ijk,\,pq} \,\right) \, .
\end{alignat}

There is an issue of completeness for the Lagrangian
\eqref{lag}, since further terms could be added to it, in principle
at least. Considerations of this type were already made in
\cite{fms} for the one-family case, with the conclusion that no
significant additions were actually possible. The reason was that
this type of terms would contribute to the field equation
proportionally to the constraint $\cC$, and would thus be
ineffective by virtue of the field equation for the Lagrange
multiplier $\b$. In the mixed-symmetry case, however, there are more
possibilities, simply because not all double traces of $\vf$ are
subject to constraints for general mixed-symmetry fields. A
potentially interesting term of a new type is for instance
\be \label{redefbeta}
\bra S^{\,m}{}_{(\,i}\, S^{\,n}{}_{j\, |} \ Y_{\{2,2\}}\, T_{|\, kl\,)}\,T_{mn} \, \cA \comma \cC_{\,ijkl} \ket
\, ,
\ee
where the $S^{\,i}{}_{j}$ operators, a key novelty of the $N$-family
case, are defined in Appendix \ref{app:MIX}. Their commutators close
on a $gl(N)$ algebra, and their net effect for $i
\neq j$ is to displace indices from one family to another. Thanks
to the presence of the $S^{\,i}{}_{j}$ operators, the double trace
above can carry a $\{2,2\}$ ``window'' projection, that is not
proportional to the constraint tensors and yet can appear in a
combination that is totally symmetric in $(ijkl)$. As a result,
terms of this type do give a contribution to the $\vf$ equation that
does not vanish when the equation for the Lagrange multiplier is
enforced. They are still irrelevant, however, since their net effect
is a mere redefinition of the multipliers by gauge invariant
quantities. More generally, such a wide freedom to redefine the
multipliers can be generalized to include terms which are not
necessarily themselves gauge invariant. This has interesting
consequences both for the Lagrangian and for the field equations,
including the possibility of working from the beginning with gauge
invariant Lagrange multipliers, an option which calls for a
reshuffling of the terms in $\cL$, and will be discussed in Section
\ref{sec:lagrangianb}.

\vskip 24pt

%%%%%%%%%%%%%%%%%%%%%%%%%%%%%%%%%%%%%%%%%%%%%%

\scss{The field equations}\label{sec:motion2b}

%%%%%%%%%%%%%%%%%%%%%%%%%%%%%%%%%%%%%%%%%%%%%%

Varying the Lagrangian of eq.~\eqref{lag} and
using repeatedly the trace rules collected in Appendix
\ref{app:bose} yields the equation of motion for the gauge
field $\varphi$,
\be \label{ep}
\begin{split}
E_\vf \, & : \ \cA \, - \, \12 \ \h^{ij} \, T_{ij} \, \cA \, + \, \frac{1}{36} \ \h^{ij} \, \h^{kl} \left(\, 2\,T_{ij}\,T_{kl} \,-\, T_{i\,(\,k}\,T_{l\,)\,j} \,\right) \cA  \\
& + \, \frac{1}{8} \ \h^{ij} \, \pr^{\,k}\pr^{\,l} \, \cC_{\,ijkl} \, - \, \frac{1}{96} \ \h^{ij} \, \h^{kl} \, \pr^{\,m}\pr^{\,n} \left(\, 2\,T_{ij}\,\cC_{\,klmn} \, - \, T_{i\,(\,k}\,\cC_{\,l\,)\,jmn} \,\right) \\
& + \, \12 \ \h^{ij} \, \h^{kl} \, \cB_{\,ijkl} \, = 0 \, ,
\end{split}
\ee
where
\be
\begin{split} \label{bijkl}
\cB_{\,ijkl} \, & \equiv \, \b_{\,ijkl} \, - \, \frac{1}{12} \, \Big\{\, T_{(\,ij} \, \pr_{\,k\,}\pr_{\,l\,)}
\, \vf \, - \, 3 \, \Box \, \pr_{\,(\,i} \, \a_{jkl\,)} - \, \pr^{\,m} \, \pr_{\,(\,i\,}\pr_{\,j}\,\a_{\,kl\,)\,m}
\,\Big\} \\
& + \ \frac{1}{144} \ \h^{mn} \, \Big\{\, 2 \,
T_{mn}\,T_{(\,ij}\,\pr_{\,k\,}\pr_{\,l\,)} \, \vf \, -
\, T_{m\,(\,i\,|}\,T_{n\,|\,j}\,\pr_{\,k\,}\pr_{\,l\,)} \, \vf \,\Big\} \\
& - \ \frac{1}{96} \ \h^{mn} \, \Box \, \Big\{\, 3 \, T_{mn} \,
\pr_{\,(\,i}\,\a_{\,jkl\,)} \, - \, 2
\, T_{m\,(\,i}\,\pr_{\,j}\,\a_{\,kl\,)\,n} \, + \, T_{(\,ij}\,\pr_{\,k}\,\a_{\,l\,)\,mn} \,\Big\} \\
& + \ \frac{1}{96} \ \h^{mn} \, \pr^{\,p} \, \Big\{\, T_{mn} \,
\pr_{\,(\,i\,}\pr_{\,j}\,\a_{\,kl\,)\,p} \, -
\, T_{m\,(\,i}\,\pr_{\,j\,}\pr_{\,k}\,\a_{\,l\,)\,np} \, + \, T_{(\,ij}\,\pr_{\,k\,}\pr_{\,l\,)}\,\a_{\,mnp}
\,\Big\} \, .
\end{split}
\ee
Notice that for fully symmetric tensors eq.~\eqref{ep} reduces to
\be
E_{\vf}  \, : \, \cA\, - \, \12 \, \h\, \cA^{\, \pe} \, +
\, \fr{1}{4} \, \h \, \pr^{\, 2} \, \cC \,
+ \, \h^{\, 2} \, \cB \, = \, 0 \, ,
\ee
where $\pr^{\,2}= \frac{1}{2}\, \pr \, \pr$ and $\eta^{\,2}=
\frac{1}{2} \, \eta \, \eta$ are defined according to the rules of
\cite{fs3,fms}, and the reader may easily sort out these terms
in the original expression.

The considerations of the previous section have an interesting link
with the nature of the $\cB_{ijkl}$. Indeed, the space-time tensors
collected in (\ref{bijkl}) are invariant under the gauge
transformations of eqs.~\eqref{gauge},
\eqref{compensator} and
\eqref{mult}, as in the one-family case of \cite{fs3,fms},
\emph{only} if the gauge transformation (\ref{mult}) is
restricted, in its $\eta$ -- dependent terms, to the $\{4,2\}$
projection, thus eliminating all $\{5,1\}$ contributions, which is
the case precisely if the $\a_{\,ijk}$ are expressed in terms of the
independent compensators $\Phi_i$ as in eq.~\eqref{newcomp}. Let us
also stress that $\cB_{\,ijkl}$ varies under the new gauge
transformation
\eqref{betalsym} as
\be \label{Bshift}
\delta \, \cB_{\,ijkl} \,=\, \eta^{mn}\, L_{\,ijkl,\,mn} \, ,
\ee
but the $\vf$ field equation is properly gauge invariant
nonetheless, since as we have pointed out the three $\eta$'s that
would accompany $L_{\,ijkl,\,mn}$ in the variation of eq.~\eqref{ep}
simply can not build a $\{5,1\}$ projection. Indeed, this argument
also shows that the combination $\eta^{ij}\,
\eta^{kl}\,\cB_{\,ijkl}$, that appears in the $\vf$ equation, is in
any case invariant under the complete transformation of
eq.~\eqref{mult}. This clearly reflects the gauge symmetry of
eq.~\eqref{ep}, that the Lagrangian \eqref{lag} guarantees
regardless of the possibility, discussed in the previous subsection,
of introducing the Lagrange multipliers $\ell_{ijk}$ in order to
allow for independent $\alpha_{ijk}$.

In addition, varying in (\ref{lag}) the Lagrange multipliers yields
the double-trace constraints
\be
E_\beta \, : \ \frac{1}{8} \ \cC_{ijkl} \, = \, 0 \label{ebeta}
\, .
\ee
On the other hand, the field equations for the compensators
$\Phi_{i}$ are
\be
\begin{split}
E_{\,\Phi} \, & : \ \h^{jk} \Big\{- \, \frac{1}{12} \ T_{(\,ij} \, \pr_{\,k\,)} \, \cA \, + \, \frac{1}{144} \ \h^{lm} \left(\, 2 \, T_{lm}\,T_{(\,ij}\,\pr_{\,k\,)} \, \cA \, - \, T_{l\,(\,i\,|}\,T_{m\,|\,j}\,\pr_{\,k\,)}\, \cA \,\right) \\
& + \, \frac{1}{8} \ \Box \, \pr^{\,l} \, \cC_{\,ijkl} \, + \, \frac{1}{48} \ \pr^{\,l}\pr^{\,m} \, \pr_{\,(\,i}\, \cC_{\,jk\,)\,lm} \\
& - \, \frac{1}{192} \ \h^{lm} \, \Box \, \pr^{\,n} \left(\, 3 \,T_{lm}\,\cC_{\,ijkn} + T_{(\,ij}\,\cC_{\,k\,)\,lmn } \,-\, 2\, T_{l\,(i\,}\,\cC_{\,jk\,)\,mn} \,\right) \\
& - \, \frac{1}{384} \ \h^{lm} \pr^{\,n}\pr^{\,p} \left(\, T_{lm}\,\pr_{\,(\,i}\, \cC_{\,jk\,)\,np} + T_{(\,ij}\, \pr_{\,k\,)}\, \cC_{\,lmnp} \,-\, T_{l\,(\,i}\, \pr_{\,j} \, \cC_{\,k\,)\,mnp} \,\right) \\
& + \, \12 \Big(\, \pr^{\,l} \, \cB_{\,ijkl} \, + \, \frac{1}{2} \
\h^{lm} \left(\, \pr_{\,(\,i}\, \cB_{jk\,)\,lm} \, - \,
\pr_{\,(\,l}\, \cB_{\,m\,)\,ijk} \,\right) \, \Big) \, \Big\} = \, 0
\, \label{ephi}
.
\end{split}
\ee

The rewriting of the Lagrangian \eqref{lag} based on the
redefinitions in eqs.~\eqref{lcunc} and \eqref{stueckmult} is
particularly convenient to derive these results. Indeed, varying
eq.~\eqref{stuecklag} one obtains
\be \label{varc-unc}
\begin{split}
& \d \, \cL_C \, = \, \bra \d \, \phi \comma E_{\phi} \ket \, + \,
\bra \d \, \g_{\,ijkl} \comma (E_{\,\g})_{\,ijkl} \ket \, = \, \bra
\d \, \vf \comma E_{\,\phi} \ket \, + \, \bra \d \, \b_{\,ijkl}
\comma (E_{\,\g})_{\,ijkl} \ket \\
& + \, \bra \d \, \Phi_{\,i} \comma \pr_{\,i} \, E_{\phi} \, + \,
\pr^{\,j}\pr^{\,k}\pr^{\,l} \, (E_{\,\g})_{\,ijkl} \, + \, \frac{1}{3}
\, \h^{\,jk\,}\pr^{\,l}\pr^{\,m} \left(\, 2\, \pr_{\,i}\,
(E_{\,\g})_{\,jklm} \, - \, \pr_{\,(\,j}\,(E_{\,\g})_{\,k\,)\,ilm}
\,\right) \\
& - \, \frac{1}{9} \, \h^{\,jk} \, \pr^{\,l}\pr^{\,m}\pr^{\,n} \left(\, 2\, T_{jk} \, (E_{\,\g})_{\,ilmn} \, - \, T_{i\,(\,j}\, (E_{\,\g})_{\,k\,)\,lmn}\,\right) \ket\, ,
\end{split}
\ee
so that
\begin{alignat}{2}
& E_{\vf} & & = \, E_\phi \nonumber \, ,\\
& E_{\Phi} & & = \, \pr_{\,i} \, E_{\phi} \, + \,
\pr^{\,j}\pr^{\,k}\pr^{\,l} \, (E_{\,\g})_{\,ijkl} \, + \, \frac{1}{3}
\, \h^{\,jk\,}\pr^{\,l}\pr^{\,m} \left(\, 2\, \pr_{\,i}\,
(E_{\,\g})_{\,jklm} \, - \, \pr_{\,(\,j}\,(E_{\,\g})_{\,k\,)\,ilm}
\,\right)\, ,\nonumber \\
& & & - \, \frac{1}{9} \, \h^{\,jk} \, \pr^{\,l}\pr^{\,m}\pr^{\,n} \left(\, 2\, T_{jk} \, (E_{\,\g})_{\,ilmn} \, - \, T_{i\,(\,j}\, (E_{\,\g})_{\,k\,)\,lmn}\,\right) \nonumber \\
& E_\beta & & = \, E_\g \, ,
\label{eqsshift}
\end{alignat}
where in these expressions $\phi$ and the $\g_{ijkl}$ are to be
expressed in terms of $\vf$ and the $\b_{ijkl}$ via
eqs.~\eqref{lcunc} and
\eqref{stueckmult}. In this fashion one can obtain all field
equations starting from the $\vf$-dependent terms of eqs.~\eqref{ep}
and \eqref{ebeta}. In particular, the $\cB_{ijkl}$ tensors can be
obtained from
\be
\cB_{\,ijkl} \, = \, \g_{\,ijkl} \, - \, \frac{1}{12} \
\pr_{\,(\,i\,}\pr_{\,j}\, T_{kl\,)} \, \phi \, + \, \frac{1}{144} \
\h^{mn} \, \pr_{\,(\,i\,}\pr_{\,j\,|} \left(\, 2\, T_{|\,kl\,)}\,
T_{mn} \,-\, T_{m\,|\,k}\,T_{l\,)\,n} \,\right) \phi\, ,
\label{bijkl2}
\ee
that after the shift \eqref{lcunc} becomes manifestly gauge
invariant since this property holds for both $\g_{ijkl}$ and $\phi$.
The apparent contradiction with the considerations made after
eq.~\eqref{bijkl} is resolved observing that eq.~\eqref{bijkl2}
describes a different form of the $\cB_{ijkl}$, that in general is
\emph{not} fully expressible in terms of $\vf$, the $\a_{ijk}$ and
the $\b_{ijkl}$. The choice \eqref{bijkl} for the $\cB_{ijkl}$ can
be recovered, however, if the $\eta$ part of the gauge
transformation of the $\b_{ijkl}$, and consequently of the shift
\eqref{stueckmult} of the $\g_{ijkl}$, is restricted to its $(4,2)$
component, the only relevant one on account of the shift symmetry
\eqref{betalsym}.

The field equations for the compensators can be derived from
eqs.~\eqref{eqsshift}, which provide a condition that is tantamount
to the conservation of external currents coupling to $\vf$.
Furthermore, as in the symmetric case of
\cite{fms}, it is also possible to rewrite the Lagrangian in the form
\be
\cL \, = \, \12 \ \bra \vf \comma E_\vf \ket \, + \, \12 \ \bra \Phi_i \comma
(E_\Phi)_{\,i} \ket \, + \, \12 \ \bra \b_{\,ijkl} \comma
(E_\b)_{\,ijkl} \ket \, ,
\ee
taking into account the fact that the $\phi$ terms present in
$\cB_{ijkl}$ are the adjoints of the $\cC_{ijkl}$ terms in
eq.~\eqref{ep}, as one can recognize varying eq.~\eqref{stuecklag}.

Let us conclude this section by stressing that these field equations
were simply obtained varying the Lagrangians of eq.~\eqref{lag} with
respect to gauge fields, compensators and Lagrange multipliers. This
was possible since, in our formulation, all these fields are
\emph{unconstrained}. On the other hand, when deriving the field
equations of the Labastida theory one is confronted with gauge
fields $\vf$ that are subject to the double-trace constraints
\eqref{labac}. This was indeed noticed in \cite{labastida}, where the author
however seems to conclude that the problem does not present itself
because he apparently expected his constrained Einstein-like tensors
\be\label{labaeinstein}
\cE_{\,\vf} \, = \, \cF \, - \, \12 \, \h^{ij}
\, T_{ij} \, \cF \, +
\, \frac{1}{36} \, \h^{ij} \, \h^{kl} \left(\, 2 \, T_{ij} \, T_{kl} \, - \,
T_{i\,(\,k} \, T_{l\,)\,j} \,\right) \cF
\ee
to have vanishing symmetrized double traces, as in Fronsdal's case.
However, the $S^{\,i}{}_j$ operators bring about a surprise, since
already for two families one finds
\be
T_{(\,ij}\,T_{kl\,)}\, \cE_{\,\vf} \, = \, - \,
\frac{1}{36} \, S^{\,m}{}_{(\,i}\, S^{\,n}{}_{j\,|} \left(\, 2\,
T_{|\,kl\,)}\,T_{mn} \, - \, T_{m\,|\,k}\,T_{l\,)\,n}  \,\right) \cF
\, , \label{s_symdt}
\ee
and as a result the actual Lagrangian equations \emph{do} require a
projection. We shall see an explicit example of this fact in Section
\ref{sec:examples2b} in the case of a reducible
rank-$(4,2)$ tensor.

Alternatively, the structure of the unconstrained formulation suggests to recast even the variation
\be
\d\, \cL \, = \, \bra \d \, \vf \comma \cE_{\,\vf}  \ket
\ee
 of the constrained
Labastida Lagrangian defining its field equations in the form
\be
\cE_{\,\vf} \, + \, \frac{1}{2} \ \eta^{ij}\, \eta^{kl}\, \cB_{\,ijkl} \, = \, 0 \, , \label{eqfull}
\ee
with an additional contribution $\cB_{ijkl}$ that is
\emph{totally symmetric} in its family indices, simply because such a term
would be orthogonal to $\d \vf$ in view of eq.~\eqref{labac}. Notice
that this is precisely the type of expression that one would obtain
gauge fixing eq.~\eqref{ep}. Now, once combined with its symmetrized
double traces, the modified field equation \eqref{eqfull} would
rebuild the non-trivial projection precisely because, according to
eq.~\eqref{s_symdt}, the symmetrized double traces of $\cE_\vf$ do
not vanish identically beyond one family.

\vskip 24pt

%%%%%%%%%%%%%%%%%%%%%%%%%%%%%%%%%%%%%%%%%%%%%%%%%%%%%%%%%%%%%%%%

\scss{On-shell reduction to ${\cal F}=0$}\label{sec:reduction2b}

%%%%%%%%%%%%%%%%%%%%%%%%%%%%%%%%%%%%%%%%%%%%%%%%%%%%%%%%%%%%%%%%

We can now turn to examine whether eq.~(\ref{ep}) can be reduced to the
non-Lagrangian Labastida form
\be \label{labeq}
\cF \, = \, 0 \, ,
\ee
with $\cF$ defined in (\ref{fronsdal}). We will show that \eqref{labeq}
can always be recovered either directly from the equations of motion or after using
additional local symmetries that emerge in special cases. It is perhaps the case
to elaborate on the meaning of the procedure, so as to clarify the spirit of the ensuing
analysis.

Eq.~\eqref{labeq} has the net effect of reducing the $gl (D)$
tensors appearing in the Lagrangians to their $so (D - 2)$
counterparts. However, for particular classes of $gl (D)$ tensors,
the relevant $so (D - 2)$ representations may simply not exist. And
indeed, a general result of representation theory (see, for
instance, the first reference in \cite{branching},
\textsection~$10$-$6$) implies that, for $so (n)$ groups, if the
total number of boxes in the first two columns of a tableau exceeds
$n$, the corresponding transverse tensor vanishes. According to this
theorem, one can thus conclude that in $D = 2$ eq.~\eqref{labeq} can
never describe propagating degrees of freedom (beyond the scalar
case), while in $D = 3$ only the vector representation $(1, 0)$
would be non trivial on-shell. In a similar fashion, in $D = 4$
non-vanishing $so (2)$ polarizations are only carried by symmetric
tensors of type $(s, 0)$ and by the $2$-form (dual to a scalar),
while in $D = 5$ one can also add to the latter cases the $(s, 1)$
tensors. Finally, for $D \geq 6$ all $gl (D)$ representations  would
reduce via \eqref{labeq} to non-trivial representations of $so (D -
2)$.

It should be stressed, however, that even the cases with no propagating degrees of freedom
are of some interest, since they include a number of topological models that generalize the rich and instructive example of two-dimensional gravity. And indeed a neat pattern of theories of this type will emerge from our analysis, together with other cases that are specific of mixed symmetry fields, as we shall illustrate in detail in what follows.

In order to introduce the general features of our procedure, let us begin by recalling
how the same problem was dealt with in \cite{fs3,fms} for the
simpler case of symmetric tensors. The Lagrangian of
eq.~\eqref{lagone} yields the single-family counterparts of the
field equations (\ref{ep}), (\ref{ebeta}) and (\ref{ephi}). These
include the double-trace constraint
\be
\cC \, \equiv \, \varphi^{\;\prime\prime} \,-\, 4\, \partial \cdot \alpha
\,-\,
\partial\, \alpha^{\;\prime}\, = \, 0  \, , \label{betaone}
\ee
the field equation for the single Lagrange multiplier
$\beta_{1111}$ present in this class of models, in the following
simply called $\beta$ as in
\cite{fs3,fms}. In a similar fashion, after making use of
eq.~\eqref{betaone} the field equation for the physical gauge field
$\vf$ takes the form
\be  \label{eomsym}
\cA \, - \, \12 \ \h \, \cA^{\, \pe} \, + \, \frac{1}{2}\ \h \, \h \, \cB \, = \, 0 \, ,
\ee
where
\be
\cB \, \equiv \, \b \, - \, \12 \ \prd \prd \vf^{\, \pe} \, + \, \Box \, \prd \a \, +
\, \12 \ \pr \, \prd \prd \a
\ee
is the gauge-invariant completion of $\beta$.

In order to reduce \eqref{eomsym} to the non-Lagrangian equation
$\cA = 0$, one can begin by noticing that $\cA$ becomes
\emph{doubly traceless} once the constraint
\eqref{betaone} is enforced. Taking multiple traces of
(\ref{eomsym}) one can then generate equations involving only $\cB$
and its traces that imply that the whole tensor $\cB$ vanishes on
shell \cite{fs3}. As a result, it is generally a simple matter to
turn eq.~\eqref{eomsym} into $\cA=0$, and finally into $\cF=0$, with
a proper gauge choice eliminating the compensator $\a$.
Alternatively, one could decompose (\ref{eomsym}) into a set of
relations involving the irreducible $o(1,d-1)$ components of the
tensors appearing in (\ref{eomsym}), letting for
instance \footnote{Here $\cB^{\, [k]}_{\,T}$ denotes the traceless
part of the $k$-th trace of $\cB$, up to an overall coefficient. In
addition, as in
\cite{fs3,fms}, $\eta^k$ denotes the product of $k$ Minkowski metric
tensors, symmetrized via the minimum possible number of terms, so
that for example $\eta \, \eta = 2\, \eta^2$.}
\be \label{tracebsym}
\cB \, = \, \cB_{\,T} \, + \, \h \, \cB^{\, \pe}_{\,T} \, +
\, \dots \, + \, \h^{\, k} \, \cB^{\, [k]}_{\,T} \, + \, \dots \ . \ee
In view of the double tracelessness of $\cA$, eq.~(\ref{eomsym})
would then imply the system
\be
\begin{cases}
\cA_{\,T} \,=\, 0  \ , \\[5pt]
(\, D-6+2\,s\, )\, \cA^{\, \pe}_{\,T} \, = \, 0 \ , \\[5pt]
\cB^{\, [k]}_{\,T} \, = \, 0 \ , \hspace{5cm} k \, = \, 0 \, , \dots , [\fr{s - 4}{2} ] \
, \label{decompsym}
\end{cases}
\ee
that would lead in all cases to the two results one is after, namely
\be
\cA \, = \, 0 \ , \qquad \cB\, = \, 0 \, ,\ee
but for a single exception.

This corresponds to two-dimensional linearized gravity, and is
characterized by the values $s=2$ and $D=2$. It actually requires
some discussion, since in this case the pre-factor in the second
equation vanishes. What happens, however, is a familiar story:
$\cA^{\,
\pe}_{\,T}$ is left undetermined, but the field equation
\eqref{eomsym} acquires at the same time the additional gauge symmetry
\be
\d\, \vf_{\m\n} \, = \, \h_{\m\n} \, \Omega\, .
\ee
This is the linearized Weyl symmetry, that as is well known plays a
key role in String Theory. More precisely, the Einstein-Hilbert
action is a total divergence in two dimensions, and as a result the
Einstein tensor vanishes, so that \emph{any} shift of the form
\be
\delta\, \vf_{\m\n} \, = \, \x_{\,\m\n}
\ee
is actually a symmetry, that in this case can nonetheless be
parametrized via Weyl shifts and linearized diffeomorphisms. We are
stressing the role of the Weyl symmetry since, as we shall see
shortly, it affords important generalizations for higher-spin fields
of mixed symmetry.

All these features of the reduction have direct counterparts in our
present, more general setting. For tensors of mixed symmetry one can
proceed along similar lines, but several new features emerge,
together with technical complications that require more
sophisticated tools. To begin with, not all double traces of $\cA$
are removed in general, even when the constraint equations $\cC_{\,
i j k l} = 0$ are enforced. As a result, the symmetrized double
traces of the Einstein-like tensors do not vanish at two families,
and one
\emph{can not} separate $\cA$ and the $\cB_{\, i j k l}$ by simply taking multiple
traces of eq.~\eqref{ep}. In addition, the $\cB_{\, i j k l}$
\emph{do not} fully vanish on-shell. Rather, as we shall see, some
of their components are not even determined, but still do not
contribute to the $\vf$ field equation, due to the shift symmetry
discussed in section
\ref{sec:motion2b}. More generally, the key novelties of two-family
gauge fields reflect the
\emph{non-Abelian} nature of the $S^{\,i}{}_j$
operators of Appendix \ref{app:MIX}, that do not appear in the
Lagrangians and in the corresponding field equations but are
ubiquitous in their traces.

As for symmetric tensors, let us begin by considering the equation
of motion (\ref{ep}) for the gauge field $\vf$,
\be \label{motion_noC}
E_{\, \vf} \, \equiv \, \cA \, - \, \12 \, \h^{\,ij}\, \cA^{\,\pe}{}_{ij} \, + \, \frac{1}{12} \, \h^{\,ij}\, \h^{\,kl} \,
\cA^{\,\pe\pe}{}_{ij\,,\,kl} \, + \, \12 \, \h^{\,ij}\, \h^{\,kl} \, \cB_{\,ijkl} \, = \,
0\, ,
\ee
where we are enforcing the constraints (\ref{ebeta}), and where
for convenience we are introducing the shorthand notation
\be \label{aprimsec}
\begin{split}
&\cA^{\,\pe}{}_{ij} \, = \, T_{ij} \, \cA \, , \\
&\cA^{\,\pe\pe}{}_{ij\,,\,kl} \, = \, \frac{1}{3} \, \left(\, 2\, T_{ij}\, T_{kl} \, -
\, T_{i\,(\,k}\, T_{l\,)\,j} \,\right) \cA \, .
\end{split}
\ee
Notice that in the two-family case $\cA^{\,\pe\pe}{}_{ij\,,\,kl} $
is effectively traceless, because all triple traces of $\cA$ can be
related to symmetrized double traces and thus vanish on account of
the constraint equations \eqref{aconstrdt} and \eqref{ebeta}, as can be seen making use of
the results in Appendix \ref{app:idsb}.

In order to proceed, it is useful to disentangle the traceless part
of the $\cB_{ijkl}$ tensors from their traces, via a decomposition
similar to the one in \eqref{tracebsym}, and thus letting
\be
\cB_{\,ijkl}\,=\,\cB^{\, (T)}{}_{ijkl}\,+\, \h^{\,
mn}\,\widetilde{\cB}_{\,ijkl,\,mn} \, .
\ee
The scalar product provides an interesting tool to this effect,
since
\be \label{killb}
\bra \widetilde{\cB}_{\,ijkl,\,mn} \comma T_{ij}\,T_{kl}\,T_{mn}\, E_\vf \ket \, \sim \, \left\|\, \h^{\,ij}\,\h^{\,kl}\,
\h^{\, mn}\, \widetilde{\cB}_{\,ijkl,\,mn} \,\right\|^{\,2}
\ee
shows that the expression on the right-hand side vanishes on-shell. While this does not imply directly that the $\widetilde{\cB}_{\,ijkl,\,mn}$ tensors vanish, it does
prove that they certainly decouple from $\cA$ and its non-vanishing
traces, so that the equations of motion reduce to
\be
\begin{split}  \label{motion_noCT}
& \cA \, - \, \12 \, \h^{\,ij}\, \cA^{\,\pe}{}_{ij} \, + \, \frac{1}{12} \, \h^{\,ij}\, \h^{\,kl} \,
\cA^{\,\pe\pe}{}_{ij\,,\,kl} \, + \, \12 \, \h^{\,ij}\, \h^{\,kl} \, \cB^{\, (T)}{}_{\,ijkl} \, = \, 0 \, ,  \\[5pt]
& \h^{\,ij}\,\h^{\,kl}\, \h^{\, mn}\, \widetilde{\cB}_{\,ijkl,\,mn} \, = \, 0 \, .
\end{split}
\ee
The second, in particular, sets to zero the $\{6\}$ and $\{4,2\}$
projections in the family indices of $\widetilde{\cB}_{ijkl,\,mn}$,
leaving undetermined its $\{5,1\}$ component, since the product of
three $\h$'s does not admit this last projection. This is just the
expected behavior, since we have seen in Section
\ref{sec:motion2b} that the trace parts of the $\cB_{ijkl}$
transform under the additional gauge symmetry
\eqref{Bshift}, and consequently can be set to zero
via the corresponding shift.

On the other hand, taking one trace of the first of
eqs.~\eqref{motion_noCT} gives
\be \label{trace_motion}
(\,D-2\,) \, E_{\,ij} \, + \, S^{\,k}{}_{(\,i}\, E_{\,j\,)\,k} \,
\eq
\, \cO\,[\,D-2\,]{}_{\,ij}{}^{\,kl} \, E_{\,kl} \, = \, 0 \, ,
\ee
which defines the ${\cal O}\,[\, \l\, ]$ operators, and where
\be \label{Eij}
E_{\,ij} \, \eq \, \cA^{\,\pe}{}_{ij} \, - \, \frac{1}{3}\,
\h^{\,kl}\, \cA^{\,\pe\pe}{}_{ij\,,\,kl} \, -
\, 2 \, \h^{\,kl}\, \cB^{\, (T)}{}_{\,ijkl} \, .
\ee
Notice that in deriving this expression we are taking into account
that eq.~\eqref{aconstrdt}, together with eq.~\eqref{ebeta}, forces
the double trace of $\cA$ to coincide with its $\{2,2\}$ projection
defined in \eqref{aprimsec}.

From eq.~\eqref{trace_motion} one thus obtains that, outside the
kernel of $\cO\,[\,D-2\,]$ (that is empty for $D
\geq 6$, as we shall see in the next section),
\be \label{eijinvert}
\cA^{\,\pe}{}_{ij} \, - \, \frac{1}{3}\, \h^{\,kl}\, \cA^{\,\pe\pe}{}_{ij\,,\,kl} \, -
\, 2 \, \h^{\,kl}\, \cB^{\, (T)}{}_{\,ijkl} \, = \, 0 \, .
\ee
A further trace of this relation then yields
\be \label{eijdouble}
(\,D-3\,) \, \cA^{\,\pe\pe}{}_{ij\,,\,kl}  \, + \,
S^{\,m}{}_{(\,i}\, \cA^{\,\pe\pe}{}_{j\,) \, m\,,\,kl} \, +
\, 6 \, \left[\, D \, \cB^{\, (T)}{}_{ijkl}  \, + \, S^{\,m}{}_{(\,i}\, \cB^{\, (T)}{}_{j\,)\, mkl} \right]\, = \, 0 \, ,
\ee
from which one can conclude that both $\cA^{\,\pe\pe}{}_{ij\,,\,kl}$
and $\cB^{\, (T)}{}_{ijkl}$ vanish in general, up to a few
exceptional cases that we shall soon identify. Substituting these in
\eqref{eijinvert} gives
\be \label{}
\cA^{\,\pe}{}_{ij} \, = \, 0 \,  ,
\ee
so that eq.~\eqref{motion_noC} finally becomes
\be
\cA \, = \, 0 \, ,
\ee
that after shifting away the compensators is tantamount to the
reduction to the Labastida form \eqref{labeq}.

As we already stressed, the current analysis leaves some loose ends
whenever eqs.~\eqref{trace_motion} and
\eqref{eijdouble} are not invertible, so that
either $\cA^{\, \pe}$ or $\cA^{\, \pe\pe}$ are not determined. The
next section is thus devoted to a systematic study of the kernel of
the relevant operators. This will also clarify the nature of some
poles that we shall find in Section \ref{sec:cur42b} in the current
exchanges of a rank-$(4,2)$ field. The final upshot of the
discussion will be that interesting patterns of Weyl-like symmetries
emerge in sporadic low-dimensional cases, precisely as needed to
gauge away the undetermined quantities.

\vskip 12pt

%%%%%%%%%%%%%%%%%%%%%%%%%%%%%%%%%%%%%%%%%%%%%%%%%%%%%%%%%%%%%%%%%%%%%%%%%%%%%%%%%%%%%%%%%%%%%%

\scsss{Weyl-like symmetries} \label{sec:Weyl}

%%%%%%%%%%%%%%%%%%%%%%%%%%%%%%%%%%%%%%%%%%%%%%%%%%%%%%%%%%%%%%%%%%%%%%%%%%%%%%%%%%%%%%%%%%%%%%

Let us therefore analyze the kernel of the operator ${\cal O}$ that
first appeared in eq.~\eqref{trace_motion},
\be \label{operatorO}
\cO\,[\,D-2\,]_{\,ij}{}^{kl} \, = \,
\fr{D\, -\, 2}{2} \, \d_i{}^{\, (\, k} \, \d^{\, l\, )}{}_j \,  + \, \12 \,
S^{\, (\,k}{}_{(\,i}\, \d^{\, l\,)}{}_{j\,)} \, ,
\ee
whose structure is determined by the properties of the $S^{\,i}{}_j$
operators. In the two family case, there are actually four such
operators, $S^{\,1}{}_1$, $S^{\,1}{}_2$, $S^{\,2}{}_1$ and
$S^{\,2}{}_2$, that generate a $gl(2)$ algebra. Their properties can
be exhibited most conveniently introducing two commuting vectors,
$u^{\,\m}$ and $v^{\,\n}$, to cast generic two-family fields in the
form
\be
\vf\,(u,v) \, = \, \frac{1}{s_{\,1}\,! \ s_{\,2}\,!} \ u^{\, \m_1} \ldots \, u^{\, \m_{s_1}}
\ v^{\, \n_1} \ldots \, v^{\, \n_{s_2}}\,
\vf_{\,\m_1 \ldots \, \m_{s_1} ; \ \n_1 \ldots \,
\n_{s_2}} \, . \label{phiuv}
\ee
In this notation
\begin{alignat}{2}
S^{\,1}{}_1 & = \, u \cdot \frac{\pr}{\pr \, u} \ , \qquad\qquad
&S^{\,1}{}_2 & = \, u \cdot \frac{\pr}{\pr \, v} \ , \nonumber \\
S^{\,2}{}_1 & = \, v \cdot \frac{\pr}{\pr \, u} \ , \qquad\qquad
&S^{\,2}{}_2 & = \, v \cdot \frac{\pr}{\pr \, v} \ .
\end{alignat}
so that the pairs $\left(u,\frac{\pr}{\pr \, u}\right)$ and
$\left(v,\frac{\pr}{\pr \, v} \right)$ play the role of the
conventional creation and annihilation operators for a couple of
quantum-mechanical oscillators. In particular, $S^{\,1}{}_1$ and
$S^{\,2}{}_2$ are diagonal on all fields of the form \eqref{phiuv},
and simply count their space-time indices, according to
\be
S^{\,1}{}_1\, \vf(u,v) \, = \, s_1 \, \vf(u,v) \ , \quad
S^{\,2}{}_2\,
\vf(u,v) \, = \, s_2 \, \vf(u,v)\, .
\ee
On the other hand, $S^{\,2}{}_1$ and $S^{\,1}{}_2$ are not diagonal,
but
\begin{alignat}{2}
& S^{\,1}{}_2 \, \vf\,(u,v) &&=\, \frac{u^{\, \m_1} \ldots \, u^{\,
\m_{s_1+1}}
\ v^{\, \n_1} \ldots \, v^{\, \n_{s_2-1}}}{(s_{\,1}+1)\,! \
(s_{\,2}-1)\,!}\ \
\vf_{\,( \, \m_1 \ldots \, \m_{s_1} ; \ \m_{s_1+1}\,) \, \n_1 \ldots \,
\n_{s_2-1}} \, , \\
& S^{\,2}{}_1 \, \vf\,(u,v) &&=\, \frac{u^{\, \m_1} \ldots \, u^{\,
\m_{s_1-1}}
\ v^{\, \n_1} \ldots \, v^{\, \n_{s_2+1}}}{(s_{\,1}-1)\,! \
(s_{\,2}+1)\,!}\ \
\vf_{\, \m_1 \ldots \, \m_{s_1-1} (\, \n_{1} ; \ \n_2 \ldots \,
\n_{s_2+1}\,)}\, ,
\end{alignat}
so that they effectively displace one index from one set to the
other, raising or lowering the corresponding eigenvalues of
$S^{\,1}{}_1$ and $S^{\,2}{}_2$. Hence, we can not refrain from
turning to the suggestive notation
\be
L_+ \, = \, S^{\,1}{}_2 \ , \quad L_- \, = \, S^{\,2}{}_1 \ , \quad
L_3 \, = \, \frac{1}{2} \left(\, S^{\,1}{}_1 \, - \, S^{\,2}{}_2\,
\right) \, ,\label{Ang_mom}
\ee
since these three operators clearly satisfy the very familiar
angular-momentum commutation relations
\be
[\, L_3 \, , \, L_\pm \, ]\, = \, \pm \, L_\pm \ , \qquad [\,L_+ \,
, \, L_- \, ]\, = \, 2 \, L_3 \ . \label{Ang_momalg}
\ee
Moreover, as usual
\be
L_- L_+ = \mathbf{L}^2 - L_3^2 - L_3 \, .\label{totang}
\ee
This is almost Schwinger's oscillator trick as reviewed for example
in \cite{sakurai}, although the present construction relies on a
pair of vectors $u^{\,\m}$ and $v^{\,\n}$, and as a result builds a
number of distinct irreducible multiplets characterized by different
symmetry properties. The origin of this fact can be appreciated
considering the product of $p$ $u$'s and $q$ $v$'s, that clearly
gives rise to a number of components on which the permutation group
of space-time indices acts irreducibly. These components reflect the
decomposition of $\{p\}\otimes\{q\}$, and a standard property of
Young diagrams implies that $L_+$ annihilates terms of the type
\be
Y_{\{p\,,\,q\}} \ u^{\,\m_1} \ldots u^{\,\m_p} \, v^{\,\n_1}
\ldots v^{\,\n_q}
 \, , \label{chaintop}
\ee
simply because it forces a symmetrization beyond a given line
\footnote{In Section \ref{sec:irreducible} we shall see how to adapt
the construction to gauge fields satisfying conditions of this type,
that we shall term irreducible. Strictly speaking, however, one
could also consider more general types of $gl(D)$-irreducible fields
that are not associated to the endpoints of chains. From this
algebraic vantage point, at two families these fields are
simultaneous eigenstates of ${\bf L}^2$ and $L_z$, and thus also of
$L_-L_+$.}. Here $Y_{\{p,q\}}$ denotes the Young projector
corresponding to the $\{p,q\}$ diagram, whose first row has the
minimal possible length in the present example. For the combination
\eqref{chaintop}, that as we have said is annihilated by $L_+$, the
eigenvalue $l$ of the ``total angular momentum'' $\mathbf{L}^2$
coincides with the eigenvalue $m\, =
\, \frac{p-q}{2}$ of $L_3$, while a whole multiplet of \emph{irreducible} polynomials in
$u$ and $v$ with lower values of $m$ can be built from
\eqref{chaintop} by successive applications of $L_-$.

Using the previous observations, for a rank-$(s_1,s_2)$ field $\vf$ it is possible to recast the system
defined by eq.~\eqref{trace_motion} in the form
\begin{align} \label{system_motion}
& (\,D+2\,s_1-6\,)\, E_{11} \, + \, 2 \, L_- \, E_{12} \, = \, 0 \, , \nn \\[2pt]
& (\,D+s_1+s_2-4\,)\, E_{12} \, + \, L_+ \, E_{11} \, + \, L_- \, E_{22} \, = \, 0 \, , \\[2pt]
& (\,D+2\,s_2-6\,)\, E_{22} \, + \, 2 \, L_+ \, E_{12} \, = \, 0 \,
. \nn
\end{align}
In order to study the non-trivial solutions of
\eqref{system_motion} it is necessary to
decompose the unknowns into their irreducible $gl(D)$ components,
independent expressions on which permutations of \emph{space-time}
indices act irreducibly. This step is essential to arrive at an ordinary ``square'' algebraic system.
Actually, this procedure splits the system of equations into a chain
of independent sub-systems for the irreducible components of
$E_{11}$, $E_{12}$ and $E_{22}$. These correspond to eigenstates of
the operator $\mathbf{L}^2$ introduced in eq.~\eqref{totang} so
that, for instance, the decomposition of $E_{12}$ can be presented
in the form
\be \label{dec_irr}
E_{12} \, = \, \sum_{n \, = \, 0}^{s_2\,-\,1} \, (E_{12})^{\{s_1+\,
s_2\,-\,n \,-\,2,\, n\}} \, ,
\ee
where each term in the sum can be related to components annihilated
by $L_+$:
\be
\begin{split}
& (E_{12})^{\{s_1+\,s_2\,-\,n-\,2,\,n\}} \, = \, (L_-)^{\, s_2 - n} \, (\widehat{E}_{12})^{\{s_1+\,s_2\,-\,n\,-\,2,\,n\}} \, , \\
&L_+ \, (\widehat{E}_{12})^{\{s_1+\,s_2\,-\,n\,-\,2,\,n\}} \, \equiv \,
0 \, .
\end{split}
\ee
Notice that, for all irreducible components of $E_{12}$ originating from a given field $\vf$, $m$ takes a fixed value, that is related to $s_1$ and $s_2$ according to
\be
m \, = \, \frac{s_1 \, - \, s_2}{2} \, ,
\ee
while the value of $\ell$ for the irreducible component
$(E_{12})^{\{s_1+\,s_2\,-\,n\,-\,2,\,n\}}$ is
\be
\ell \, = \, \frac{s_1\,+\,s_2}{2} \, - \, (\,n+1\,)\, ,
\ee
so that
\be
\frac{s_1\,-\,s_2}{2} \, \le \, \ell \, \le \, \frac{s_1\,+\,s_2}{2} \, - \, 1 \, ,
\ee
where the allowed range is spanned by values of $\ell$ that differ
by integers. Being eigenstates of $\mathbf{L}^2$, the various
irreducible components of $E_{12}$ are also eigenstates of
$L_-\,L_+$, so that
\be \label{l_minusl_plus}
L_-\, L_+ \, (E_{12})^{\{s_1\, +\,s_2 \, - \, n \, - \, 2 \, , n\}} \, = \, (\,s_1 \, - \,n \,)\, (\,s_2-n-1\,)\,
(\,E_{12}\,)^{\{s_1+\,s_2\,-\,n\,-\,2,\,n\}} \, ,
\ee
where we have used the standard $sl(2)$ formula
\be \label{sl2form}
L_- \,L_+ \, V_{\,\ell \,,\,m}\,=\, (\,\ell - m\,)\,(\,\ell+m+1\,)\, V_{\,\ell \,,\, m} \, .
\ee

In order to proceed with the analysis of \eqref{system_motion}, it
will prove convenient to distinguish various cases corresponding to
various possible values of $s_2$, while taking into account that, as
we have stressed, the $E_{ij}$ bear fewer indices than the original
field $\vf$. The ensuing discussion is inevitably somewhat
technical, and the reader may wish to skip it, confining the
attention to the two tables at the end of it, where the results are
collected. The naive reduction to the Labastida form indeed fails
only for a number of low-dimensional theories that in any case would
reduce to trivial $SO(D-2)$ representations. As we stressed at the
beginning of this section, these systems are nonetheless rather
interesting, since they generalize the familiar example of
two-dimensional gravity, that brought about many surprises over the
years. Here are anyway the relevant cases $(s_1
\geq s_2)$:
\begin{description}

 \item[$\bf s_2 = 0$:] This is the symmetric case of \cite{fs3, fms}, and the system
 either does not exist (for $s_1 = 0, 1$) or degenerates to the first equation (for $s_1 \geq 2$)
 without $E_{12}$. In all cases the equations of motion are invertible in any dimension $D$, with the already
 recalled exception of two-dimensional gravity.

 \item[$\bf s_2 = 1$:] In this case $E_{22}$ does not exist, and one is left at most with
 the first two equations, so that the system reduces to
\be
\begin{split} \label{system_s1}
& (\,D+2\,s_1 \,-\, 6\,)\, E_{11} \, + \, 2 \, L_- \, E_{12} \, = \, 0 \, ,  \\[2pt]
& (\,D+s_1 - 3\,)\, E_{12} \, + \, L_+ \, E_{11} \,  = \, 0 \, , \\[2pt]
\end{split}
\ee
In addition, with $s_2=1$ $E_{12}$ only bears indices of the first
family, and as a result contains only an irreducible component, the
fully symmetric one. Hence, the first equation sets to zero the
``hooked'' $\{s_1-2,1\}$ component of $E_{11}$ whenever its
coefficient does not vanish, while in general the symmetric
components give rise to a $2 \times 2$ system of equations.
Furthermore, for the only available component of $E_{12}$ the values
of $\ell$ and $m$ coincide, so that it is annihilated by $L_-L_+$,
and this observation will prove convenient in the detailed analysis
of eq.~\eqref{system_s1} presented in the following.

For $s_1 = 1$ only the second equation is actually present, and in
$D = 2$ it degenerates to an identity, so that $E_{12}$ is left
undetermined.

For $s_1 = 2$ and $D = 1$ the second equation reduces to $L_+ \,
E_{11} \,  = \, 0$. Thus, applying $L_+$ to the first equation and
using eqs. \eqref{Ang_momalg} and
\eqref{totang} gives $E_{12} \,  = \, 0$, and finally $E_{11} \,  = \, 0$.

For $s_1 = 2$ and $D = 2$ the first equation gives $L_- \, E_{12} \,
=
\, 0$, while acting with $L_-$ on the second gives
\be
 L_- \, L_+ \, E_{11} \,  = \, 0 \, .
\ee
On the other hand, since $l$ can only assume the same value as $m$
($E_{11}$ is a vector in this case), this relation reduces to an identity, and
the equations of motion are not invertible.

For $s_1 = 2$ in $D \geq 3$, or for $s_1 \geq 2$ and any value of
$D$, one must consider the full system \eqref{system_s1}. Eliminating
$E_{11}$ via the first equation and substituting in the second gives
\be
(\,D \, + \, 2s_1 \, - \, 4\,) \, (\,D \, + \, s_1 \, - \, 5\,) \,
E_{12} \, = \, 0 \, ,
\ee
that always implies $E_{12} \, = \, 0$, but for three special cases.
These are $s_1 = 4$ in $D = 1$, $s_1 = 3$ in $D = 2$ and $s_1 = 2$
in $D = 3$, where the value of $E_{12}$ is left undetermined by the
equations of motion.

 \item[$\bf s_2 \geq 2$:]
In this case one must consider the full set of three equations, but
for $s_1 = 2$, $s_2 = 2$ in $D = 2$ eq.~\eqref{system_motion}
simplifies to
\begin{align} \label{system_simpl}
&  2 \, L_- \, E_{12} \, = \, 0 \, , \nn \\[2pt]
& 2\, E_{12} \, + \, L_+ \, E_{11} \, + \, L_- \, E_{22} \, = \, 0 \, , \\[2pt]
&  \, 2 \, L_+ \, E_{12} \, = \, 0 \, , \nn
\end{align}
so that $E_{12}$ only contains its $\{1, 1\}$ irreducible component.
Since in this case $E_{11}$ and $E_{22}$ are symmetric, one can also
conclude that $E_{12} = 0$. On the other hand, the same equation
does not determine the other variables, $E_{11}$ and $E_{22}$, but
only provides a linear relation between them, and thus the solution
is not determined.

For $s_1 > 2$ and $s_2 = 2$, still in $D = 2$, the system becomes
\begin{align} \label{system_simpl2}
&  (\,2\, s_1 \, - \, 4\,) \, E_{11} \, + \, 2 \, L_- \, E_{12} \, = \, 0 \, , \nn \\[2pt]
& s_1\, E_{12} \, + \, L_+ \, E_{11} \, + \, L_- \, E_{22} \, = \, 0 \, , \\[2pt]
&  \, 2 \, L_+ \, E_{12} \, = \, 0 \, , \nn
\end{align}
so that one can still conclude that $E_{12}$ is irreducible, and
thus only contains the component with $l = m$. On the other hand, in
this case one can determine $E_{11}$ and $E_{22}$. In fact, applying
$L_+$ to the first equation gives
\be
L_+ \, E_{11} \, = \, -  \, E_{12} \, ,
\ee
that upon substitution in the second equation provides a relation
between $L_- \, E_{22}$ and $E_{12}$. Applying $L_+$ to this
relation one finally obtains
\be
2 \, L_3 \, E_{22} \, + \, L_- \, L_+ \, E_{22} \, = \, 0 \, .
\ee
Since $E_{22}$ is a symmetric tensor of rank $s_1 - 2$ the last
equation reads
\be
2 \, s_1  \, E_{22} \, = \, 0 \, ,
\ee
so that all $E_{i j }$ tensors actually vanish.

Finally, for $s_1 \geq 2$ and $s_2 \geq 2$ in $D \neq 2$, and for
$s_1 > 2$ and $s_2 > 2$ in $D = 2$, one must consider the full
system
\eqref{system_motion}. It is then convenient to solve it directly by
substitution, letting
\be \label{e11e22}
\begin{split}
& E_{11} \, = \, - \, \frac{2}{D+2\,s_1-6} \, L_- \, E_{12} \, ,  \\
& E_{22} \, = \, - \, \frac{2}{D+2\,s_2-6} \, L_+ \, E_{12} \, ,
\end{split}
\ee
and thus reducing the second of \eqref{system_motion} to a set of
relations for the irreducible components of $E_{12}$ via the
decomposition
\eqref{dec_irr},
\be \label{singleE12}
4 \, \left(\, s_1 \, + \, s_2 \,  - \, n \, + \, \fr{D}{2} \, - \, 3 \,\right) \, \left(\,2
\, - \, n \, - \, \fr{D}{2} \,\right) \,
(E_{12})^{\{s_1+\,s_2\,-\,n\,-\,2,\,n\}} \, = \, 0 \, ,
\ee
where
\be
n \, =  \, 0, \, \dots , \, s_2 \, - \, 1 \, .
\ee
As a first observation, eq.~\eqref{singleE12} implies that
non-trivial solutions of \eqref{system_motion} do not exist in odd
dimensions, $D = 2 k + 1$. Furthermore, neither of the two equations
\be
\begin{split}
&s_1 \, + \, s_2 \,  - \, n \, + \, \fr{D}{2} \, - \, 3 \, = \, 0 \, , \\
& 2 \, - \, n \, - \, \fr{D}{2} \,  = \, 0 \, ,
\end{split}\label{singled4}
\ee
can be satisfied for $D \geq 6$. In $D=4$ the second equation  gives
$n \, = \, 0$, that corresponds to the symmetric component of
$E_{12}$, while in $D = 2$ there are non-trivial solutions for $n =
1$, that selects the $\{s_1 + s_2 - 3, 1\}$ component of $E_{12}$.
In both cases, because of
\eqref{e11e22}, the available components of $E_{11}$ and $E_{22}$
are the same as those of $E_{12}$.
\end{description}

It should be appreciated that, in all cases where the system
\eqref{system_motion} sets to zero the $E_{ij}$ tensors, it is still
possible that the equations of motion be not invertible at the level
of the double trace. At two families the relevant equations to this
effect are given in \eqref{eijdouble}, and in order to perform this analysis
it is convenient to split them into their three projections in the
family indices: the $\{2,2\}$, the $\{4\}$ and  the $\{3, 1\}$. In
particular, the $\{2,2\}$ projection only contains
$\cA^{\,\pe\pe}{}_{ij\,,\,kl} $ and reads
\be \label{projA22}
(\,D-3\,) \, \cA^{\,\pe\pe}{}_{ij\,,\,kl}  \, + \, \12 \, \left[\,
S^{\,m}{}_{(\,i}\, \cA^{\,\pe\pe}{}_{j\,) \, m\,,\,kl} \, + \,
S^{\,m}{}_{(\,k}\, \cA^{\,\pe\pe}{}_{l\,) \, m\,,\,ij} \,
\right]\, = \, 0 \, ,
\ee
while only the tensors $\cB^{\, (T)}{}_{ijkl}$ appear in the
$\{4\}$:
\be \label{projB4}
D \, \cB^{\, (T)}{}_{ijkl}  \, + \, \12 \, S^{\,m}{}_{(\,i}\,
\cB^{\, (T)}{}_{jkl\, ) \, m} \, = \, 0 \, .
\ee
Finally, the $\{3,1\}$ projection,
\be \label{projAB31}
S^{\,m}{}_{(\,i}\, \cA^{\,\pe\pe}{}_{j\,) \, m\,,\,kl} \, - \,
S^{\,m}{}_{(\,k}\, \cA^{\,\pe\pe}{}_{l\,) \, m\,,\,ij} \, + \, 6
\, \left[\, S^{\,m}{}_{(\,i}\, \cB^{\, (T)}{}_{j\,) \, mkl} \, - \,
S^{\,m}{}_{(\,k}\, \cB^{\, (T)}{}_{l\,) \, mij} \, \right] \, = \, 0
\, ,
\ee
relates the two types of tensors.

Let us now recall that the $\{2,2\}$ Young projection defining
$\cA^{\,
\pe
 \pe}$ implies the relation
\be
\cA^{\,\pe\pe}{}_{11\,,\,22} \, + \, 2 \, \cA^{\,\pe\pe}{}_{12\,,\,12} \, = \, 0 \, ,
\ee
together with the vanishing of all other components. As a result,
for a two-family tensor $\vf$ of rank $(s_1, s_2)$,
\eqref{projA22} reduces to the single equation
\be \label{trace_app}
(\,D+s_1+s_2-7\,)\, \cA^{\,\pe\pe}{}_{11\,,\,22} \, = \, 0 \, ,
\ee
so that $\cA^{\,\pe\pe}$ vanishes in general, the only exceptions
being provided by the $(4, 2)$ case in $D = 1$, the $(3, 3)$ case in
$D = 1$, the $(3, 2)$ case in $D = 2$ and the $(2, 2)$ case in $D =
3$. On the other hand, even in the latter cases, before making
definite statements one must take into account the conditions
originating from the other two projections in eqs.~\eqref{projB4}
and \eqref{projAB31}.

Since the number of possibilities is relatively small, one can
analyze the behavior of the solutions explicitly for the cases of
interest. It is thus possible to verify that in the $(3,2)$ case in
$D = 2$ the double trace of $\cA$ actually vanishes. On the other
hand, in $D = 1$ the only effective irreducible component of any
tensor is the symmetric one, whose equations, as we know, are
invertible, while the other possibilities simply do not exist. In
conclusion, the only case in which the equations of motion leave the
double trace of $\cA$ undetermined is the $(2, 2)$ in $D = 3$ so
that, in particular, the $D=1$ pole in the propagator of a rank-$(4,
2)$ field that we shall find in Section \ref{sec:cur42b} lacks a
true physical meaning.

The relevant results are summarized in Table \ref{table2fam}: the
first pair of columns lists the \emph{reducible} $gl (D)$-tensors
whose traces are left undetermined by the equations of motion. In a
similar fashion, the second pair of columns lists the corresponding
cases of
\emph{reducible} $gl (D)$-tensors for which $\cA^{\, \pe \pe}$ has a
similar behavior.
\begin{table}[htb]
\begin{center}
\begin{tabular}{||c||c|c||c|c||}
\cline{2-5}
\multicolumn{1}{c|| }{} & \multicolumn{2}{|c|| }{$\cA^{\, \pe}$} & \multicolumn{2}{|c|| }{$\cA^{\, \pe \pe}$}  \\
\hline
\hline
$D$  &  $s_1$  &  $s_2$   &  $s_1$  &  $s_2$ \\
\hline
\hline
$2$  &  $2$  &  $0$ &  &  \\
\hline
$2$ &  $1$  &  $1$ &   &    \\
  \hline
$2$ &  $2$  &  $1$ &   &    \\
  \hline
$2$  &  $3$  &  $1$ &   & \\
 \hline
$2$  &  $2$  &  $2$ &   & \\
\hline
$2$  &  $\geq 3$  &  $\geq 3$ &   & \\
\hline
$3$  &  $2$  &  $1$ &  $2$  &  $2$ \\
\hline
$4$  &  $\geq 2$  &  $\geq 2$ &   & \\
\hline
\end{tabular}
\end{center}
\caption{$gl(D)$-\emph{reducible} fields with Weyl-like symmetries} \label{table2fam}
\end{table}

Indeed, it is possible to be slightly more precise, identifying the
\emph{irreducible} components of $\vf$ whose field equations can not
be brought directly to the Labastida form. For $\cA^{ \, \pe
\pe}$ this is straightforward, since the only relevant case can be
associated to a rank-$(2 ,2)$ field, whose double trace is fully
contained in its irreducible $\{2, 2\}$ component. In order to
obtain the same type of information for $\cA^{\,\pe}$, as a first
step one should sort out the irreducible components of $\cA^{\,
\pe}$ that are not fixed by the dynamics. To this end, one
should notice that its traceless part corresponds to the traceless
part of the $E_{\, ij}$ tensors, whose undetermined irreducible
components were already identified in the previous discussion.

For instance, for the $(s_1, s_2)$ case in $D = 4$, with both
families containing at least two indices, we saw after
eq.~\eqref{singled4} that the symmetric component of $E_{\, 12}$ was
in the kernel of $\cO$. Therefore, the corresponding irreducible
component of $\h
\, \cA^{ \,\pe}$, and then of $\cA$, might be the
symmetric one, or the $\{s_1 + s_2 -1, 1\}$ or the $\{s_1 + s_2 - 2,
2\}$. On the other hand, we know that the symmetric case is always
invertible in $D =4$, while an equivalent conclusion can be drawn
for the $\{s, 1\}$-component, on account of the results presented in
Table \ref{table2fam}. Hence, the lack of invertibility of the
equations of motion must be ascribed to the $\{s_1 + s_2 - 2,
2\}$-component. In a similar fashion, when both families contain
more than two indices in $D = 2$, the $\{s_1 + s_2 - 3, 1\}$
component of $\cA^{\, \pe}$ is not determined. By a similar
reasoning, it is then possible to conclude that the undetermined
traces are contained in the $\{s_1 + s_2 - 3, 3\}$-component, since
the other two available ones, $\{s_1 + s_2 - 1, 1\}$ and $\{s_1 +
s_2 - 1, 1\}$, were already excluded by our previous discussion. In
the other cases the same type of analysis, together with a more
explicit computation concerning $(2, 2)$ fields in $D=2$ that we
shall return to later, lead finally to the results collected in
Table
\ref{table2fam_irr}.

\begin{table}[htb]
\begin{center}
\begin{tabular}{||c||c|c||c|c||}
\cline{2-5}
\multicolumn{1}{c|| }{} & \multicolumn{2}{|c|| }{$\cA^{\, \pe}$} & \multicolumn{2}{|c|| }{$\cA^{\, \pe \pe}$}  \\
\hline
\hline
$D$  &  $s_1$  &  $s_2$   &  $s_1$  &  $s_2$ \\
\hline
\hline
$2$  &  $2$  &  $0$ &   &  \\
\hline
$2$ &  $2$  &  $1$ &   &     \\
\hline
$2$  &  $3$  &  $1$ &   & \\
\hline
$2$  &  $s$  &  $3$ & & \\
\hline
$3$  &  $2$  &  $1$ &  $2$  &  $2$ \\
\hline
$4$  &  $s$  &  $2$ &   & \\
\hline
\end{tabular}
\end{center}
\caption{$gl(D)$-\emph{irreducible} fields with Weyl-like symmetries} \label{table2fam_irr}
\end{table}

We have thus completed the classification of the Lagrangians whose
field equations are not directly reducible to the Labastida form,
and now we would like to elaborate on their meaning.

According to a theorem that we already recalled at the beginning of this section,
all fields appearing in Table \ref{table2fam_irr} would correspond
to vanishing irreducible representations of $so(D-2)$, if the Labastida equation
\eqref{labeq} were to hold. On the other  hand, the
example of two-dimensional gravity suggests that a novel class of
symmetries might emerge in the previous
cases, and indeed, the very fact that these do not fix some of the
traces of $\cA$ implies that it is possible to redefine arbitrarily
the corresponding quantities.

Moreover, in even closer analogy with two-dimensional
gravity, at least for a subclass of the theories of Table
\ref{table2fam_irr} the Lagrangian could well be a total derivative.
More precisely, referring for simplicity to the constrained case,
where the Lagrangians take the simple form
\be
\cL \, = \, \12 \, \bra \vf \comma \cE_{ \, \vf} \ket \, ,
\ee
with $ \cE_{ \, \vf}$ Labastida's Einstein-like tensor
\eqref{labaeinstein}, it is simple to see that some theories can be
topological \emph{only if} the tensor defining the equations of
motion vanishes identically:
\be \label{topoB}
E_{\, \vf} \, : \, \cE_{ \, \vf} \, + \, \12 \, \h^{\, ij} \,
\h^{\, kl} \, \cB_{\, ijkl} \, \equiv \, 0\, .
\ee
Whenever $\cA$ does not vanish identically \ft{This actually happens
in particularly degenerate cases, for instance for a vector in $D=1$
or for a two form in $D=2$. However, this behavior is quite rare, as
can be seen for instance working in de Donder gauge, so that $\cA$
reduces to $\Box
\, \vf$. The point is that only in low-enough dimensions de
Donder conditions and the irreducibility properties
suffice to force $\Box \, \vf$ to vanish.}, eq.~\eqref{topoB} can
only hold in the cases collected in Table
\ref{table2fam_irr}, the only ones in which the condition
$E_{\, \vf} = 0$ does not imply $\cA = 0$. It is then possible to
see directly that for irreducible fields of type $\{1\}$ in $D = 1$,
$\{2\}$  and $\{1, 1\}$ in $D = 2$ the Einstein-like tensor does vanish
identically.

In addressing the remaining possibilities, it will be convenient to
treat separately two-column tensors. These types of fields with only
two columns possess in fact the peculiar feature, stressed in
\cite{labastida1}, that \emph{no constraints} arise in their case,
for both the gauge field and the gauge parameters, so that in a
sense they are closer to spin-$2$ fields than to ordinary
higher-spin ones. In particular, their equations of motion do not
contain the $\cB_{ijkl}$ tensors, so that the condition that the
theory be topological rests solely on the vanishing of the
Einstein-like tensor $\cE_{\, \vf}$.

For this class of fields, we can actually go beyond two families and
provide a characterization of a wider class of topological theories,
observing that the Lagrangians for $N$-families, that we shall
discuss in Section
\ref{sec:generalb}, when restricted to $\{p,q\}$ two-column fields can be
recast in the compact form (see for instance \cite{dmh} and the first paper in \cite{othernew})
\be \label{lagschou}
\cL \, \sim \, \d^{\,[\m_1 \, \dots \, \m_{p + q + 1}]\, }_{\, [\n_1 \, \dots \, \n_{p + q + 1}]\,}\,
\pr_{\, \m_1} \vf_{\,\m_2}{}^{\n_1}{}_{; \, \dots}{}_{; \, \m_{q + 1}}{}^{\n_q}{}_{; \,
\m_{q+2} \, ; \,\dots \, ; \, \m_{p + 1}}
\pr^{\,\n_{q + 1}} \vf_{\,\m_{p + 2}}{}^{\n_{q + 2}}{}_{; \, \dots}{}_{; \, \m_{p + q + 1}}{}^{\n_{p + 1}}{}^{; \,
\n_{p+2} \, ; \,\dots \, ; \, \n_{p + q + 1}} \, ,
\ee
where we are now using explicit space-time indices, and where the
symbol
\be
\d^{\, [\m_1 \, \dots \, \m_{p + q + 1}]\, }_{\, [\n_1 \, \dots \, \n_{p + q + 1}]\,}
\ee
denotes the usual product of antisymmetrized Kronecker delta's. One
can prove that the expression in eq. \eqref{lagschou} is
proportional to the Labastida Lagrangian observing that, because of
the antisymmetrizations, it defines a gauge-invariant quadratic polynomial containing $\Box \, \vf$, and recalling that
the Labastida Lagrangian is the unique polynomial with
these features, up to total derivatives. It is then possible to
recognize rather directly that all expressions of the form
\eqref{lagschou} vanish identically whenever the total number of indices belonging to the two families is such that
\be
p \, + \, q \,  \leq \, D \, .
\ee

Summarizing, in a given space-time dimension $D$ all theories describing
irreducible two-column fields whose total number of indices is not larger than $D$ are topological, provided the representation exist.

In order  to complete the discussion, one should also analyze the
cases with more than two columns present in Table
\ref{table2fam_irr}, namely the $\{s \geq 3, 2\}$ in $D = 4$, the
$\{s, 3\}$ and the $\{3, 1\}$ in $D = 2$. We verified that for a
$\{3, 1\}$ field the equations of motion do not vanish identically,
so that the model is not topological. It would be interesting to
come to a definite conclusion also in the other cases.

At any rate, whenever the field equations can not be directly
reduced to the Labastida form $\cA \, = \, 0$, we can now show that
a wider gauge
symmetry emerges. Consequently, the portion of $\cA$ that is not set
to zero directly by the field equations can be eliminated by a
choice of gauge. In order to discuss the origin of these shift
symmetries, it is convenient to first perform the partial gauge
fixing that removes the $\Phi_i$ compensators. Moreover, one
can notice that the equation of motion \eqref{motion_noC} always
sets to zero the traceless part of $\cF$.

A shift symmetry that does not affect the traceless part of the
Fronsdal-Labastida tensor $\cF$ can be identified starting from a
variation of the gauge field $\vf$ of the form
\be \label{shiftphi}
\d \, \vf \, = \, \h^{\,ij} \, \O_{\,ij} \, ,
\ee
that generalizes to the mixed-symmetry case the linearized version
of the Weyl symmetry of two-dimensional gravity. The corresponding
variation of the Fronsdal-Labastida tensor \eqref{fronsdal} reads
\be
\d \, \cF \, = \, \h^{\,ij} \, \cF_{\,ij}\,(\,\O\,) \, + \, \12 \, \pr^{\,i}\pr^{\,j} \left[\, (\,D-2\,)\,
\O_{\,ij} \, + \, S^{\,k}{}_{(\,i}\, \O_{\,j\,)\,k} \,\right] \, ,
\ee
where $\cF_{ij}(\,\O\,)$ is the Fronsdal-Labastida tensor for
$\O_{ij}$. Hence, if the parameters satisfy the relations
\be \label{shift_cond}
(\,D-2\,)\, \O_{\,ij} \, + \, S^{\,k}{}_{(\,i}\,
\O_{\,j\,)\,k}
\, =
\, 0
\ee
the corresponding shift of the $\cF$ tensor is simply
\be \label{shiftF}
\d \, \cF \, = \, \h^{\,ij} \, \cF_{\,ij}\,(\,\O\,) \, .
\ee
The reader will not fail to notice that the condition
\eqref{shift_cond} involves precisely the operator
$\cO$ introduced in eq.~\eqref{operatorO} at the beginning of this
section. Eq.~\eqref{shift_cond} also simplifies the constraint on
the $\O_{ij}$ parameters induced by the request of gauge invariance
for the Labastida constraints \eqref{labac}, since on account of
eq.~\eqref{shift_cond} the condition
\be
T_{(\,ij}\, T_{kl\,)}\, \d\, \vf \, = \, T_{(\,ij\,|}\,
\cO\,[\,D-2\,]_{\,|\,kl\,)}{}^{mn} \, \O_{\,mn} \, + \, \h^{\,mn}\,
T_{(\,ij}\, T_{kl\,)}\, \O_{\,mn} \, = \, 0
\ee
takes the simple form
\be \label{omegaconstr}
T_{(\,ij}\, T_{kl\,)} \, \O_{\,mn} \, = \, 0 \, .
\ee

If eq.~\eqref{shift_cond} is satisfied, the Einstein-like tensor
\eqref{labaeinstein} varies as
\be
\begin{split}
\d \, \cE_{\, \vf} \, & = \, - \, \12 \left\{\, \h^{ij} \, - \, \frac{1}{3}\, Y_{\{2,2\}} \left(\, \h^{ij}\,\h^{kl} \,\right) T_{kl} \,\right\} \left\{\, (\,D-2\,)\, \cF_{\,ij}\,
(\,\O\,) \, + \, S^{\,k}{}_{(\,i}\,
\cF_{\,j\,)}\,(\,\O\,) \, \right\} \\
& - \, \12 \, \h^{ij}\,\h^{kl}\, Y_{\{4\}}\, T_{ij}\,\O_{\,kl}\,
\, +
\, \frac{1}{12} \, \h^{ij}\,\h^{kl}\,\h^{mn} \left(\, Y_{\{2,2\}}
T_{ij}\,T_{kl}
\,\right) \O_{\,mn}\, ,
\end{split}
\ee
but the $S^{\,i}{}_{j}$ operators commute with the
Fronsdal-Labastida operator, and therefore eq.~\eqref{shift_cond}
implies that
\be \label{condFshift}
(\,D-2\,)\, \cF_{\,ij}\,(\,\O\,) \, + \, S^{\,k}{}_{(\,i}\,
\cF_{\,j\,)}\,(\,\O\,) \, = \, 0\, ,
\ee
so that at two-families one is left with
\be
\d \, \cE_{\, \vf} \, = \, - \, \12 \, \h^{ij}\,\h^{kl}\, Y_{\{4\}}\, T_{ij}\,\O_{\,kl}\, \, + \,
\frac{1}{12} \, \h^{ij}\,\h^{kl}\,\h^{mn}\, Y_{\{4,2\}} \left(\, Y_{\{2,2\}}\,
T_{ij}\,T_{kl} \,\right) \O_{\,mn} \, .
\ee
Note that this identifies a symmetry, since the remainder can be
adsorbed in a redefinition of the $\cB_{ijkl}$ in the gauge
fixed-version of eq.~\eqref{motion_noC}, that translates into a
shift of the $\b_{ijkl}$ Lagrange multipliers. Equivalently, this variation would not alter the Lagrangian of the constrained theory, simply because these transformations contract to zero in the scalar products against a field $\vf$ subject to the Labastida constraints \eqref{labac}.

We have thus displayed Weyl-like symmetries of the field equations
for the models listed in the first two columns of Table
\ref{table2fam}. By a direct calculation, one can also identify
precisely which irreducible components are shifted, thus recovering
the results of Table \ref{table2fam_irr}. For instance, whenever
eq.~\eqref{e11e22} applies, in order to satisfy
eq.~\eqref{shift_cond} the shift
\eqref{shiftphi} must take the form
\be \label{shiftO12}
\d \, \vf \, = \, - \, \frac{2}{D+2\,s_1-6}\,\h^{11}\, L_-\, \O_{12} \, + \, 2 \, \h^{12} \, \O_{12} \, - \,
\frac{2}{D+2\,s_2-6}\,\h^{22}\, L_+\, \O_{12}\, ,
\ee
with a non trivial doubly traceless $\O_{12}$, that can only exist
for the models of Table \ref{table2fam}. As we already explained in
the preceding pages, eq.~\eqref{shiftO12} affects a single
$gl(D)$-irreducible component of $\vf$. We can now see, in fact,
that the corresponding $\d \, \vf$ is an eigenvector of $L_-\,L_+$,
as can be recognized taking into account the relations
\begin{align}
& [\, L_-\,L_+ \comma \h^{11} \,] \, = \, 2\, \h^{12}\, L_+ \, , \nn \\
& [\, L_-\,L_+ \comma \h^{12} \,] \, = \, 2\, \h^{12}\, + \, \h^{11} \, L_- \, + \, \h^{22}\, L_+ \, ,\\
& [\, L_-\,L_+ \comma \h^{22} \,] \, = \, 2\, \h^{22} \, + \, 2\,
\h^{12}\, L_-  \nn \, .
\end{align}
Therefore in $D=2$ and, as we have seen in eq.~\eqref{singled4},
with an $\{s_1+s_2-3,1\}$-projected parameter $\O_{12}$ one finds
\be
L_- \,L_+ \, \d\, \vf \, = \, (\,s_1-2\,)\,(\,s_2-3\,)\, \d\, \vf\,
,
\ee
so that $\d\,\vf$ is indeed irreducible. Comparing this result with
\eqref{sl2form} then shows that the shift affects only the
$\{s_1+s_2-3,3\}$ component of a rank-$(s_1,s_2)$ gauge field $\vf$.
In a similar fashion, in $D=4$ and with a symmetric parameter
$\O_{12}$ one finds
\be
L_- \,L_+ \, \d\, \vf \, = \, (\,s_1-1\,)\,(\,s_2-2\,)\, \d\, \vf \, ,
\ee
so that $\d\,\vf$ can only affect the $\{s_1+s_2-2,2\}$ component.
With a similar procedure it is also possible to show that only the
$\{3,1\}$ irreducible component of the reducible rank-$(2,2)$ field
in $D=2$ presents a shift symmetry, thus concluding the analysis
that leads to Table \ref{table2fam_irr}.

Finally, the presence in Table \ref{table2fam_irr} of a $\{2,2\}$
field in $D = 3$ rests on a new gauge symmetry for $\vf$. Indeed,
one can consider a shift of the form
\be \label{shiftdouble}
\d \, \vf \, = \, \h^{\,ij}\, \h^{\,kl} \, \O_{\,ij\,,\,kl} \, ,
\ee
under which the Fronsdal-Labastida tensor transforms as
\be
\d \, \cF \, = \, \h^{ij}\, \h^{kl} \, \cF_{\,ij\,,\,kl}\,(\,\O\,) \, + \, \h^{ij}\, \pr^{\,k}\pr^{\,l} \left[\, (\,D-3\,)\,
\O_{\,ij \comma kl} \, + \, S^{\,m}{}_{(\,k}\, \O_{\,l\,)\,m \comma ij} \,\right] \, .
\ee
If the parameters $\O_{\,ij\,,\,kl}$ satisfy
\be \label{eqdoubleshift}
(\,D-3\,)\, \O_{\,ij \comma kl} \, + \, S^{\,m}{}_{(\,i}\,
\O_{\,j\,)\,m \comma kl} \, = \, 0 \, ,
\ee
only the double trace of $\cF$ is affected, since
\be
\d \, \cF \, = \, \h^{ij}\, \h^{kl} \, \cF_{\,ij\,,\,kl}\,(\,\O\,)
\ee
while, as we have already seen in eq.~\eqref{condFshift},
eq.~\eqref{eqdoubleshift} implies
\be \label{condFshift2}
(\,D-3\,)\, \cF_{\,ij \comma kl}\,(\,\O\,) \, + \,
S^{\,m}{}_{(\,i}\,
\cF_{\,j\,)\,m \comma kl}\,(\,\O\,) \, = \, 0 \, .
\ee
In general, taking into account eq.~\eqref{condFshift2}, the
variation of the Einstein tensor under \eqref{shiftdouble} would
become
\be
\begin{split}
\d \, \cE_{\,\vf} \, & = \, - \, \12 \, \h^{ij}\, \h^{kl}\, \h^{mn}\, T_{ij}\, \cF_{\,kl \comma mn}\,(\,\O\,) \, + \,
\frac{2}{3} \left(\, Y_{\{2,2\}}\, \h^{ij}\,\h^{kl} \,\right) \h^{mn} \, T_{ij}\, \cF_{\,kl \comma mn}\,(\,\O\,) \\
& + \, \frac{1}{12} \left(\, Y_{\{2,2\}}\, \h^{ij}\,\h^{kl}
\,\right) \h^{mn}\, \h^{pq}\, T_{ij}\,T_{kl}\, \cF_{\,mn \comma pq}\,(\,\O\,) \,
,
\end{split}
\ee
that could be canceled by a suitable shift of the $\cB_{\,ijkl}$
tensors. To this end, one should recall that with only two families
the product of three or more $\h$'s can be always rewritten with two
of them explicitly symmetrized, as explained in Appendix
\ref{app:idsb}. Eq.~\eqref{eqdoubleshift} takes a form that is very
close to that of eq.~\eqref{eijdouble}, and actually the two differ
only due to the presence of the $\cB_{\,ijkl}$ tensors. In the
$\{2,2\}$ case the latter are actually not present, so that the two
equations coincide, and one can conclude that the presence of double
traces that are not fully determined reflects the new Weyl-like
gauge symmetry of eq.~\eqref{shiftdouble}. Actually, in the
$\{2,2\}$ case the analysis is far simpler, since the $\O_{\,ij
\comma kl}$ are scalars and thus no traces are available. For this
reason, the double-trace constraints are also identically satisfied.

%%%%%%%%%%%%%%%%%%%%%%%%%%%%%%%%%%%%%%%%%%%%%%%%%%%%%%%%%%%%%%%%%%%%%%%%%%%%%%%%%%%%%

\scss{Examples: reducible tensors of ranks $(s,1)$ and $(4,2)$}\label{sec:examples2b}

%%%%%%%%%%%%%%%%%%%%%%%%%%%%%%%%%%%%%%%%%%%%%%%%%%%%%%%%%%%%%%%%%%%%%%%%%%%%%%%%%%%%%

In the previous sections, Lagrangians and field equations were
presented in a concise notation capable of encompassing results
valid for all types of two-family bosonic fields, independently of
the Lorentz group labels that they carry. We thus feel it
appropriate to complement the analysis with explicit details on a
few types of models that stand out for their relative simplicity and
nonetheless can illustrate some key subtleties of the construction.
To this end, we shall also return momentarily to the standard
notation with space-time indices. Hopefully, this discussion will
also make the contents of the other sections more concrete and more
accessible for the interested readers. In this spirit, at time we
shall take the freedom to repeat and stress, in a more concrete
context, some points already made in the previous sections.

\vskip 12pt
%%%%%%%%%%%%%%%%%%%%%%%%%%%%%%%%%%%%%%%%%%%%%%%%%%%%%%%%%%%%%%%%%%%%%%%%%%%%%%%%%%%%%

\scsss{Lagrangians and field equations}\label{sec:Lagexamples2b}

%%%%%%%%%%%%%%%%%%%%%%%%%%%%%%%%%%%%%%%%%%%%%%%%%%%%%%%%%%%%%%%%%%%%%%%%%%%%%%%%%%%%%

In order to proceed, it is convenient to resort to a shorthand
notation for the possible traces. Here a trace with respect to a
pair of indices belonging to the same family will be denoted by a
``prime'' with a suffix that identifies the family, so that for
instance
\be
\h^{\, \m_1 \m_2} \, \vf_{\m_1 \m_2 \m_3 \m_4;\, \n_1 \n_2} \, \equiv \,
\vf^{\, \pe_{\m}}{}_{\m_3 \m_4 ;\, \n_1 \n_2} \, .\label{tracemu}
\ee
Two-family fields, however, admit a different type of trace that
brings together two indices belonging to the two different families.
This will be denoted by a ``hatted'' prime, so that for instance
\be
\h^{\, \m_1 \n_1} \, \vf_{\m_1 \m_2 \m_3 \m_4;\, \n_1 \n_2} \, \equiv \,
\vf^{\, \hpe}{}_{\m_2 \m_3 \m_4;\, \n_2} \, .\label{tracemixed}
\ee

In Section \ref{sec:irreducible} we shall explain how to adapt the
theory to the case of Young-projected fields, but here we keep
restricting our attention to unprojected fields, as in the previous
section. Our first class of examples, in particular, concerns
unprojected tensors of rank $(s,1)$, so that the gauge fields are of
the form
\be
\vf \, \equiv \, \vf_{\, \m_1 \ldots\, \m_s ; \, \n} \, ,
\ee
while the corresponding gauge transformations read
\be
\d \, \vf \, = \, \pr^{\,1} \, \L_{\,1} \, + \, \pr^{\,2} \, \L_{\,2} \, \equiv \, \pr_{\,(\, \m_1} \, \L^{(1)}{}_{\, \m_2
\ldots\, \m_s \, ) \, ; \, \n} \, + \pr_{\, \n} \, \L^{(2)}{}_{\m_1 \ldots\, \m_s} \, .
\ee
The $\Phi_{\,i}$ are now the two compensators
\bea
\Phi_{\,1} & \equiv & \Phi^{\,(1)}{}_{\!\m_1 \ldots \, \m_{s-1} ; \, \n} \, , \nonumber \\
\Phi_{\,2} & \equiv & \Phi^{\,(2)}{}_{\!\m_1 \ldots \, \m_s} \, ,
\eea
that enter the Lagrangians only via their symmetrized traces
\bea
\a_{\,111} & = & T_{11}\,\Phi_{\,1} \ \equiv \ \a^{\,(1)}{}_{\!\m_1 \ldots \, \m_{s-3} \, ; \, \n} \,  , \nonumber \\
\a_{\,112} & = & \frac{1}{3} \, \left(\, T_{11}\,\Phi_{\,2} \, + \, 2\,T_{12}\,\Phi_{\,1} \,\right) \ \equiv \ \a^{\,(2)}{}_{\!\m_1 \ldots \, \m_{s-2}} \,
,
\eea
and there are finally two Lagrange multipliers,
\bea
\b_{\,1111} & \equiv & \b^{\,(1)}{}_{\!\m_1 \ldots \, \m_{s-4} ; \, \n}   \, , \nonumber \\
\b_{\,1112} & \equiv & \b^{\,(2)}{}_{\!\m_1 \ldots \, \m_{s-3}}   \, .
\eea

In the conventional space-time notation, the Lagrangians \eqref{lag}
for this class of fields read
\be \label{lag_s1}
\begin{split}
\cL \, & = \, \12 \ \vf^{\,\m_1 \ldots\, \m_s ;\, \n} \left(\, \cA_{\,\m_1 \ldots\, \m_s ;\, \n} \, - \,
\12 \ \h_{\, (\, \m_1 \m_2} \,
\cA^{\, \pe_\m}{}_{\m_3 \ldots\, \m_{s} \,) ;\, \n} \,
- \, \12 \ \h_{\, \n \,(\, \m_1} \,
\cA^{\, \hpe}{}_{\m_3 \ldots\, \m_{s}\,)} \,\right) \\
& - \frac{3}{4} \left( s \atop 3 \right) \, \a^{\,(1)\, \m_1 \ldots\, \m_{s-3} ;\,\n} \ \pr^{\,\l} \, \cA^{\,\pe_\m}{}_{\l\,\m_1 \ldots\, \m_{s-3} ;\,\n} \\
& - \, \frac{1}{4} \left( s \atop 2 \right) \, \a^{\,(2)\, \m_1 \ldots\, \m_{s-2}} \, \pr^{\,\l} \left(\, \cA^{\,\pe_\m}{}_{\m_1 \ldots\, \m_{s-2} ;\,\l} \, + \, 2 \, \cA^{\,\hpe}{}_{\l\,\m_1 \ldots\, \m_{s-2}} \,\right) \\
& + \, 3 \left( s \atop 4 \right) \b^{\,(1)\, \m_1 \ldots\, \m_{s-4}
;\,\n} \ \cC^{\,(1)}{}_{\!\m_1 \ldots\, \m_{s-4} ;\,\n} + \,
3 \left( s \atop 3 \right) \b^{\,(2)\, \m_1 \ldots\,
\m_{s-3}} \ \cC^{\,(2)}{}_{\!\m_1 \ldots\, \m_{s-3}} \, ,
\end{split}
\ee
where the two $\cC^{\,(i)}$ tensors are the space-time
manifestations of $\cC_{1111}$ and $\cC_{1112}$:
\begin{alignat}{2} \label{c_s1}
& \cC^{\,(1)}{}_{\!\m_1 \, \ldots \, \m_{s-4} ; \, \n} & & = \, \h^{\,\m_1\m_2}\,\h^{\,\m_3\m_4} \left(\, \vf_{\, \m_1 \ldots\, \m_s ; \, \n} \, - \, \pr_{\,(\, \m_1}\Phi^{\,(1)}{}_{\!\m_2 \ldots\, \m_s \, ) \, ; \, \n} \, - \, \pr_{\, \n}\, \Phi^{\,(2)}{}_{\m_1 \ldots\, \m_s} \,\right) \, , \nonumber \\
& \cC^{\,(2)}{}_{\!\m_1 \, \ldots \, \m_{s-3}} & & = \, \h^{\,\m_1\m_2}\,\h^{\,\m_3\n} \left(\, \vf_{\, \m_1 \ldots\, \m_s ; \, \n} \, - \, \pr_{\,(\, \m_1}\Phi^{\,(1)}{}_{\! \m_2 \ldots\, \m_s \, ) \, ; \, \n} \, - \, \pr_{\, \n}\, \Phi^{\,(2)}{}_{\m_1 \ldots\, \m_s} \,\right) \, .
\end{alignat}

Let us stress that the Lagrangians of eq.~\eqref{lag_s1} contain
more terms involving compensators or multipliers with respect to
their counterparts for the symmetric case proposed in
\cite{fs3,fms}. Still, the Einstein-like tensor only contains simple
traces of $\cA$, because in this class of models the two available
double traces,
\bea
T_{11}\,T_{11}\,\cA & \equiv & \cA^{\,\pe_\m \,\pe_\m}{}_{\m_1 \ldots \, \m_{s-4} \, ; \, \n} \, , \nonumber \\
T_{11}\,T_{12}\,\cA & \equiv & \cA^{\,\pe_\m \,\hpe}{}_{\m_1
\ldots \, \m_{s-3}} \, ,
\eea
are actually \emph{symmetrized} double traces, that as such do not
enter eq.~\eqref{lag}. In this respect the $(s,1)$ fields are a bit
too simple to display all novelties of the mixed symmetry case, but
nonetheless, as we shall see shortly, they can convey interesting
lessons.

The equations of motion for the two lagrange multipliers
$\b^{\,(1)}$ and $\b^{\,(2)}$ read
\be
\begin{split}
&\cC^{\, (1)}{}_{\!\m_1 \ldots\, \m_{s - 4} ;\, \n} \,  = \, 0 \, , \\
&\cC^{\, (2)}{}_{\!\m_1 \ldots\, \m_{s - 3}} \, = \, 0 \, ,
\label{eqc_s1}
\end{split}
\ee
and reduce the $\vf$ field equation to
\be \label{eoms1}
\begin{split}
E_{\, \vf} \, & = \, \cA_{\,\m_1 \ldots\, \m_s ;\, \n} \, - \, \12
\,
\h_{\,(\,\m_1\m_2}\,\cA^{\, \pe_{\m} }{}_{\!\ldots\, \m_s \, ) ;\, \n} \,
- \, \12 \, \h_{\,\n \,(\, \m_1} \, \cA^{\, \hpe}{}_{\ldots\, \m_s\,)} \\
& + \, \h_{\, (\,\m_1 \m_2}\,\h_{\,\m_3
\m_4}\,\cB^{\,(1)}{}_{\!\ldots\,\m_s \,);\, \n} \, + \,
\h_{\, \n \, ( \, \m_1} \h_{\, \m_2 \m_3} \, \cB^{\, (2)}{}_{\!\ldots\,\m_s\,)} \, = \, 0 \, ,
\end{split}
\ee
where the two $\cB^{\,(i)}$ tensors are gauge invariant completions
of the two Lagrange multipliers. We shall refrain from discussing
further the field equations for the compensators that, as we have
seen in eq.~\eqref{eqsshift}, simply guarantee the conservation of
external currents.

Aside from the previous class of models, it will prove instructive
to also discuss in some detail the case of unprojected tensors of
rank $(4,2)$,
\be
\vf \, \equiv \, \vf_{\m_1\m_2\m_3\m_4 ;\, \n_1\n_2} \, ,
\ee
since in this relatively simple setting \emph{not all} double traces
are automatically symmetrized. In this case the gauge
transformations are
\be
\d \, \vf \, = \, \pr^{\,1} \, \L_{\,1} \, + \, \pr^{\,2} \, \L_{\,2} \, \equiv \,
\pr_{\,(\, \m_1} \, \L^{(1)}{}_{\! \m_2\m_3\m_4 \, )
\, ; \, \n_1\n_2} \, + \pr_{\, (\, \n_1\,|} \, \L^{(2)}{}_{\m_1 \m_2\m_3\m_4 ;\,|\,\n_2 \,) } \, ,
\label{redgauge42}
\ee
and there are two compensators, as is the cases for all two-family
fields, that in the present example have the following Lorentz
structure:
\bea
\Phi_{\,1} & \equiv & \Phi^{\,(1)}{}_{\!\m_1 \m_2\m_3 ; \, \n_1\n_2} \, , \nonumber \\
\Phi_{\,2} & \equiv & \Phi^{\,(2)}{}_{\!\m_1 \m_2\m_3\m_4 ;\,\n} \, .
\eea
A general two-family field would allow four distinct symmetrized
traces of the $\Phi_{\,i}$, and thus four distinct $\alpha_{ijk}$.
In this respect, the present example is still a bit degenerate,
since one can at most remove two indices from the second family, so
that one of the symmetrized traces is impossible and only three
$\a_{ijk}$ fields exist:
\bea
\a_{\,111} & = & T_{11}\,\Phi_{\,1} \ \equiv \ \a^{\,(1)}{}_{\m \, ; \, \n_1\n_2} \,  , \nonumber \\
\a_{\,112} & = & \frac{1}{3} \, \left(\, T_{11}\,\Phi_{\,2} \, + \, 2\,T_{12}\,\Phi_{\,1} \,\right) \ \equiv \ \a^{\,(2)}{}_{\m_1 \m_2 ;\,\n} \, , \\
\a_{\,122} & = & \frac{1}{3} \, \left(\, T_{22}\,\Phi_{\,1} \, + \, 2\,T_{12}\,\Phi_{\,2} \,\right) \ \equiv \ \a^{\,(3)}{}_{\m_1\m_2\m_3} \, . \nonumber
\eea

In a similar fashion, the present $(4,2)$ example only allows three
symmetrized double traces, and thus the three Labastida constraints
\bea
&& T_{11}\,T_{11} \,\vf \ \, \equiv \ \, \vf^{\,\pe_\m \,\pe_\m}{}_{\n_1\n_2} \, = \, 0 \, , \nonumber\\
&& T_{11}\,T_{12} \,\vf \ \, \equiv \ \, \vf^{\,\pe_\m \, \hpe}{}_{\m\, ; \, \n} \, = \, 0  \, , \\
&& \left(\, T_{11}\,T_{22} \, + \, 2 \, T_{12}\,T_{12} \,\right)
\,\vf \ \, \equiv \ \,
 \vf^{\,\pe_\m \,\pe_\n}{}_{\m_1\m_2} \, + \, 2 \, \vf^{\,\hpe \,\hpe}{}_{\m_1\m_2} \, = \, 0  \, , \nonumber
\eea
rather than the five available for generic two-family fields, but
the interesting novelty with respect to the previous class of models
is the existence of a single non-symmetrized double trace, that
therefore plays an interesting role both in the Lagrangian and in
the corresponding field equations. This non-symmetrized double trace
bears family indices subject to a $\{2,2\}$ projection, is present
for all generic two-family models and in the present example reads
\be
\left(\, T_{11}\,T_{22} \, - \, T_{12}\,T_{12}\,\right) \vf \ \, \equiv \ \, \vf^{\,\pe_\m \,\pe_\n}{}_{\m_1\m_2} \, - \,
\vf^{\,\hpe \,\hpe}{}_{\m_1\m_2}  \, .
\ee
The structure of the Labastida constraints for this model clearly
reflects itself in the constraint tensors $\cC$ and in the Lagrange
multipliers $\b$ that couple to them:
\bea
\b_{\,1111} & \equiv & \b^{\,(1)}{}_{\n_1\n_2}  \, , \nonumber \\
\b_{\,1112} & \equiv & \b^{\,(2)}{}_{\m\,;\, \n}  \, , \\
\b_{\,1122} & \equiv & \b^{\,(3)}{}_{\m_1\m_2}  \, . \nonumber
\eea

The Lagrangian for unprojected fields of rank $(4,2)$ thus reads
\be
\begin{split}
\cL \, & = \, \12 \ \vf^{\, \m_1 \ldots\, \m_4 ;\, \n_1\n_2} \, (\cE_{\,\vf})_{\, \m_1 \ldots\, \m_4 ;\, \n_1\n_2} \\
& - \, 3 \, \a^{\,(1)\, \m\, ; \, \n_1\n_2} \, \pr^{\,\l} \left\{\, \cA^{\,\pe_\m}{}_{\l\,\m\,;\,\n_1\n_2} \, - \, \frac{1}{6} \ \h_{\,\n_1\n_2} \left(\, \cA^{\, \pe_{\m} \pe_{\n}} - \, \cA^{\, \hpe\, \hpe} \,\right){}_{\!\m\,\l}  \,\right\} \\
& - \, 3 \, \a^{\,(2)\, \m_1\m_2\, ; \, \n} \, \pr^{\,\l} \left\{ \left(\, \cA^{\,\pe_\m}{}_{\m_1\m_2;\,\n\l} + \, 2 \, \cA^{\,\hpe}{}_{\l\,\m_1\m_2;\,\n} \,\right) + \, \frac{1}{6} \ \h_{\,\n\,(\,\m_1\,|} \left(\, \cA^{\, \pe_{\m} \pe_{\n}} - \, \cA^{\, \hpe\, \hpe} \,\right){}_{\!|\,\m_2\,)\,\l}  \,\right\} \\
& - \, \a^{\,(3)\, \m_1\m_2\m_3} \, \pr^{\,\l} \left\{ \left( \cA^{\,\pe_\n}{}_{\l\,\m_1\m_2\m_3} + \, 2 \, \cA^{\,\hpe}{}_{\m_1\m_2\m_3;\,\l} \right) - \, \frac{1}{6} \ \h_{\,(\,\m_1\m_2\,|}\! \left( \cA^{\, \pe_{\m} \pe_{\n}} - \, \cA^{\, \hpe\, \hpe} \,\right){}_{\!|\,\m_3\,)\,\l}  \,\right\} \\
& + \, 3 \, \b^{\,(1)\,\n_1\n_2} \, \cC^{\,(1)}{}_{\n_1\n_2} \, + \,
24 \,  \b^{\,(2)\,\m\,;\,\n} \, \cC^{\,(2)}{}_{\m\,;\,\n} \, + \, 18
\,  \b^{\,(3)\,\m_1\m_2} \, \cC^{\,(3)}{}_{\m_1\m_2} \, ,
\label{l42expl}
\end{split}
\ee
with
\be
\begin{split} \label{einst4,2}
\cE_\vf \, & = \, \cA_{\,\m_1 \ldots\, \m_4 ;\, \n_1\n_2} \\
& - \, \12 \ \h_{\,(\, \m_1 \m_2} \, \cA^{\,
\pe_{\m}}{}_{\m_3\m_4\,) ;\, \n_1\n_2} \, - \, \12 \
\h_{\,(\,\n_1\,|\,(\,\m_1} \, \cA^{\, \hpe}{}_{\m_2\m_3\m_4\,)
;\,|\,\n_2\,)} \, - \, \12 \
\h_{\, \n_1 \n_2} \, \cA^{\, \pe_{\n}}{}_{\m_1 \ldots\, \m_4} \\
& + \, \fr{1}{18} \left(\, 2\ \h_{\, \n_1 \n_2} \,
\h_{\,(\,\m_1\m_2\,|} \, - \, \h_{\, \n_1 \,(\, \m_1}
\h_{\,\m_2\,|\,\n_2} \,\right) \left(\, \cA^{\, \pe_{\m} \pe_{\n}} -
\, \cA^{\, \hpe\, \hpe} \,\right){}_{\!|\,\m_3\m_4\,)} \, .
\end{split}
\ee
In the following we shall often use the shorthand notation
\be
a_{\, \m_1 \m_2} \, \equiv \, \cA^{\, \pe_{\m} \pe_{\n}}{}_{\, \m_1
\m_2 }\, -
\, \cA^{\, \hpe \, \hpe}{}_{\, \m_1 \m_2 } \, \label{smalla}
\ee
to denote this $\{2,2\}$-projected trace of $\cA$ that appears in
$\cE_\vf$. The equations for the Lagrange multipliers are simply
\begin{alignat}{2} \label{eqc4,2}
& \cC^{\,(1)}{}_{\n_1\n_2} \!& & = \, 0 \, , \nonumber \\
& \cC^{\,(2)}{}_{\m\,;\, \n} \!& & = \, 0 \, , \\
& \cC^{\,(3)}{}_{\m_1\m_2} \!& & = \, 0 \, , \nonumber
\end{alignat}
and after enforcing them the equation for the gauge field $\vf_{\m_1
\ldots\, \m_4 ;\,
\n_1\n_2}$ becomes
\be \label{eq4,2uncon}
\begin{split}
E_\vf \, = \, \cE_\vf \, & + \, \h_{\,(\,\m_1\m_2}\,\h_{\,\m_3\m_4\,)}\, \cB^{\,(1)}{}_{\n_1\n_2} \, + \, \h_{\,(\,\n_1\,|\,(\,\m_1}\,\h_{\,\m_2\m_3}\, \cB^{\,(2)}{}_{\,\m_4\,)\,;\,|\,\n_2\,)} \\
& + \left(\, \h_{\,\n_1\n_2}\,\h_{\,(\,\m_1\m_2\,|} \, + \,
\h_{\,\n_1\,(\,\m_1}\h_{\,\m_2\,|\,\n_2}\, \right)
\cB^{\,(3)}{}_{|\,\m_3\m_4\,)} \, = \, 0 \, .
\end{split}
\ee

In order to appreciate the relative simplicity of the previous
results, it is worth comparing eq.~\eqref{eq4,2uncon} with the
corresponding Labastida equation. This, as we have stressed, differs
from the result quoted in \cite{labastida}, since it must include
some higher traces that are needed in order that it satisfy the same
double trace constraints as the Labastida gauge field. The subtlety
draws its origin from the presence of the non-vanishing double trace
\eqref{smalla}, that also brings about non-vanishing \emph{symmetrized}
double traces of the Einstein tensor,
\be \label{symdt4,2}
\begin{split}
&\cE_{\,\vf}^{\, \pe_{\m} \pe_{\m}}{}_{\n_1\n_2} \, = \, - \, \fr{4}{9} \ a_{\, \n_1\n_2} \, , \\
&\cE_{\,\vf}^{\, \pe_{\m} \hpe}{}_{\m\,;\,\n} \, = \, \fr{1}{9} \ a_{\, \m \n} \, , \\
&\cE_{\,\vf}^{\, \pe_{\m} \pe_{\n}}{}_{\m_1\m_2} \, + \, 2 \
\cE_{\vf}^{\, \hpe\,
\hpe}{}_{\m_1\m_2}  \, = \, - \, \fr{2}{9} \ a_{\, \m_1 \m_2} \, ,
\end{split}
\ee
where the $\{2,2\}$ projected trace $a_{\,\m_1\m_2}$ can appear on account of
eq.~\eqref{s_symdt}.

For completeness, we have presented these results for the
unconstrained theory, but at any rate their Labastida counterparts
can be simply recovered from the previous expressions removing the
$\Phi_{\,i}$. As we anticipated, the proper Lagrangian Labastida
equation is indeed more complicated than one would naively expect,
and reads
\be \label{proj4,2}
\begin{split}
E_{\vf} \, = \, \cE_\vf \, & + \, \fr{4}{3\, (3\, D^2 - 8)} \
\h_{\,(\,\m_1\m_2}\,\h_{\,\m_3\m_4\,)} \, a_{\, \n_1 \n_2} \,
- \, \fr{1}{3\, (3 \, D^2 - 8)} \ \h_{\,(\,\n_1\,|\,(\,\m_1}\,  \h_{\,\m_2 \m_3}\, a_{\,\m_4\,)\,|\,\n_2\,)} \\
& +\, \fr{2}{9 \, (3\, D^2 - 8)} \ \left(\, \h_{\, \n_1 \n_2} \,
\h_{\,(\,\m_1\m_2\,|} \, + \, \h_{\, \n_1 \,(\,
\m_1}\h_{\,\m_2\,|\,\n_2} \,\right)  \, a_{\,|\, \m_3 \m_4\,)} \, =
\, 0 \, ,
\end{split}
\ee
where the coefficients depend explicitly on the space-time
dimension. This result should be compared with the relatively simple
and universal form of eq.~\eqref{eq4,2uncon}. Let us stress that for
higher-rank fields the relative complication of the constrained
theory would be even more sizable, because all possible combinations
of the non-vanishing double trace with reshuffled indices would
contribute.

We can now illustrate explicitly the on-shell reduction of these
types of fields. In the ensuing discussion, we shall often make use
of the shorthand notation
\be \label{rhok}
\rho_{\,k} \, \equiv \, D \, + \, k \, ,
\ee
where $D$ denotes the space-time dimension.

\vskip 12pt

%%%%%%%%%%%%%%%%%%%%%%%%%%%%%%%%%%%%%%%%%%%%%%%%%%%%%%%%%%%%%%%%%%%%%%%%%%%

\scsss{On-shell reduction} \label{sec:s1b}

%%%%%%%%%%%%%%%%%%%%%%%%%%%%%%%%%%%%%%%%%%%%%%%%%%%%%%%%%%%%%%%%%%%%%%%%%%%

As we have stressed, the constraints \eqref{eqc_s1} force the two
types of double traces of the kinetic tensor $\cA_{\,\m_1 \ldots\,
\m_s ;\,\n}$ of a generic $(s,1)$ field $\vf_{\m_1 \ldots\, \m_s ;
\,
\n}$ to vanish, so that these fields entail no new subtleties,
in this respect, when compared to their fully symmetric
counterparts. They are thus particularly convenient to illustrate
how the $\cB_{ijkl}$ fields are not fully determined, and how
nonetheless this peculiarity does not affect the field equation for
$\vf$. We can conveniently begin by describing the reduction
procedure for a relatively simple rank-six tensor of $(5,1)$ type,
before moving to analyze generic $(s,1)$ fields.

After enforcing the constraints \eqref{eqc_s1}, the equation of
motion \eqref{eoms1} for a $(5, 1)$-field $\vf_{\m_1
\ldots\, \m_5 ;\, \n}$ becomes
\be \label{eom51}
\begin{split}
E_{\, \vf} \, & = \, \cA_{\,\m_1 \ldots\, \m_5 ;\, \n} \, - \, \12 \
\h_{\,(\,\m_1\m_2}\,\cA^{\, \pe_{\m} }{}_{\m_3 \m_4 \m_5 \, ) ;\, \n} \,
- \, \12 \ \h_{\,\n \,(\, \m_1} \, \cA^{\, \hpe}{}_{\m_2 \m_3 \m_4 \m_5\,)} \\
& + \, \h_{\, (\,\m_1 \m_2}\,\h_{\,\m_3 \m_4}\,\cB^{\,(1)}{}_{\m_5
\,);\, \n} \, + \,
\h_{\, \n \, ( \, \m_1} \, \h_{\, \m_2 \m_3} \, \cB^{\, (2)}{}_{\m_4\m_5\,)} \, = \, 0 \, .
\end{split}
\ee
In this relatively simple model, and actually in the whole family of
$(s,1)$ examples that we shall shortly discuss, one can compute
explicitly all possible traces of (\ref{eom51}) in order to extract
the conditions on the $\cB_{ijkl}$ tensors, that in this case are
the pair $\cB^{\, (1)}{}_{\mu;\,\nu}$ and $\cB^{\,
(2)}{}_{\mu_1\mu_2}$, as we saw in Section
\ref{sec:Lagexamples2b}. Taking the two independent double traces of (\ref{eom51})
yields
\begin{alignat}{2} \label{double51}
& E_{\, \vf}^{\, \pe_{\m} \pe_{\m}} & & : \quad \rho_2 \,
\cB^{\,(1)}{}_{\m \,;\, \n} \, + \, 4 \, \cB^{\, (2)}{}_{\m \n} \, + \, 2 \, \h_{\, \m \n}\, \cB^{\, (2) \, \pe_{\m}} \, = \, 0 \ , \nonumber \\
& E_{\, \vf}^{\, \pe_{\m} \hpe} & & : \quad \cB^{\, (1)}{}_{(\m_1 ;\,
\m_2)} \, + \, \rho_{\,4} \, \cB^{\, (2)}{}_{\m_1 \m_2} \, + \, \h_{\m_1
\m_2} \left(\, \cB^{\, (1)\, \hpe } \, + \, \cB^{\, (2) \, \pe_{\m}}
\,\right) \, = \, 0 \, ,
\end{alignat}
while there is only one possible triple trace,
\be
E_{\, \vf}^{\, \pe_{\m} \pe_{\m} \hpe} \, : \quad \cB^{\,(1) \, \hpe}
\, + \, 2 \,  \cB^{\,(2) \, \pe_{\m}} \, = \, 0\, .
\ee
Clearly, one can at most relate the traces of $\cB^{\,( 1)}$ and
$\cB^{\,( 2)}$ to one another, so that in terms of $\cB^{\,(2){\,
\pe_\m}}$ the system \eqref{double51} finally becomes
\begin{alignat}{2} \label{six1}
& E_{\, \vf}^{\, \pe_{\m} \pe_{\m}}{}_{\!(\,\m \, ; \, \n\,)} & & :
\quad \frac{\rho_2}{2} \,
\cB^{\,(1)}{}_{(\,\m\,;\,\n\,)} \, + \, 4 \, \cB^{\,(2)}{}_{\m\n} \, + \, 2 \, \h_{\m\n} \, \cB^{\,(2) \pe_{\m}}  \, = \, 0 \, , \nonumber \\
& E_{\, \vf}^{\, \pe_{\m} \pe_{\m}}{}_{\!
[\,\m \,;\, \n\,]}  & &  : \quad \cB^{\,(1)}{}_{[\,\m\,;\,\n\,]} \, = \, 0 \, ,  \\
& E_{\, \vf}^{\, \pe_{\m} \hpe}{}_{\m_1 \m_2} & & : \quad
\cB^{\,(1)}{}_{\,(\,\m_1 ;\, \m_2\,)} \, + \, \rho_{\,4} \,
\cB^{\,(2)}{}_{\m_1\m_2} \, - \, \h_{\m_1\m_2} \,
\cB^{\,(2)\,\pe_{\m}}\, = \, 0 \, . \nonumber
\end{alignat}
The solution of eqs.~\eqref{six1} is therefore
\begin{alignat}{2} \label{sol51}
&\cB^{\, (1)}{}_{\m \,;\, \n}  & & = \, - \, \frac{2}{\rho_0} \ \eta_{\m\n} \, \cB^{\, (2) \, \pe_{\m}} \ , \nonumber \\
&\cB^{\, (2)}{}_{\m_1 \m_2} & & = \, \frac{1}{\rho_0} \ \eta_{\m_1 \m_2}\, \cB^{\, (2) \, \pe_{\m}} \ .
\end{alignat}

These results display quite clearly two features of the reduction
process that are foreign to the symmetric case:
\begin{itemize}
 \item in general, the field equations do not force all
 components of $\cB_{ijkl}$ to vanish. In this case, for instance, only the
 traceless parts of $\cB^{\, (1)}$ and $\cB^{\, (2)}$ are forced to vanish,
 while their traces are simply subject to a linear relation. As a result, one can take, say, $\cB^{\, (2) \, \pe_{\m}}$ as an independent
 variable, but as we shall see shortly this undetermined quantity does not enter the resulting equation of motion for $\vf$;
 \item eq.~(\ref{double51}) allows two independent projections that are to be treated
 separately. In general, there are as many independent conditions as the irreducible projections of
 all traces of the original $\vf$-equation. This is an explicit
 example of the decomposition \eqref{dec_irr}.
\end{itemize}

It is instructive to see explicitly how the undetermined trace in
(\ref{six1}) disappears from the final result. Substituting
eqs.~\eqref{sol51} in eq.~\eqref{eom51} one indeed obtains for the
reduced $\vf$-equation
\be
\begin{split}\label{eom51f}
E_{\, \vf} \, & \equiv \, \cA_{\,\m_1 \m_2 \m_3 \m_4 \m_5 ;\, \n} \,
-
\,
\12 \,
\h_{\,(\,\m_1\m_2}\,\cA^{\, \pe_{\m} }{}_{\m_3 \m_4 \m_5 \, ) ;\, \n} \,
- \, \12 \, \h_{\,\n \,(\, \m_1} \, \cA^{\, \hpe}{}_{\m_2 \m_3 \m_4 \m_5\,)} \\
& - \, \frac{2}{\rho_0} \, \h_{\, (\,\m_1 \m_2}\,\h_{\,\m_3
\m_4}\,\h_{\m_5 \,)\, \n}\, \cB^{\,(2) \, \pe_{\m}} \, + \,
\frac{2}{\rho_0} \, \h_{\, \n \, ( \, \m_1} \, \h_{\, \m_2 \m_3} \, \h_{\m_4\m_5\,)} \, \cB^{\,(2) \, \pe_{\m}} = 0 \ ,
\end{split}
\ee
where the reader should notice that an overall factor 2 was
generated in the second $\cB$ term, precisely as needed in order
that $\cB^{\,(2)}$ disappear from the resulting expression. This
overall factor draws its origin from the standard symmetrization,
that overcounts by a factor two the minimal number of permutations
needed to define a symmetric pair of $\eta$ tensors. In other words,
one is facing a manifestation of the rule according to which
$\eta\,\eta =
\,2\,\eta^{\,2}$, that was spelled out for the symmetric case
in \cite{fms}. Having eliminated $\cB_{ijkl}$ from the $\vf$
equation, the reduction of \eqref{eom51f} to $\cA_{\,\m_1
\ldots \, \m_5;\, \n} \, = \, 0$ can finally proceed as in the constrained case.

This result can be understood comparing the overall number of
possible projections for the successive traces of the $\vf$ field
equation with the number of unknown tensorial quantities arising
from $\cB_{ijkl}$. One could also count, more conveniently, the
available traces of the various quantities, that here are simply
bound to allow two independent projections if the $\n$ index is
still present and one otherwise. In the previous example, for
instance, we had to face a single triple trace involving two
distinct unknowns, the traces of $\cB^{\,(1)}$ and $\cB^{\,(2)}$. In
addition, we faced two distinct double traces involving the two
distinct tensors $\cB^{\,(1)}$ and $\cB^{\,(2)}$. We thus ended up
with one undetermined quantity, say $\cB^{\,(2)\,
\prime_\m}$, that however did not affect the reduced equation for
the gauge field $\vf_{\m_1 \ldots\, \m_5;\,\n}$.

Moving on to the general case, one can note that an $(s,1)$ field
$\vf_{\m_1 \ldots\, \m_s;\,\n}$ admits two independent types of
$k$-th traces, for $k<\left[\frac{s+1}{2}\right]$. In the first,
here denoted by $\vf^{[k]}$, all traced indices belong to the first
set, so that the $\n$ index is left untouched, while in the second,
here denoted by $\vf^{[k-1]\,
\hpe}$, one of the traces involves the $\n$ index of the
second family. It is also convenient to distinguish two cases when
analyzing the equations starting from the highest trace, the $n$-th
one, with $n=\left[\frac{s+1}{2}\right]$. If $s$ is odd the $n$-th
trace is unique and yields a scalar equation, but as in the
$\{5,1\}$ examples this involves
\emph{two} unknowns, the two highest traces $\cB^{\,(1) [n-3] \, \hpe}$ and
$\cB^{\,(2) [n-2]}$ of $\cB^{\,(1)}$ and $\cB^{\,(2)}$. On the other
hand, if $s$ is even there are two distinct $n$-th traces leading to
two distinct vector equations, but these involve
\emph{three} independent unknowns, $\cB^{\,(1) [n-3] \, \hpe}$,
$\cB^{\,(1) [n-2]}$ and $\cB^{\,(2) [n-2]}$. Moving backwards to the
lower traces of the $\vf$ field equation, at every step the new
unknowns, three lower traces, $\cB^{\,(1) [k-3] \, \hpe}$,
$\cB^{\,(1) [k-2]}$ and $\cB^{\,(2) [k-2]}$, exceed by one unit the
available traces of the $\vf$ field equation, $E_\vf{}^{[k-1]
\, \hpe}$ and $E_\vf{}^{\,(1) [k]}$. The
last step, however, is slightly different, since the two double
traces of the $\vf$ field equation match at this order the two new
entries, $\cB^{\, (1)}$ and $\cB^{\, (2)}$. All in all, for an
$(s,1)$ field one thus finds an excess of
$\left[\frac{s-3}{2}\right]$ unknowns. Still, in the general case
the $\vf$ field equation can be cast in a form that does not involve
any of these undetermined quantities. This can be seen from the
explicit solution for $\cB^{\,(1)}$ and $\cB^{\,(2)}$, that can be
obtained in closed form for generic $\{s,1\}$ fields. For brevity,
here we limit ourselves to quoting the final results,
\be
\begin{split}
& \cB^{\,(1)}{}_{\m_1 \ldots\, \m_{s-4} ;\, \n} \, = \, 2 \ \sum_{l\,=\,0}^{n-3} \ \frac{(-1)^{\,l}}{l+1} \, \frac{1}{\prod_{\,i\,=\,0}^{\,l}
\, \rho_{\,2(2n-5-i)}} \ \h_{\n \, \,(\, \m_1} \, \h^{\,l}{}_{\m_2 \ldots\, \m_{2l+1}}\, \cB^{\,(2)\,[\,l+1\,]}{}_{\ldots\, \m_{s-4}\,)}\, , \\
& \cB^{\,(2)}{}_{\m_1 \ldots\, \m_{s-3}} \, = \, \sum_{l\,=\,0}^{n-3} \ \frac{(-1)^{\,l}}{\prod_{\,i\,=\,0}^{\,l}
\, \rho_{\,2(2n-5-i)}} \ \h^{\,l+1}{}_{(\,\m_1 \ldots\, \m_{2(l+1)}} \, \cB^{\,(2)\,[\,l+1\,]}{}_{\ldots\,
\m_{s-3}\,)} \, ,\label{b12general}
\end{split}
\ee
where $\eta^{\,l}$ denotes a symmetrized product of $l$ Minkowski
metrics with the minimal number of terms, consistently with the
notation of Appendix \ref{app:MIX} and of \cite{fms}. Substituting
eqs.~\eqref{b12general} in the $\vf$ equation of motion finally
gives
\bea
0 &\!\! = \!\!& \cA_{\,\m_1 \ldots\, \m_s ;\, \n} \, - \, \12 \
\h_{\, (\, \m_1 \m_2} \,
\cA^{\, \pe_{\, \m}}{}_{\!\! \ldots\, \m_{s} \,) ;\, \n} \,
- \, \12 \ \h_{\, \n \,(\, \m_1} \,
\cA^{\, \hpe}{}_{\ldots\, \m_{s}\,)} \\
&\!\! + \!\!& \sum_{l\,=\,0}^{n-3} \
\frac{(-1)^{\,l}}{\prod_{\,i\,=\,0}^{\,l}
\, \rho_{\,2(2n-5-i)}} \, \left\{ \, \frac{2}{l+1} \binom{l+2}{2} \, - \, (l+2) \,
\right\}
\, \h_{\n \, \,(\, \m_1} \, \h^{\,l+2}{}_{\m_2 \ldots\, \m_{2l+5}}\, \cB^{\,(2)\,[\,l+1\,]}{}_{\ldots\, \m_s\,)}\, . \nonumber
\eea
As we have stressed, the theory does not fully determine the traces
of the $\cB^{\, (i)}$ tensors, but the reader should appreciate that
they all cancel out manifestly in this expression. Hence, in this
class of models, this novel phenomenon does not hamper the reduction
of the $\vf$ field equation, that after all ought to be possible for
physical reasons.

Now we would like to elaborate again on the origin of this peculiar
feature, that actually presents itself for all mixed-symmetry fields
and whose origin can be traced to the symmetry principle discussed
in Section
\ref{sec:lagrangian2b}. Indeed, as we have seen, aside from some degenerate cases of low ranks, the
Lagrangians of mixed-symmetry fields possess a symmetry allowing
shifts of the Lagrange multipliers in $\cB_{ijkl}$ according to
\be
\delta \, \b_{ijkl} \, = \, \eta^{\,mn} \, L_{\,ijkl,\,mn} \ , \label{shbsym}
\ee
where $L_{\,ijkl,\,mn}$ is $\{5,1\}$ projected. Clearly, these
transformations also shift the $\cB_{\,ijkl}$ by the same amount, as
in eq.~\eqref{Bshift}. As we pointed out in Section
\ref{sec:lagrangian2b}, this symmetry draws its origin from the nature of the $\cC_{ijkl}$ constraints,
that start with $T_{ij}\,T_{kl}\,\vf$. The point is that a further
trace simply can not give rise to a $\{5,1\}$ projection when
combined with the others accompanying $\vf$, as the reader can
verify, due to the presence of identical $T_{ij}$ tensors. Let us
stress, however, that the whole $\cC_{\,ijkl}$ constraints share
this property of the term involving the gauge field $\vf$
\emph{only if} the $\alpha_{\,ijk}$ are expressed in terms of the
independent compensators $\Phi_{\,i}$, as should be clear from
eq.~\eqref{C2}, but not if one were to work with independent
$\a_{\,ijk}$.

The symmetry \eqref{shbsym} has a simple origin: as we have
stressed, while nicely covariant with respect to the family
structure, the Labastida constraints \emph{are not} independent,
since higher traces of two different constraints can give rise to
the same condition. For instance, referring to the previous example
one can see that the two tensors $\cC^{\, (1)}$ and $\cC^{\, (2)}$
are not fully independent, since indeed
 \be
 \cC^{\, (1)\, \hpe}{}_{\m_1 \ldots\, \m_{s-5}} \, = \, \cC^{\, (2)\, \pe_\m}{}_{\m_1 \ldots\, \m_{s-5}} \,
 .
 \ee

In this respect, the symmetry
\eqref{shbsym} is precisely as needed to remove all left-over
components of $\cB^{\, (1)}$ and $\cB^{\, (2)}$. Indeed,
specializing eq.~\eqref{shbsym} to the case of rank--$\,(5,1)$ $\vf$
fields gives
\be
\begin{split}
& \d \, \cB_{\,1111} \, = \, 2\, \eta^{\,12}\, L_{\,1111,\,12} \ ,\\
& \d \, \cB_{\,1112} \, = \, \eta^{\,11}\, L_{\,1112,\,11} \ ,
\end{split}
\ee
where in this case the $\{5,1\}$ projection requires that the two
gauge parameters above satisfy the relation
\be
L_{\,1111,\,12} \, = \, - \, 2 \, L_{\,1112,\,11} \ . \label{rel51}
\ee
Translating these results in the conventional space-time notation
then yields
\be
\begin{split}
& \d \, \cB^{\, (1)}{}_{\m\,;\,\n} \, = \, - \, 2 \, \eta_{\m\n} \, L \ ,\\
& \d \, \cB^{\, (2)}{}_{\m_1\m_2} \, = \, \eta_{\m_1\m_2} \, L \ ,
\end{split}
\ee
where $L$ denotes the single scalar parameter left over by
eq.~\eqref{rel51}. Clearly, these transformations allow to eliminate
the undetermined quantities prior to the substitution in
eq.~\eqref{eom51}.

For more general tensors, the reduction of the field equations to
the Labastida form proceeds along similar lines, albeit with higher
technical complications that are inevitable when working in an
explicit space-time notation. It is instructive to dwell upon some
of these, and therefore we now turn to illustrate the case of
$(4,2)$ tensors, where some non-vanishing double traces of $\cA$ are
present even after enforcing the $\cC_{ijkl}$ constraints. Our aim
will be, again, to reach the non-Lagrangian Labastida equation $\cF
= 0$ starting from (\ref{eq4,2uncon}). However, an intermediate
stage of our analysis will be also of some interest, since it will
recover the proper Lagrangian form of the Labastida equation
\eqref{proj4,2}, that differs from the result presented in
\cite{labastida} by the addition of some terms needed to project the
Einstein-like tensor in such a way that its symmetrized double
traces vanish identically. The analysis of this example terminates
with the derivation of the current-current exchange for this type of
field, an interesting application of the formalism that we refrained
from presenting for the previous class of fields for brevity.

In order to accomplish our task, we can proceed by direct
substitutions, as in the $(s,1)$ cases. Starting from
\eqref{eq4,2uncon}, let us therefore begin by considering the three
symmetrized double traces
\be \label{syst4,2}
\begin{split}
&E_{\,\vf}^{\, \pe_{\m} \pe_{\m}} \, = \, - \, \fr{4}{9} \, a_{\,
\n_1
\n_2} \, + \, \rho_0 \, \rho_2 \, \cB^{\, (1)}{}_{\n_1 \n_2} \, + \, 4 \, \rho_2
\, \cB^{\, (2)}{}_{(\,\n_1 ;\, \n_2 \,)} \, + \,8 \,  \cB^{\, (3)}{}_{\n_1 \n_2} \\
&\phantom{E_{\vf}^{\, \pe_{\m} \pe_{\m}} \, = \,} + \, 2\, \rho_{4} \, \h_{\, \n_1 \n_2} \, \cB^{\, (3)\, \pe_{\m}} \, = \, 0 \, , \\
&E_{\,\vf}^{\, \pe_{\m} \hpe} \, = \, + \, \fr{1}{9} \, a_{\, \m \n}
\, + \, \rho_2 \, \cB^{\, (1)}{}_{\m \n} \, + \, \rho_2 \, \rho_4 \,
\cB^{\, (2)}{}_{\m\,;\, \n} \, + \, 2 \,  \cB^{\, (2)}{}_{(\,\m\,;\, \n\,)} \,
+ \, (\,3\, \rho_3 \, + \, 1\,) \, \cB^{\, (3)}{}_{\m \n} \\
& \phantom{E_{\vf}^{\, \pe_{\m} \hpe} \, = \,} + \, \rho_4 \, \h_{\,
\m \n} \, \cB^{\, (2)\, \hpe}
\, + \, \rho_4 \, \h_{\, \m \n} \,
\cB^{\, (3)\, \pe_{\m}}\, = \, 0 \, ,\\
&E_{\,\vf}^{\, \pe_{\m} \pe_{\n}} \, + \, 2 \, E_{\vf}^{\, \hpe \,
\hpe}  \, = \, - \, \fr{2}{9} \, a_{\, \m_1 \m_2} \, + \, 4 \, \cB^{\,
(1)}{}_{\m_1 \m_2} \, + \, 2\, (\,3\, \rho_3 \, + \, 1\,) \, \cB^{\, (2)}{}_{(\,\m_1;\,\m_2\,)} \\
& \phantom{E_{\vf}^{\, \pe_{\m} \pe_{\n}}} + \, (\,3\,
\rho_{\,3}^{\, 2} \, + \, 1\,) \, \cB^{\, (3)}{}_{\m_1 \m_2} \, + \,
\rho_4 \, \h_{\, \m_1 \m_2} \, \left(\,\cB^{\, (1)\,
\pe_{\n}} \, + \, 4 \,  \cB^{\, (2)\, \hpe}  \, + \,
\cB^{\, (3)\, \pe_{\m}} \,\right) \, = \, 0 \, ,
\end{split}
\ee
and the two triple traces of \eqref{eq4,2uncon},
\be
\begin{split}
& E_{\,\vf}^{\, \pe_{\m} \pe_{\m} \pe_{\n}}  \, =  \,  \rho_0 \,
\rho_2
\, \cB^{\, (1)\, \pe_{\n}} \, + \, 8 \, \rho_2 \, \cB^{\, (2) \, \hpe}
\, + \,2 \, \rho_2^{\,2} \,  \cB^{\, (3)\, \pe_{\m}}
\, = \, 0 \, , \\
& E_{\,\vf}^{\, \pe_{\m} \hpe \, \hpe} \, = \, \rho{\,_2} \, \cB^{\,
(1)\,
\pe_{\n}} \, + \, 2 \, \rho_2 \, \rho_3 \, \cB^{\, (2) \, \hpe} \, + \,
\rho_2 \, \rho_ 5 \,  \cB^{\, (3)\, \pe_{\m}}
\, = \, 0 \, .
\end{split}
\ee
As in eq.~\eqref{smalla}, $a_{\m_1\m_2}$ denotes the
$\{2,2\}$-projected double trace of $\cA$, that as we stressed is
not subject to any constraints. Notice that the last two equations
suffice to express, for instance, the traces of $\cB^{\, (1)}$ and
$\cB^{\, (2)}$ in terms of the third one, as
\be
\begin{split}
&\cB^{\, (1)\, \pe_{\n}}  \, = \, - \, 2 \, \cB^{\, (3)\, \pe_{\m}}  \, , \\
&\cB^{\, (2) \, \hpe} \, = \, - \, \12 \, \cB^{\, (3)\, \pe_{\m}}
\, . \label{tensB4,21}
\end{split}
\ee
Substituting this result in the system (\ref{syst4,2}) it is simple
to see that, again, only the \emph{traceless} parts of the
$\cB^{\,(i)}$ tensors survive. Aside from $\cB^{\,(2)}$, whose
antisymmetric part is set to zero by the equations, all tensors
involved are actually symmetric. As a result, one is finally led to
a system for the symmetric parts, consisting of four equations for
five unknowns:
\be
\begin{split}
&\rho_0 \, \rho_2 \, \cB^{\, (1)}_{\,T}{}_{\n_1\n_2} \,
+ \, 4 \, \rho_2 \, \cB^{\, (2)}_{\,T}{}_{(\,\n_1;\,\n_2\,)} \, + \,8 \,  \cB^{\, (3)}_{\,T}{}_{\n_1\n_2} \,  = \, \fr{4}{9} \ a_{\,\n_1\n_2} \, , \\
& \rho_2 \, \cB^{\, (1)}_{\,T}{}_{\m\n}\, + \, \12\, (\,\rho_2 \,
\rho_4 \, + \, 4\,)\,
\cB^{\, (2)}_{\,T}{}_{(\,\m\,;\,\n\,)} \,
+ \, (\,3\, \rho_3 \, + \, 1\,) \, \cB^{\, (3)}_{\,T}{}_{\m\n} \, = \, - \, \fr{1}{9} \ a_{\,\m\n} \,,\\
& 4 \, \cB^{\, (1)}_{\,T}{}_{\m_1\m_2} \, + \, 2\, (\,3\, \rho_3 \,
+ \, 1\,) \, \cB^{\, (2)}_{\,T}{}_{(\,\m_1;\,\m_2\,)} \, + \, (\,3\,
\rho_3^{\, 2} \, + \, 1\,) \, \cB^{\, (3)}_{\,T}{}_{\m_1\m_2} \, =
\,
\fr{2}{9} \ a_{\,\m_1\m_2} \, \, .
\end{split}
\ee
These equations can be used to relate the traceless part of each of
the $\cB^{\, (i)}$ to the non-vanishing double trace tensor $a$,
according to
\be \label{tensB4,2}
\begin{split}
&\cB^{\, (1)}_{\,T}{}_{\n_1\n_2} \, = \, \fr{4}{3 \, (\,3\, D^{\,2} - 8\,)} \ a_{\,\n_1\n_2} \, ,\\
&\cB^{\, (2)}_{\,T}{}_{(\,\m\,;\,\n\,)} \, = \, - \, \fr{2}{3 \, (\,3\, D^{\,2} - 8\,)} \ a_{\,\m\n} \, ,\\
&\cB^{\, (3)}_{\,T}{}_{\m_1\m_2} \, = \, \fr{2}{9 \, (\,3\, D^{\,2}
- 8\,)} \ a_{\,\m_1\m_2} \, .
\end{split}
\ee

Substituting in the unconstrained $\vf$ field equation
\eqref{eq4,2uncon} one reaches the correct \emph{Lagrangian} form of the Labastida
equation \eqref{proj4,2}. At this point one can also set to zero the
left-over tensor $a$. To this end, one needs to combine
\eqref{tensB4,2} with the single equation that was not considered so far,
which is also the real novelty of this example, the
$\{2,2\}$-projected double trace of the original field equation:
\be \label{badtrace}
\begin{split}
E_{\,\vf}^{\,
\pe_{\m}
\pe_{\n}} - \, E_{\,\vf}^{\,
\hpe \,
\hpe} \, \equiv \ & \frac{3\,\r_1\r_{-4}+2}{18} \ a_{\,\m_1\m_2} \, - \, 2 \left(\, \cB^{\,(1)}{}_{\m_1\m_2} \, - \, \cB^{\,(2)}{}_{(\,\m_1;\,\m_2\,)} \, + \, \cB^{\,(3)}{}_{\m_1\m_2} \,\right) \\
& + \, \r_1 \, \h_{\,\m_1\m_2} \left(\, \cB^{\,(1)\, \pe_\n} \, - \,
2 \, \cB^{\,(2)\, \hpe} \, + \, \cB^{\,(3)\, \pe_\m} \,\right) \, =
\, 0 \, .
\end{split}
\ee
Having set to zero the left-over tensor $a$, the traceless parts of
the $\cB^{\, (i)}$ tensors also vanish on account of
eqs.~\eqref{tensB4,2}, and the $\vf$ equation of motion becomes
\be
\cA_{\,\m_1 \ldots\, \m_4 ;\, \n_1\n_2} - \, \12 \, \h_{\,(\, \m_1 \m_2} \, \cA^{\, \pe_{\m}}{}_{\m_3\m_4\,) ;\, \n_1\n_2}
- \, \12 \, \h_{\,(\,\n_1\,|\,(\,\m_1} \, \cA^{\,
\hpe}{}_{\ldots\,\m_4\,) ;\,|\,\n_2\,)}\, - \, \12 \,
\h_{\, \n_1 \n_2} \, \cA^{\, \pe_{\n}}{}_{\m_1 \ldots\, \m_4} = \, 0 \, ,
\ee
from which the final reduction to $\cA \, = \, 0$ can be attained in
a few steps, at least for $D \geq 6$.

As in the previous $(s,1)$ examples, the trace parts of the
$\cB^{\,(i)}$ are simply proportional to one another and decouple
from the dynamical equations. It is interesting to dwell further
upon this fact, so as to display an explicit realization of the
phenomenon discussed in Section \ref{sec:motion2b}. Indeed, in this
case the parameters $L_{ijkl,mn}$ of eqs.~\eqref{shbsym} are subject
to the two constraints
\be
\begin{split}
& Y_{\{6\}} \, L_{\,ijkl,\,mn} \, \sim \, L_{(\,ijkl,\,mn\,)} \, = \, 0 \, ,  \\
& Y_{\{4,2\}} \, L_{\,ijkl,\,mn} \, \sim \, 12 \, L_{\,ijkl,\,mn} \,
- \, 3
\left(\,L_{\,m\,(\,ijk,\,l\,)\,n} + L_{\,n\,(\,ijk,\,l\,)\,m} \,\right) + \,  2 \, L_{mn\,(\,ij,\,kl\,)} \, = \, 0 \, ,
\end{split}
\ee
that in our $(4,2)$ case reduce to
\be
\begin{split}
&L_{\,1111,\,22}\,+\, 8\, L_{\,1112,\,12} \,+\, 6 \,
L_{\,1122,\,11} \,=\, 0 \, ,
\\
& L_{\,1111,\,22} \,-\, 2 \, L_{\,1112,\,12} \,+ \, L_{\,1122,\,11} \,=\, 0 \, .
\end{split}
\ee
Making use of the previous relations to express everything in terms
of $L_{1111,22}$, one is finally led to the transformations
\be
\begin{split}
& \delta \, \cB_{\,1111} \,=\, \eta^{\,22}\ L_{\,1111,\,22} \, , \\
& \delta \, \cB_{\,1112} \,=\, \frac{1}{2} \ \eta^{\,12}\ L_{\,1111,\,22} \, , \\
& \delta \, \cB_{\,1122} \,=\, -\ \frac{1}{2} \ \eta^{\,11}\
L_{\,1111,\,22} \, ,
\end{split}
\ee
that in the conventional space-time notation read
\begin{alignat}{2}
& \delta\, \cB^{\,(1)}{}_{\n_1\n_2} & & = \, \eta_{\,\n_1\n_2} \ L \, ,\nonumber \\
& \delta\, \cB^{\,(2)}{}_{\m\n} & & = \, \frac{1}{4} \ \eta_{\,\m\n} \ L \, , \\
& \delta\, \cB^{\,(3)}{}_{\m_1\m_2} & & = \, - \ \frac{1}{2} \
\eta_{\,\m_1\m_2} \ L \, , \nonumber
\end{alignat}
since $L_{1111,22}$ is actually a scalar parameter in the $(4,2)$
model, here denoted by $L$ for brevity. Comparing with
eq.~\eqref{tensB4,21}, one can see directly that the three traces of
the $\cB^{\,(i)}$ can be gauged away by these transformations, which
provides again a different perspective on their decoupling from the
field equation for $\vf$.

\vskip 12pt

%%%%%%%%%%%%%%%%%%%%%%%%%%%%%%%%%%%%%%%%%%%%%%%%%%%%%%%%%%%%%%%%%%%%%%%%%%%%%%

\scsss{Current exchanges for rank-$(4, 2)$ gauge fields}\label{sec:cur42b}

%%%%%%%%%%%%%%%%%%%%%%%%%%%%%%%%%%%%%%%%%%%%%%%%%%%%%%%%%%%%%%%%%%%%%%%%%%%%%%

The results of the previous section afford an interesting
generalization, along the lines of \cite{fronsdal,fms}, when
external currents are present. Let us stress that our unconstrained
formulation requires that the external currents be conserved, a
feature to be contrasted with their partial conservation in the
constrained Fronsdal or Labastida constructions. Referring to the
$(4,2)$ case, our aim is now to invert the field equations in the
presence of external sources in order to derive an explicit
expression for the massless current exchange. Many of the results
will follow the pattern already visible in
\cite{fms}, with an important novelty related to the non-trivial role
of double traces, that after all already emerged in the previous
discussion of the source-free case.

Let us begin by considering the field equation in the presence of an
external source ${\cal J}$,
\be \label{eomJ4,2k}
\begin{split}
E_{\,\vf} \, = \, \cE_\vf \, & + \, \h_{\,(\,\m_1\m_2}\,\h_{\,\m_3\m_4\,)}\, \cB^{\,(1)}{}_{\n_1\n_2} \, + \, \h_{\,(\,\n_1\,|\,(\,\m_1}\,\h_{\,\m_2\m_3}\, \cB^{\,(2)}{}_{\,\m_4\,)\,;\,|\,\n_2\,)} \\
& + \left(\, \h_{\,\n_1\n_2}\,\h_{\,(\,\m_1\m_2\,|} \, + \,
\h_{\,\n_1\,(\,\m_1}\,\h_{\,\m_2\,|\,\n_2}\, \right)
\cB^{\,(3)}{}_{|\,\m_3\m_4\,)} \, = \, \cJ_{\,\m_1 \ldots\,
\m_4;\,\n_1\n_2} \, .
\end{split}
\ee
As in the previous section, its two independent triple traces, that
in this more general setting give
\begin{alignat}{2}
& E_{\,\vf}^{\, \pe_{\m} \pe_{\m} \pe_{\n}} & &:  \quad  \rho_0 \,
\cB^{\, (1)\, \pe_{\n}} \, + \, 8 \, \cB^{\, (2) \, \hpe} \, + \,2
\, \rho_2 \,  \cB^{\, (3)\, \pe_{\m}}
\, = \, \fr{1}{\rho_2} \, \cJ^{\, \pe_{\m} \pe_{\m} \pe_{\n}} \, , \nonumber \\
& E_{\,\vf}^{\, \pe_{\m} \hpe \, \hpe} & &: \quad \cB^{\, (1)\,
\pe_{\n}}
\, + \, 2 \, \rho_3 \, \cB^{\, (2) \, \hpe} \, + \, \rho_ 5 \,
\cB^{\, (3)\, \pe_{\m}}
\, = \, \fr{1}{\rho_2} \, \cJ^{\, \pe_{\m} \hpe \, \hpe} \, ,
\end{alignat}
provide a convenient starting point for the analysis. One is thus
facing, again, a system of equations that is not fully determined,
but allows nonetheless to relate two of the traces of $\cB_{ijkl}$
to the third and to the external currents, according to
\be
\begin{split}
&\cB^{\, (2) \, \hpe} \, = \,- \, \12 \,  \cB^{\, (3)\, \pe_{\m}}
\, -\, \fr{1}{2\, \rho_4 \, \rho_{-1} \, \rho_2} \, \left(\,\cJ^{\, \pe_{\m} \pe_{\m} \pe_{\n}} - \,
\rho_0 \, \cJ^{\, \pe_{\m} \hpe \, \hpe}\,\right) \, , \\
& \cB^{\, (1)\, \pe_{\n}} \, = \, - \, 2 \,  \cB^{\, (3)\, \pe_{\m}}
\, + \,
\fr{1}{\rho_4 \, \rho_{-1} \, \rho_2} \, \left(\,\rho_3 \, \cJ^{\, \pe_{\m} \pe_{\m} \pe_{\n}} - \,
4 \, \cJ^{\, \pe_{\m} \hpe \, \hpe} \,\right) \, .
\label{triplsolvJ}
\end{split}
\ee
As we saw in the previous section, the shift symmetry of
eq.~\eqref{shbsym} actually allows to gauge away the first terms
above. The other terms involving the external current, however, can
not be eliminated and contribute to the current exchanges.

Making use of eqs.~\eqref{triplsolvJ} in the non-homogeneous version
of the relations \eqref{syst4,2} for the double traces of $E_\vf$
finally leads to the system
\be \label{syst4,2J}
\begin{split}
&E_{\,\vf}^{\, \pe_{\m} \pe_{\m}} \, : \, - \, \fr{4}{9} \ a_{\,
\n_1
\n_2} \, + \, \rho_0 \, \rho_2 \, \cB^{\, (1)}_{\,T}{}_{\n_1 \n_2} \, + \, 4 \, \rho_2
\, \cB^{\, (2)}_{\,T}{}_{(\,\n_1 ;\, \n_2 \,)} \, + \,8 \,  \cB^{\, (3)}_{\,T}{}_{\n_1 \n_2} \\
&\phantom{E_{\vf}^{\, \pe_{\m} \pe_{\m}} \,} = \, \cJ^{\, \pe_{\m} \pe_{\m}}{}_{\n_1\n_2} \, - \, \fr{1}{\rho_0} \ \h_{\,\n_1 \n_2} \, \cJ^{\, \pe_{\m} \pe_{\m} \pe_{\n}} \, , \\
&E_{\,\vf}^{\, \pe_{\m} \hpe} \, : \, \fr{1}{9} \ a_{\, \m \n}
 + \, \rho_2 \, \cB^{\, (1)}_{\,T}{}_{\m \n}  + \, \frac{\rho_2 \, \rho_4 + 4}{2} \,
 \cB^{\, (2)}_{\,T}{}_{(\,\m\,;\, \n\,)} + \, \rho_2 \, \rho_4 \, \cB^{\, (2)}_{\,T}{}_{[\,\m\,;\, \n\,]}
+ \, (\,3\, \rho_3 + 1\,) \, \cB^{\, (3)}_{\,T}{}_{\m \n} \\
& \phantom{E_{\vf}^{\, \pe_{\m} \hpe} \,} = \, \cJ^{\, \pe_{\m}
\hpe}{}_{\m\,;\,\n} \, - \, \fr{1}{\rho_0}
\ \h_{\, \m \n} \,  \cJ^{\, \pe_{\m} \hpe \, \hpe} \, ,\\
&E_{\,\vf}^{\, \pe_{\m} \pe_{\n}} + \, 2 \, E_{\vf}^{\, \hpe \,
\hpe}  \, : \, - \, \fr{2}{9} \ a_{\, \m_1 \m_2} \, + \, 4 \, \cB^{\,
(1)}_{\,T}{}_{\m_1 \m_2} \, + \, 2\, (\,3\, \rho_3 \, + \, 1\,) \, \cB^{\, (2)}_{\,T}{}_{(\,\m_1;\,\m_2\,)} \\
& \phantom{E_{\vf}^{\, \pe_{\m} \pe_{\n}}} + \, (\,3\,
\rho_{\,3}^{\, 2} \, + \, 1\,) \,
\cB^{\, (3)}_{\,T}{}_{\m_1 \m_2} \, = \left(\, \cJ^{\, \pe_{\m} \pe_{\n}}\! + 2 \,
\cJ^{\, \hpe \, \hpe} \,\right){}_{\!\m_1\m_2}
\, - \, \fr{1}{\rho_0} \ \h_{\,\m_1 \m_2} \left(\,
\cJ^{\, \pe_{\m} \pe_{\m} \pe_{\n}}\! + 2 \, \cJ^{\, \pe_{\m} \hpe \, \hpe}\,\right) ,
\end{split}
\ee
while the non-homogeneous counterpart of eq.~\eqref{badtrace} is
\be \label{badtraceJ}
\begin{split}
E_{\,\vf}^{\,
\pe_{\m}
\pe_{\n}} - \, E_{\,\vf}^{\,
\hpe \,
\hpe} \, & : \, \frac{3\,\r_1\r_{-4}+2}{18} \ a_{\,\m_1\m_2} \, - \, 2 \left(\, \cB^{\,(1)}_{\,T}{}_{\m_1\m_2} \, - \,
\cB^{\,(2)}_{\,T}{}_{(\,\m_1;\,\m_2\,)} \, + \, \cB^{\,(3)}_{\,T}{}_{\m_1\m_2} \,\right) \\
& = \left(\, \cJ^{\, \pe_{\m} \pe_{\n}} - \, \cJ^{\, \hpe \, \hpe}
\,\right){}_{\!\m_1\m_2}
\, - \, \fr{1}{\rho_0} \ \h_{\,\m_1 \m_2} \left(\,
\cJ^{\, \pe_{\m} \pe_{\m} \pe_{\n}} - \, \cJ^{\, \pe_{\m} \hpe \, \hpe}\,\right) \,
.
\end{split}
\ee
As in the previous section, we have separated the traceless parts of
the $\cB^{\,(i)}$, and correspondingly the right-hand sides involve
the traceless combinations
\be \label{doublejnew}
\begin{split}
& j^{\,(1)}{}_{\n_1\n_2} \, = \, \cJ^{\, \pe_{\m} \pe_{\m}}{}_{\n_1\n_2} \, - \, \fr{1}{\rho_0} \ \h_{\,\n_1 \n_2} \, \cJ^{\, \pe_{\m} \pe_{\m} \pe_{\n}} \, , \\
& j^{\,(2)}{}_{\m\n} \, = \, \cJ^{\, \pe_{\m} \hpe}{}_{\m\,;\,\n} \,
- \, \fr{1}{\rho_0}
\ \h_{\, \m \n} \,  \cJ^{\, \pe_{\m} \hpe \, \hpe} \, , \\
& j^{\,(3)}{}_{\m_1\m_2} \, = \, \left(\, \cJ^{\, \pe_{\m}
\pe_{\n}}\! + 2 \, \cJ^{\, \hpe \, \hpe} \,\right){}_{\!\m_1\m_2}
\, - \, \fr{1}{\rho_0} \ \h_{\,\m_1 \m_2} \left(\,
\cJ^{\, \pe_{\m} \pe_{\m} \pe_{\n}}\! + 2 \, \cJ^{\, \pe_{\m} \hpe \, \hpe}\,\right) \, , \\
& j^{\,(4)}{}_{\m_1\m_2} \, = \, \left(\, \cJ^{\, \pe_{\m}
\pe_{\n}}\! - \, \cJ^{\, \hpe \, \hpe} \,\right){}_{\!\m_1\m_2}
\, - \, \fr{1}{\rho_0} \ \h_{\,\m_1 \m_2} \left(\,
\cJ^{\, \pe_{\m} \pe_{\m} \pe_{\n}}\! - \, \cJ^{\, \pe_{\m} \hpe \, \hpe}\,\right) \, , \\
\end{split}
\ee
so that one is actually dealing with a triply traceless effective
current. As pointed out in the discussion of the homogeneous
equations, the system \eqref{syst4,2J} effectively splits into a $3
\times 3$ system for the symmetric components of the $\cB^{\,(i)}$
and a single equation for the antisymmetric component of
$\cB^{\,(2)}$. Eqs.~\eqref{syst4,2J} and \eqref{badtraceJ} can then
be solved, so that, for instance
\be \label{sol4a}
\begin{split}
a_{\, \m_1 \m_2} \, & = \, \frac{12}{\rho_{-4} \rho_{-1} \rho_0
\rho_2} \left\{\, \cJ^{\, \pe_{\m} \pe_{\m}}{}_{\!\m_1\m_2}
- \, \cJ^{\, \pe_{\m} \hpe}{}_{\m_1\m_2} + \, \frac{1}{3}\,\left(\, \cJ^{\, \pe_{\m} \pe_{\n}} + \, 2 \,\cJ^{\, \hpe \, \hpe} \,\right){}_{\!\m_1\m_2}  \right\} \\
& + \,  \frac{2\,(\,3\,D^{\,2} - 8\,)}{\rho_{-4} \rho_{-1} \rho_0
\rho_2} \left(\, \cJ^{\, \pe_{\m} \pe_{\n}} - \,\cJ^{\, \hpe \, \hpe} \,\right){}_{\!\m_1\m_2} \, - \, \frac{6}{\rho_{-4} \rho_{-1} \rho_2} \ \h_{\, \m_1 \m_2 } \left(\, \cJ^{\, \pe_{\m} \pe_{\m} \pe_{\n}} \, - \,  \, \cJ^{\, \pe_{\m} \hpe \, \hpe} \,\right) \, .
\end{split}
\ee
Rather than displaying the corresponding solutions for the $\cB^{\,
(i)}$ directly in terms of the external current $\cJ_{\m_1\ldots\,
\m_4;\,\n_1\n_2}$, it is instructive to elaborate on their form prior
to eliminating $a$, since these connect directly with the current
exchange amplitudes for the Labastida formulation, that are
determined by the properly projected field equation
\eqref{proj4,2}. Eqs.~\eqref{syst4,2J} lead to
\be \label{tensB4,2J}
\begin{split}
&\cB^{\, (1)}_{\,T}{}_{\n_1\n_2} \, = \, \fr{4}{3 \, (3 D^2 - 8)} \ a_{\,\n_1\n_2} \, + \, C \left\{\, (3D^3+24D^2+40D-16) \, j^{\,(1)}{}_{\n_1\n_2} \right.\\
&\phantom{\cB^{\, (1)}_{\,T}{}_{\n_1\n_2} \, = \,} \left. -\,8\,(3D^2+12D+4)\, j^{\,(2)}{}_{\n_1\n_2} \, + \, 16\,\r_1 \, j^{\,(3)}{}_{\n_1\n_2} \,\right\} \, , \\
&\cB^{\, (2)}_{\,T}{}_{(\,\m\,;\,\n\,)} \, = \, - \, \fr{2}{3 \, (3 D^2 - 8)} \ a_{\,\m\n} \, + \, 2 \, C \left\{\, - \,(3D^2+12D+4) \, j^{\,(1)}{}_{\m\n}  \right. \\
&\phantom{\cB^{\, (2)}_{\,T}{}_{(\,\m\,;\,\n\,)} \, = \,} \left. + \, (3D^3+12D^2+16D-8) \, j^{\,(2)}{}_{\m\n} \, - \, \r_2\,(3D-2) \, j^{\,(3)}{}_{\m\n} \,\right\} \, , \\
&\cB^{\, (2)}_{\,T}{}_{[\,\m\,;\,\n\,]} \, = \, \frac{1}{\r_2\r_4} \ \cJ^{\, \pe_\m \hpe}{}_{[\,\m\,;\,\n\,]}  \, , \\
&\cB^{\, (3)}_{\,T}{}_{\m_1\m_2} \, = \, \fr{2}{9 \, (3 D^2 - 8)} \ a_{\,\m_1\m_2} \, + \, C \left\{\, 8\, \r_1 \, j^{\,(1)}{}_{\m_1\m_2} \, - \, 4 \, \r_2 \, (3D-2) \, j^{\,(2)}{}_{\m_1\m_2} \right. \\
&\phantom{\cB^{\, (3)}_{\,T}{}_{\m_1\m_2} \, = \,} \left. + \, \r_2
\, (D^2 + 2D-4) \, j^{\,(3)}{}_{\m_1\m_2}\,\right\}  \, ,
\end{split}
\ee
with
\be
C \, = \, \frac{1}{\r_0\, \r_4\, \r_6\, (\,3\,D^{2}-8\,)}
\ee
and the $j^{\,(i)}$ defined as in eq.~\eqref{doublejnew}.

Substituting these expressions in (\ref{eomJ4,2k}) one obtains
\be \label{proj4,2J}
\begin{split}
E_{\,\vf} \, = \, \cE_{\,\vf} \, & + \, \fr{4}{3\, (3\, D^2 - 8)} \
\h_{\,(\,\m_1\m_2}\,\h_{\,\m_3\m_4\,)} \, a_{\, \n_1 \n_2} \,
- \, \fr{1}{3\, (3 \, D^2 - 8)} \ \h_{\,(\,\n_1\,|\,(\,\m_1}\,  \h_{\,\m_2 \m_3}\, a_{\,\m_4\,)\,|\,\n_2\,)} \\
& +\, \fr{2}{9 \, (3\, D^2 - 8)} \ \left(\, \h_{\, \n_1 \n_2} \,
\h_{\,(\,\m_1\m_2\,|} \, + \, \h_{\, \n_1 \,(\, \m_1}\,
\h_{\,\m_2\,|\,\n_2} \,\right)  \, a_{\,|\, \m_3 \m_4\,)} \, = \,
\cK_{\,\m_1 \ldots\, \m_4;\,\n_1\n_2}\, .
\end{split}
\ee
where $\cE_\vf$ is the tensor defined by eq.~\eqref{einst4,2} and
$\cK_{\,\m_1 \ldots\, \m_4;\,\n_1\n_2}$ is a constrained effective
current whose symmetrized double traces vanish. Notice that this
expression reduces to the inhomogeneous Lagrangian equation of the
Labastida formulation if the compensator is eliminated by a gauge
choice.

In terms of $\cK$, eq.~\eqref{sol4a} reduces to the simpler form
\be
a_{\,\m_1\m_2} \, = \,
\frac{2\,(\,3\,D^{\,2}-8\,)}{\r_{-4}\,\r_{-1}\,\r_0\,\r_2} \,
\left(\, \cK^{\, \pe_\m \pe_\n} - \, \cK^{\,
\hpe\,\hpe}\,\right){}_{\!\m_1\m_2}\, ,
\ee
and eliminating the various traces of $\cA$ one can finally build
the complete current-exchange amplitude
\be
\begin{split}
& \cA_{\,\m_1 \ldots\, \m_4;\,\n_1\n_2} \, = \, \cK_{\,\m_1 \ldots\, \m_4; \,\n_1\n_2} \, - \, \frac{1}{\r_{4}} \ \h_{\,(\m_1\m_2|} \, Y_{\{2,2\}} \, \cK^{\,\pe_\m}{}_{|\m_3\m_4);\,\n_1\n_2} \\
& - \frac{1}{\r_{-4}\r_0\r_6} \, \h_{\,(\m_1\m_2|} \, Y_{\{4\}} \Big\{\, (D^2-12)\, \cK^{\,\pe_\m}{}_{|\m_3\m_4);\,\n_1\n_2} -  2 \, \r_{-2} \, \cK^{\,\hpe}{}_{|\m_3\m_4)\,(\n_1;\,\n_2)} \\
& + 4 \, \cK^{\,\pe_\n}{}_{|\m_3\m_4)\,\n_1\n_2} \,\Big\} - \frac{1}{\r_0\r_4} \, \h_{\,(\m_1\m_2|} \, Y_{\{3,1\}} \Big\{\, \r_2 \, \cK^{\,\pe_\m}{}_{|\m_3\m_4);\,\n_1\n_2} - 2 \, \cK^{\,\hpe}{}_{|\m_3\m_4)\,(\n_1;\,\n_2)} \,\Big\} \\
& - \frac{1}{\r_{-4}\r_0\r_6} \, \h_{\,(\n_1|\,(\m_1|} \, Y_{\{4\}} \Big\{\, -\,\r_{-2} \, \cK^{\,\pe_\m}{}_{|\m_2\m_3;\,\m_4)\,|\n_2)} + (D^2-4)\, \cK^{\,\hpe}{}_{|\m_2\m_3\m_4);\,|\n_2)}  \\
& - \, \r_2 \, \cK^{\,\pe_\n}{}_{|\m_2\m_3\m_4)\,|\n_2)} \Big\} - \frac{1}{\r_0\r_6} \, \h_{\,(\n_1|\,(\m_1|} \, Y_{\{3,1\}} \Big\{\, - \, \cK^{\,\pe_\m}{}_{|\m_2\m_3;\,\m_4)\,|\n_2)} + \, \r_2 \, \cK^{\,\hpe}{}_{|\m_2\m_3\m_4);\,|\n_2)} \,\Big\} \\
& - \!\frac{1}{\r_{-4}\r_0\r_6} \, \h_{\,\n_1\n_2} Y_{\{4\}} \Big\{\, 24\, \cK^{\,\pe_\m}{}_{\m_1\m_2;\,\m_3\m_4}\! -  8 \, \r_2\, \cK^{\,\hpe}{}_{\m_1\m_2\m_3;\,\m_4}\! + (D^2+4D-8) \, \cK^{\,\pe_\n}{}_{\m_1\m_2\m_3\m_4} \Big\} \\
& - \! \frac{2}{3\r_{-4}\r_{-1}\r_0\r_2} \Big\{\! - 4\, \h_{\,(\m_1\m_2}\h_{\,\m_3\m_4)} \!\left( \cK^{\,\pe_\m\pe_\n}\! - \cK^{\,\hpe\,\hpe} \right){}_{\!\n_1\n_2}\! + \, \h_{\,(\n_1|\,(\m_1}\h_{\,\m_2\m_3}\!\left( \cK^{\,\pe_\m\pe_\n}\! - \cK^{\,\hpe\,\hpe} \right){}_{\!\m_4)\,|\n_2)}   \\
& - \, \frac{1}{3} \left(\, \h_{\,\n_1\n_2}\,\h_{\,(\m_1\m_2|} + \h_{\,\n_1(\m_1}\h_{\m_2|\,\n_2} \,\right)\left( \cK^{\,\pe_\m\pe_\n}\! - \cK^{\,\hpe\,\hpe} \,\right){}_{\!|\m_3\m_4)} \\
& - \, \frac{3\,D^2-8}{6} \left(\, 2 \,
\h_{\,\n_1\n_2}\,\h_{\,(\m_1\m_2|} -
\h_{\,\n_1(\m_1}\h_{\m_2|\,\n_2} \,\right)\left(
\cK^{\,\pe_\m\pe_\n}\! - \cK^{\,\hpe\,\hpe}
\,\right){}_{\!|\m_3\m_4)} \, \,\Big\} \, .
\end{split}
\ee
where the $Y_i$ are Young projectors and the coefficients that
appear on the right-hand side formally build, as was shown to be the
case for all symmetric tensors in \cite{fms}, a traceless current in
$D-2$ dimensions. Notice that these expressions present poles at the
special dimensions where an additional Weyl-like develops as
discussed in Section \ref{sec:Weyl}.

Let us conclude this section by stressing again that the analysis
was performed using the minimal Lagrangian of eq.~\eqref{lag}, while
in principle this setting could be modified by the addition of terms
involving the constraints. These further couplings, however, would
merely induce redefinitions of the Lagrange multipliers, that as
such would have no effect on the current exchanges, as the reader
can verify.

\vskip 24pt

%%%%%%%%%%%%%%%%%%%%%%%%%%%%%%%%%%%%%%%%%%%%%%%%

\scs{General bosonic fields}\label{sec:generalb}

%%%%%%%%%%%%%%%%%%%%%%%%%%%%%%%%%%%%%%%%%%%%%%%%

Having discussed in some detail two-family fields, the simplest
class of mixed-symmetry gauge bosons, we can now turn to the
Lagrangians for the general case of arbitrary numbers of index
families. In this section we begin by reconsidering the original
result of Labastida, that we approach via a different route,
dictated by the Bianchi identities and their traces. We then turn to
a detailed derivation of the unconstrained Lagrangians, and conclude
with a description of the corresponding field equations and of their
reduction to ${\cal F}=0$.

\vskip 24pt

%%%%%%%%%%%%%%%%%%%%%%%%%%%%%%%%%%%%%%%%%%%%%

\scss{The Lagrangians}\label{sec:lagrangianb}

%%%%%%%%%%%%%%%%%%%%%%%%%%%%%%%%%%%%%%%%%%%%%

The traces of the Bianchi identities are a useful tool to derive the
Lagrangians of mixed-symmetry fields with arbitrary numbers of index
families. In order to illustrate their role, let us begin by showing
how one can recover in this fashion the constrained Labastida
theory.

After enforcing the double trace constraints
\be
T_{(\,ij}\, T_{kl\,)}\, \vf \, = \, 0 \,  \label{2trnf}
\ee
the Bianchi identities take in general the form already foreseen in
the two-family case,
\be \label{bianchi_constr}
\pr_{\,i}\,\cF \, - \, \12 \, \pr^{\,j}\,T_{ij}\,\cF \, = \, 0
\ee
where, of course, here the range for the family indices is meant to
be arbitrary. Taking a trace of eq.~\eqref{bianchi_constr} gives
\be \label{bianchi_constr_tr}
\pr_{\,i} \, T_{jk}\, \cF \, - \, \12 \ \pr_{\,(\,j}\, T_{k\,)\,i} \,
\cF \, - \, \12 \ \pr^{\,l} \, T_{il}\,T_{jk}\,\cF \, = \, 0 \, ,
\ee
a relation that admits both a symmetric and a ``hooked'' $\{2,1\}$
projection for its family indices. Let us stress, however, that as
we saw already for two-family fields the symmetric projection does
not contain a divergence, but only the gradient of an expression
that vanishes \emph{identically}, since for the Labastida theory
\be
T_{(\,ij}\,T_{kl\,)} \, \cF \, = \, 0 \, .
\label{symmcomp}
\ee
Indeed, as we already observed in Section \ref{sec:lagrangian2b},
eq.~\eqref{2trnf} implies that $\cF$ satisfies the same double trace
constraints as the gauge field $\vf$, so that the symmetric
projection of eq.~\eqref{bianchi_constr_tr} does not convey any new
information, but merely recovers an algebraic property of the double
trace of $\cF$ that could be derived directly, and rather simply,
from eq.~\eqref{doubtrF}. To this end, one would only need to recall
that, for a pair of traces $T_{ij}T_{kl}$, the symmetrization over
three of the four indices induces the full symmetrization. In
contrast, the hooked $\{2,1\}$ projection of the first trace
\eqref{bianchi_constr_tr} of the Bianchi identities,
\be \label{bianchi_constr_proj}
\left( \, 2 \, \pr_{\,i}\, T_{jk} \, - \, \pr_{\,(\,j}\,T_{k\,)\,i} \, \right)
\cF \, =  \,\frac{1}{3} \ \pr^{\,l} \left(\, 2 \, T_{il}\,T_{jk} \, - \,
T_{i\,(\,j}\,T_{k\,)\,l} \, \right) \cF \, ,
\ee
plays an important role in the construction, since it relates in a
non-trivial fashion terms with divergences to others with gradients.

A further trace of eq.~\eqref{bianchi_constr_tr} gives
\be
2\,\pr_{\,i}\, T_{jk}\,T_{lm}\,\cF \, - \, \left(\,
\pr_{\,(\,j}\, T_{k\,)\,i}\,T_{lm} \, + \, \pr_{\,(\,l}\, T_{m\,)\,i}\,T_{jk}
 \,\right) \cF \, = \,
\pr^{\,n} \, T_{in}\,T_{jk}\,T_{lm}\,\cF \, ,\label{bianchittunp}
\ee
an expression that in principle would admit four distinct
projections, the symmetric $\{5\}$, the $\{4,1\}$, the $\{3,2\}$ and
the $\{2,2,1\}$. However, a $\{2,2,1\}$ projected expression obtains
simply replacing the double traces of $\cF$ on the left-hand side
with their $\{2,2\}$ projections, which is clearly possible since
the remainder vanishes on account of eq.~\eqref{symmcomp}. In other
words, the left-hand side of eq.~\eqref{bianchittunp} is actually
$\{2,2,1\}$ projected, modulo the Labastida constraints
\eqref{2trnf}. This fact has an important consequence: out of the
available projections for the double traces of the Bianchi
identities,
\emph{only one}, the $\{2,2,1\}$ projection, relates
in a non-trivial fashion divergences and gradients. This result can
be related to a general fact, whose proof is deferred to Appendix
\ref{app:idsb}, according to which \emph{all projections of multiple
traces of $\cF$ corresponding to Young diagrams with more than two
columns vanish} on account of the double trace constraint
\eqref{symmcomp}.

The pattern just identified extends to the higher traces of the
Bianchi identities. At any given order, only \emph{a single} Young
projection plays a dynamical role, and relates divergences of $p$
traces and gradients of $p+1$ traces of $\cF$, with all traces
subject to projections of the ``window'' $\{2,\ldots\, ,2\}$ type,
while the others vanish on account of eq.~\eqref{symmcomp}. The
uniqueness of this relevant type of projections at any given order
suggests to resort to the shorthand notation
\bea \label{shortnotf}
\cF^{\,\prime}_{ij} & = & T_{ij}\,\cF \, , \nonumber \\ \cF^{\,\prime\,\prime}_{ij\,;\,kl} & =
& \frac{1}{3} \, \left(\, 2\,T_{ij}\,T_{kl} \, -
\, T_{i\,(\,k}\,T_{l\,)\,j} \,\right) \cF \, , \\ & \vdots & \nonumber \\
\cF^{\,[\,p\,]}{}_{i_1 j_1,\,\ldots\,,\,i_p j_p} & = & Y_{\{2^p \}} \, T_{i_1 j_1} \, \ldots \,
T_{i_p j_p} \, \cF \, , \nonumber
\eea
where for brevity we have also let
\be
Y_{\{2^p\}} \, \equiv \, Y_{\{\,\underbrace{2,\ldots,2}_p \,\}} \, .
\ee

For a general $N$-family constrained field $\vf$, the Lagrangian is
thus necessarily of the form
\be
\cL \, = \, \frac{1}{2} \ \bra \vf \comma \cF \, + \, \sum_{p\,=\,1}^{N} \, k_{\,p} \ \h^{\,p} \, \cF^{\,[\,p\,]} \ket \, ,
\label{l.laba}
\ee
where
\be \label{contraction}
\h^{\,p} \, \cF^{\,[\,p\,]} \, \equiv \, \h^{i_1 j_1} \, \ldots \, \h^{i_{\,p} j_{\,p}} \,
\cF^{\,[\,p\,]}{}_{i_1 j_1,\,\ldots\,,\,i_{p} j_{p}} \, .
\ee
We shall see shortly that the relations between the divergence of
$\cF^{\,[\,p\,]}$ and the gradient of $\cF^{\,[\,p+1\,]}$ provided
by the left-over two-column traces of the Bianchi identities fix
uniquely the $k_{\,p}$ that guarantee the gauge invariance of $\cL$,
with the end result that
\be
k_{\,p} \, = \, \frac{(-1)^{\,p}}{p\,!\,(\,p+1\,)\,!} \,
.\label{coeflab}
\ee

This recovers the complete Labastida Lagrangian, with some minor
differences with respect to the original presentation. First, in
\cite{labastida} the higher traces of $\cF$ do not carry specific
projections like here, but we have seen that, at any given order,
only one projection actually survives the Labastida constraints.
This specification will play an important role in the unconstrained
theory. In addition, our $\eta$ tensors are conveniently rescaled,
as explained in Appendix
\ref{app:MIX}, which brings about a factor $2^p$ when compared to
\cite{labastida}. The last difference has to do with what is probably
just a misprint in
\cite{labastida}, where the factor $(-1)^{\, p}$ is not indicated.
For brevity, we shall defer the missing details of this derivation
to the ensuing discussion of the unconstrained theory, to which we
now turn. From our vantage point, both settings rest on similar
steps determined by the Bianchi identities. The key differences
concern the $\cC_{ijkl}$ constraints and the composite compensators
$\a_{ijk}(\Phi)$, that are only present in the unconstrained theory.

Unconstrained Lagrangians for generic mixed-symmetry fields can be
constructed starting from the Bianchi identities for the $\cA$
tensor,
\be \label{bianchi_gen}
\pr_{\,i}\,\cA \, - \, \12 \, \pr^{\,j}\,T_{ij}\,\cA \, = \, - \, \frac{1}{4} \
\pr^{\,j}\pr^{\,k}\pr^{\,l} \,  \cC_{ijkl} \, .
\ee
Taking $p$ traces one then obtains
\be \label{bianchi_tr}
\begin{split}
& 2 \, \pr_{\,k} \, T_{i_1 j_1}\,\ldots\,T_{i_p j_p} \, \cA \, - \,
\left(\, \pr_{\,(\,i_1} \, T_{j_1\,)\,k} \,\ldots\, T_{i_p j_p} \, +
\, \ldots \, + \, \pr_{\,(\,i_p} \, T_{j_p\,)\,k} \,\ldots\,
T_{i_{p-1} j_{p-1}} \,\right) \cA \\
& - \, \pr^{\,l} \, T_{i_1 j_1}\,\ldots\,T_{i_p j_p}\,T_{kl}\,\cA
\, =
\, - \, \12
\ T_{i_1 j_1}\,\ldots\,T_{i_p j_p} \, \pr^{\,l}\pr^{\,m}\pr^{\,n}\, \cC_{\,klmn} \, ,
\end{split}
\ee
where the left-hand side is not projected to begin with. The
constrained case, however, provided two clear indications:
\begin{itemize}
\item the relevant projections of the Bianchi identities, to all orders, arise from the
$\{2^p,1\}$ Young diagram;
\item all projections of the multiple traces of $\cA$ corresponding to Young diagrams with more than two columns
can be expressed in terms of the $\cC_{ijkl}$ constraints.
\end{itemize}

If we now construct the $\{2^p,1\}$ projection of the Bianchi
identities, the gradient term produces directly the
$\{2^{p+1}\}$-projected trace $\cA^{\,[\,p+1\,]}_{\,i_1
j_1\,,\,\ldots\,,\,i_p j_p\, , \, kl}$, while the terms bearing a
divergence need further care. The key observation is that
\be
\begin{split} \label{n+2}
& Y_{\{2^p,1\}} \left[\, 2 \, \pr_{\,k} \, T_{i_1
j_1}\,\ldots\,T_{i_p j_p} \, \cA \, - \, \left(\, \pr_{\,(\,i_1} \,
T_{j_1\,)\,k} \,\ldots\, T_{i_p j_p} \, + \, \ldots \,\right) \cA
\,\right] \\ & = \, (\,p+2\,) \ Y_{\{2^p,1\}} \, \pr_{\,k} \,
\cA^{\,[\,p\,]}{}_{\,i_1 j_1,\,\ldots\,,\,i_p j_p} \, ,
\end{split}
\ee
and is proved in Appendix \ref{app:idsb}. The factor $(p+2)$ plays a
crucial role in the following derivation. It draws its origin from
the column antisymmetrization in $p+1$ indices, that has precisely
the effect of bringing together the $p+1$ terms on the left-hand
side of eq.~\eqref{n+2}, the first of which bears an overall
coefficient $2$. In conclusion, the remaining operations recover the
$\{2^p,1\}$ projection of the divergence $\pr_{\, k}
\, \cA^{\,[\,p\,]}_{\,i_1 j_1\,,\,\ldots\,,\,i_p j_p}$, but with an
overall factor $p+2$, as indicated in eq.~\eqref{n+2}, so that the
end result is
\be \label{chain}
\begin{split}
& (\,p+2\,) \ Y_{\{2^p,1\}} \, \pr_{\,k} \, \cA^{\,[\,p\,]}{}_{i_1
j_1,\,\ldots\,,\,i_p j_p}  - \, \pr^{\,l} \,
\cA^{\,[\,p+1\,]}{}_{i_1
j_1,\,\ldots\,,\,i_p j_p\,,\,k\,l} \, = \\
& = \, - \, \12 \ Y_{\{2^p,1\}} \,
T_{i_1 j_1} \ldots \, T_{i_p j_p} \, \pr^{\,l}\pr^{\,m}\pr^{\,n}\, \cC_{\,klmn} \, .
\end{split}
\ee
It should be appreciated that these arguments bring the general case
of mixed-symmetry fields very close to the far simpler symmetric
construction of \cite{fs3,fms}, since at any order in the traces one
is left with a single type of relevant consequences of the Bianchi
identities.

Eqs.~\eqref{chain} determine completely the structure of the
unconstrained Lagrangian, as we now show starting from a trial
Lagrangian built from all traces of $\cA$ not related to the
constraints. We shall see shortly how to fix the coefficients in
order to obtain a gauge invariant result. Our starting point is thus
\be \label{einstein_gen}
\cL_0 \, = \, \frac{1}{2} \ \bra\, \vf \,\comma\, \sum_{p\,=\,0}^{N} \, k_{\,p} \ \h^{\,p}\, \cA^{\,[\,p\,]}
\,\ket\, ,
\ee
where $k_0 \, = \, 1$, and for brevity we are using the shorthand
notation of eq.~\eqref{contraction}. Up to partial integrations, the
resulting gauge variation reads
\be
\begin{split}
\d \, \cL_0 \, = \, & - \, \sum_{p\,=\,0}^{N} \, \frac{1}{2^{\,p+1}} \ \bra\, T^{\,p} \L \,\comma\, k_{\,p} \ \prd \cA^{\,[\,p\,]} \, + \, (\,p\!+\!1\,) \ k_{\,p+1} \ \pr \, \cA^{\,[\,p+1\,]} \,\ket \\
= \, & - \sum_{p\,=\,0}^{N} \, \frac{1}{2^{\,p+1}} \ \bra\, T_{i_1j_1} \ldots\, T_{i_pj_p}\, \L_{\,k} \,\comma\, k_{\,p} \ \pr_{\,k} \, \cA^{\,[\,p\,]}{}_{i_1 j_1,\,\ldots\,,\,i_p j_p} \\
& + \, (\,p+1\,)\, k_{\, p+1} \, \pr^{\,l}
\cA^{\,[\,p+1\,]}{}_{kl\,,\,i_1 j_1,\,\ldots\,,\,i_p j_p} \,\ket
\, .
\end{split}
\ee

The right entries of the scalar product admit only the
$\{3,2^{p-1}\}$ and $\{2^p,1\}$ projections, which induce
corresponding projections on the left entries, so that the gauge
variation of the Lagrangian can be turned into the form
\be
\begin{split}\label{1stvarN}
\d \, \cL_0 \, = \, & - \, \sum_{p\,=\,0}^{N} \ \frac{1}
{2^{\,p+1}} \ \bra\, Y_{\{2^p,1\}} \ T^{\,p} \L \,\comma\, k_{\,p} \
Y_{\{2^p,1\}} \ \prd \cA^{\,[\,p\,]} \, + \, (\,p\!+\!1\,) \
k_{\,p+1} \ \pr \, \cA^{\,[\,p+1\,]} \,\ket \\& - \,
\sum_{p\,=\,1}^{N} \
\frac{1}{2^{\,p+1}} \ \bra\,
Y_{\{3,\,2^{p-1}\}} \ T^{\,p} \L \,\comma\, k_{\,p} \
Y_{\{3,\,2^{p-1}\}} \ \prd \cA^{\,[\,p\,]} \, \ket \, ,
\end{split}
\ee
here displayed in a concise but hopefully still clear notation.
Notice that
\emph{no} $\{ 3,2^{p-1}\}$ component originates from the gradient
term, whose free indices belong to $\cA^{\,[\,p+1\,]}$ that is
two-column projected. Moreover, the first row of \eqref{1stvarN} is
closely related to the multiple traces of the Bianchi identities of
eq.~\eqref{chain}, that in this compact notation read
\be
(\,p+2\,) \ Y_{\{2^p,1\}} \ \prd \cA^{\,[\,p\,]} \, - \, \pr \,
\cA^{\,[\,p+1\,]} \, = \, - \, \frac{1}{2}  \ Y_{\{2^p,1\}} \ T^{\,p} \, \pr\,\pr\,\pr\, \cC \, .
\ee
Hence, if the coefficients satisfy the recursion relation
\be \label{coeff}
\frac{k_{\,p+1}}{k_{\,p}} \, = \, - \ \frac{1}{(\,p+1\,)(\,p+2\,)} \, ,
\ee
whose solution is given in eq.~\eqref{coeflab}, all $\cA$ tensors
disappear from the first line of the gauge variation
\eqref{1stvarN}, that reduces to
\be
\d \, \cL_0 \, = \, - \, \frac{1}{8} \ \bra \d \, \b \comma \cC \ket \, -
\, \sum_{p\,=\,1}^{N} \ \frac{1}{2^{\,p+1}} \ \bra\, Y_{\{3,\,2^{p-1}\}} \ T^{\,p} \L
\,\comma\, k_{\,p} \ Y_{\{3,\,2^{p-1}\}} \ \prd \cA^{\,[\,p\,]} \, \ket \, ,
\ee
where
\be
\d \, \b_{\,ijkl} \, = \, \frac{1}{2} \, \sum_{p\,=\,0}^{N} \, \frac{k_{\,p}}{p+2} \
\pr_{\,(\,i\,}\pr_{\,j\,}\pr_{\,k\,|} \, \h^{m_1n_1} \ldots\, \h^{m_pn_p} \, Y_{\{2^p,1\}}
\, T_{m_1n_1} \ldots\, T_{m_pn_p} \, \L_{\,|\,l\,)} \label{betagen}
\ee
identifies the gauge transformations of the Lagrange multipliers,
that we shall add shortly. This is a particularly compact expression
for their gauge variation, but it is also possible to move the
divergences to the right of the $\eta$'s to recover the result
displayed in Section \ref{sec:lagrangian2b}, together with
additional contributions that appear starting from three families.

The rest of the gauge variation can be canceled adding to $\cL_0$ a
sum of terms involving the composite compensators $\a_{\,ijk}(\Phi)$
of eq.~\eqref{compensator}, whose structure follows the pattern that
clearly emerged in the two-family case. Their identification,
however, requires a non-trivial identity, that will be derived in
Appendix \ref{app:idsb} and reads
\be \label{id_alpha}
\begin{split}
& \bra Y_{\{3,2^{p-1}\}} \, T_{i_1j_1} \ldots\, T_{i_pj_p} \, \L_{\,k} \comma  Y_{\{3,2^{p-1}\}} \, \pr_{\,k} \, \cA^{[\,p\,]}{}_{i_1j_1,\, \ldots\, ,\,i_pj_p}\ket \\
& = \, \frac{p}{p+2} \, \bra T_{i_1j_1} \ldots\, T_{i_pj_p} \,
\L_{\,k} \comma \pr_{\,(\,k} \, \cA^{[\,p\,]}
{}_{i_1j_1\,),\,i_2j_2\,,\, \ldots\, ,\,i_pj_p} \ket\, .
\end{split}
\ee
The symmetrization in $(i_1j_1k)$ induces a corresponding
symmetrization in the left entry of the scalar product, that is then
manifestly related to traces of the $\d\, \a_{i_1j_1k}$. As a
result, the gauge variation can be presented in the rather compact
form
\be
\d \, \cL_0 \, = \, - \, \frac{1}{8} \ \bra \d \, \b \comma \cC \ket \, - \, \frac{3}{4} \ \sum_{p\,=\,1}^{N}
\ \bra\, \d \, \a \,\comma\, \frac{p\, k_{\,p}}{p+2} \ \h^{\,p-1} \ Y_{\{3,\,2^{p-1}\}} \ \prd \cA^{\,[\,p\,]} \, \ket \, ,
\ee
so that the \emph{unconstrained} Lagrangian for unprojected
$N$-family bosonic gauge fields is finally
\be
\begin{split}
\cL \, & = \, \12 \ \bra\, \vf \,\comma\, \cA \, + \, \sum_{p\,=\,1}^{N} \,
k_{\,p} \ \h^{\,p} \, \cA^{\,[\,p\,]}\, \ket \, + \, \frac{3}{4} \,
\bra\, \a \,\comma\, \sum_{p\,=\,1}^{N} \,
\frac{p\, k_{\,p}}{p+2} \ \h^{\,p-1}\, Y_{\{3,\,2^{p-1}\}} \ \prd \cA^{\,[\,p\,]} \,\ket \\
& + \, \frac{1}{8} \, \bra \b \comma \cC \ket \, ,
\end{split}
\ee
where the $k_{\,p}$ coefficients are given in eq.~\eqref{coeflab},
or more explicitly
\be  \label{laggenb}
\begin{split}
\cL \, & = \, \12 \, \bra\, \vf \,\comma\, \sum_{p\,=\,0}^{N} \,
\frac{(-1)^{\,p}}{p\,!\,(\,p+1\,)\,!} \ \h^{i_1 j_1} \ldots \, \h^{i_{\,p} j_{\,p}} \,
\cA^{\,[\,p\,]}{}_{i_1 j_1,\,\ldots\,,\,i_{p} j_{p}} \,\ket \\
& - \frac{1}{4} \, \bra\, \a_{\,ijk} \,\comma\, \sum_{p\,=\,0}^{N-1}
\,
\frac{(-1)^{\,p}}{p\,!\,(\,p+3\,)\,!} \ \h^{i_1 j_1} \ldots \, \h^{i_{p} j_{p}} \, \pr_{\,(\,i}\,
\cA^{\,[\,p+1\,]}{}_{jk\,),\,i_1 j_1,\,\ldots\,,\,i_{p} j_{p}}
\,\ket\\
 &+ \, \frac{1}{8} \, \bra\, \b_{\,ijkl} \,\comma\, \cC_{\,ijkl}
 \,\ket\, .
\end{split}
\ee
The first few terms of this general result are
\be
\begin{split}
\cL \, & = \12 \, \bra\, \vf \comma \cA \, - \, \12 \, \h^{ij} \, \cA^{\,\prime}{}_{ij} \, + \,
\frac{1}{12} \, \h^{ij}\,\h^{kl} \, \cA^{\,\prime\prime}{}_{ij\,;\,kl} \, - \, \frac{1}{144} \,
\h^{ij}\,\h^{kl}\,\h^{mn}\, \cA^{\,\prime\prime\prime}{}_{ij\,;\,kl\,;\,mn} \, + \, \ldots \,\ket \\
& - \frac{1}{24} \, \bra\, \a_{ijk} \comma \pr_{\,(\,i}\,\cA^{\,\prime}{}_{jk\,)} \, - \frac{1}{4}
\, \h^{lm} \, \pr_{\,(\,i}\,\cA^{\,\prime\prime}{}_{jk\,)\,;\,lm} \, + \, \frac{1}{40} \, \h^{lm}\,\h^{np} \,
\pr_{\,(\,i}\,\cA^{\,\prime\prime\prime}{}_{jk\,)\,;\,lm\,;\,np} \, + \, \ldots \,\ket \\
& + \frac{1}{8} \, \bra\, \b_{\,ijkl} \comma \cC_{\,ijkl} \,\ket\, ,
\end{split}
\ee
and agree nicely with the result of Section \ref{sec:lagrangian2b}.
Once the $\a_{ijk}$ and $\b_{ijkl}$ are removed from this
expression, the remainder reproduces the result of Labastida, up to
some notational changes and up to the oscillating signs, that are
probably missing in
\cite{labastida} due to a misprint.

We can not refrain from associating to these expressions simple
generating functions for the coefficients. This can be attained
starting from the family of contour integrals
\be
\mathpzc{b}_{\,k}[z] = (\,k+1\,)\,!\, \oint_\gamma \frac{d \zeta}{2 \pi i}\ \zeta^{k} \ e^\frac{1}{\zeta}
\, e^{-z\,
\zeta} \, = \, \sum_{p\,=\,0}^{\infty} \,
\frac{(-1)^{\,p}\,(\,k+1\,)\,!\, z^{\,p}}{p\,!\,(\,p+k\,)\,!} \, = \, (\,k+1\,)\,!\ \frac{J_{k}\left(\,2\, \sqrt{\,z\,}
\,\right)}{\left(\,\sqrt{\, z\,}\,\right)^{\,{k}}}\, ,
\ee
where the contour $\g$ encircles the origin and $J_k$ denotes a
Bessel function of order $k$. One can thus write the general
Lagrangians rather concisely in the form
\be
\cL \, = \, \12 \ \bra \vf \comma \mathpzc{b}_{\,0}[\, \eta \, T\,] \ \cA \ket \, - \, \frac{1}{8} \,
\bra \a \comma \mathpzc{b}_{\,2}[\, \eta \, T\,] \ \prd T \, \cA \ket \, + \, \frac{1}{8} \, \bra \b
\comma \cC \ket \, , \label{laggenbred}
\ee
where all $\eta$'s are meant to be placed to the left of all traces,
while the latter are meant to be projected as in
eq.~\eqref{laggenb}, so that for $N$ families the sums actually
terminate after the $N$-th trace. In a similar spirit, the gauge
transformation
\eqref{betagen} of Lagrange multipliers $\b_{ijkl}$ can be written
concisely as
\be
\d\,\b_{\,ijkl}\, =\, \frac{1}{4}\ \pr_{\,(\,i\,}\pr_{\,j\,}\pr_{\,k\,|}
\, \mathpzc{b}_{\,1}[\, \eta \, T\,] \, \L_{\,|\,l\,)}\, ,
\ee
where the Young projector in eq.~\eqref{betagen} is left implicit.

As was the case for two-family fields, it is possible to present the
general Lagrangians \eqref{laggenb} in an alternative way that will
soon prove convenient to derive the field equations. To this end,
let us consider again a field $\phi$ \emph{not} subject to the
double trace constraints \eqref{labac}, but whose gauge parameters
$\L_i$ are still constrained according to \eqref{labacgaugeb}. The
corresponding Lagrangians can be cast in the form
\be
\begin{split}
\cL_C\left(\,\phi\,,\,\gamma_{ijkl}\,\right) \, & = \, \12 \, \bra\, \phi \,\comma\, \sum_{p\,=\,0}^{N} \,
\frac{(-1)^{\,p}}{p\,!\,(\,p+1\,)\,!} \ \h^{i_1 j_1} \ldots \, \h^{i_{\,p} j_{\,p}} \,
\cF^{\,[\,p\,]}(\phi){}_{\,i_1 j_1,\,\ldots\,,\,i_{p} j_{p}} \,\ket \\
& + \, \frac{1}{24} \, \bra\, \g_{\,ijkl} \,\comma\, T_{(\,ij}\,T_{kl\,)}\, \phi
 \,\ket \label{stuecklaggen} \, ,
\end{split}
\ee
that differ from the Labastida Lagrangians of \cite{labastida}
simply because gauge invariance demands the simultaneous presence of
projected traces and Lagrange multipliers $\gamma_{ijkl}$. The
latter fields enforce the double trace constraints, and their gauge
transformations are actually those given for the $\beta_{ijkl}$ in
eq.~\eqref{betagen}. The relation between the two Lagrangians of
eqs.~\eqref{laggenb} and \eqref{stuecklaggen} is then
\be
\cL\left(\,\vf\,,\,\Phi_i\,,\,\beta_{\,ijkl}\,\right)\, = \, \cL_C\left(\, \vf \,-\, \partial^{\,i}\,
\Phi_{\,i} \comma \beta_{\,ijkl}\,-\,\Delta_{\,ijkl}(\Phi)\,\right)\, ,
\label{lcuncgen}
\ee
where
\be
\Delta_{\,ijkl}(\Phi) \, = \, \frac{1}{2} \, \sum_{p\,=\,0}^{N} \, \frac{(-1)^{\,p}}{p\,!\,(\,p+2\,)\,!} \
 \pr_{\,(\,i\,}\pr_{\,j\,}\pr_{\,k\,|} \, \h^{m_1n_1} \ldots\, \h^{m_pn_p} \, Y_{\{2^p,1\}}
 \, T_{m_1n_1} \ldots\, T_{m_pn_p} \, \Phi_{\,|\,l\,)}\, ,
\label{stueckmultgen}
\ee
or more compactly
\be
\Delta_{\,ijkl}(\Phi) \, = \, \frac{1}{4}\ \pr_{\,(\,i\,}\pr_{\,j\,}\pr_{\,k\,|}
\, \mathpzc{b}_{\,1}[\, \eta \, T\,] \, \Phi_{\,|\,l\,)}\, .
\ee

Actually, even in the general multi-family setting there is a wide
freedom to conveniently redefine the Lagrange multipliers
$\b_{ijkl}$ or $\g_{ijkl}$. For instance, one could also start from
the Lagrangians
\be
\begin{split}
\widetilde{\cL}_{\,C} \left(\,\phi\,,\,\cB_{\,ijkl}\,\right) \, & = \, \12 \, \bra\, \phi \,\comma\, \sum_{p\,=\,0}^{N} \,
\frac{(-1)^{\,p}}{p\,!\,(\,p+1\,)\,!} \ \h^{i_1 j_1} \ldots \, \h^{i_{\,p} j_{\,p}} \,
\widehat{\cF}^{\,[\,p\,]}(\phi){}_{\,i_1 j_1,\,\ldots\,,\,i_{p} j_{p}} \,\ket \\
& + \, \frac{1}{24} \, \bra\, \cB_{\,ijkl} \,\comma\,
T_{(\,ij}\,T_{kl\,)}\, \phi \label{laggen_invmult}
 \,\ket \, ,
\end{split}
\ee
where the $\widehat{\cF}^{\,[\,p\,]}$ are projected traces of $\cF$
\emph{deprived} of the $\{4,2^{p-1}\}$ components of the last terms that
appear in eq.~\eqref{gentraceb}, so that
\begin{align}
& \widehat{\cF}^{\,[\,p\,]}(\phi){}_{\,i_1 j_1,\,\ldots\,,\,i_{p}
j_{p}} \, = \, Y_{\{2^p\}}\, \left\{(\,p+1\,)
\
\Box
\prod_{r\,=\,1}^p T_{i_rj_r}
\, \phi \, - \, (\,p+1\,) \sum_{n\,=\,1}^p \, \pr_{\,i_n}\pr_{\,j_n} \prod_{r\,\neq\,n}^p T_{i_rj_r} \, \phi \right.\\
& \left. - \, \pr^{\,k} \bigg[\ \pr_{\,k} \prod_{r\,=\,1}^p
T_{i_rj_r}
\,
\phi \, - \, \sum_{n\,=\,1}^p \, \pr_{\,(\,i_n}\,T_{\,j_n\,)\,k}
\prod_{r\,\neq\,n}^p T_{i_rj_r} \, \phi \ \bigg]\right\} + \, \12 \
\pr^{\,k}\pr^{\,l} \, Y_{\{2^{p+1}\}}\, T_{kl} \, \prod_{r\,=\,1}^p T_{i_rj_r} \,
\phi\,. \nonumber
\end{align}
Interestingly, the $\widehat{\cF}^{\,[\,p\,]}$ satisfy the
two-column projected Bianchi identities
\be \label{chainhat}
(\,p+2\,) \ Y_{\{2^p,1\}} \, \pr_{\,k} \,
\widehat{\cF}^{\,[\,p\,]}(\phi){}_{\,i_1 j_1,\,\ldots\,,\,i_p j_p}  - \,
\pr^{\,l} \,
\widehat{\cF}^{\,[\,p+1\,]}(\phi){}_{\,i_1
j_1,\,\ldots\,,\,i_p j_p\,,\,k\,l} \, = \, 0 \, ,
\ee
that are \emph{free from the classical anomalies} related to the
constraints. As a result, the gauge invariance of the first series
of terms in the Lagrangians \eqref{laggen_invmult} is essentially
manifest, and consequently the Lagrange multipliers, that we now
denoted directly $\cB_{ijkl}$ as was the case for their
gauge-invariant contributions to the $\vf$ field equations
\eqref{ep}, are also gauge invariant. In addition, this way of
presenting the Lagrangians makes the terms quadratic in $\phi$
manifestly self adjoint. An unconstrained Lagrangian, that however
differs from \eqref{laggenb} by a field redefinition of the
multipliers $\b_{ijkl}$, obtains rather simply in this case as
\be
\widetilde{\cL}\left(\,\vf\,,\,\Phi_i\,,\,\cB_{\,ijkl}\,\right)\, = \,
\widetilde{\cL}_{\,C}\left(\, \vf \,-\, \partial^{\,i}\,
\Phi_{\,i} \comma \cB_{\,ijkl}\,\right)\, .
\label{lcuncgenhat}
\ee
Notice that the constrained gauge invariance of the original
Lagrangian guarantees that $\widetilde{\cL}_{\,C}$ depends on the
symmetrized traces of the $\Phi_i$, and thus on the composite
compensators $\a_{ijk}$ but not on the naked $\Phi_i$, as in the
other constructions.

Let us stress, to conclude, that the terms involving the various
types of Lagrange multipliers present in eqs.~\eqref{laggenb},
\eqref{stuecklaggen} and \eqref{laggen_invmult} have the structure
that surfaced in the two-family case. The arguments presented there
still apply, so that transformations of the type
\be
\delta \, \b_{\,ijkl} \, = \, \eta^{\,mn}\, L_{\,ijkl,\,mn}
\ee
generalize to all these presentations and to arbitrary numbers of
index families the local symmetry described in Section
\ref{sec:lagrangian2b}.

\vskip 24pt

%%%%%%%%%%%%%%%%%%%%%%%%%%%%%%%%%%%%%%%%%%%%%

\scss{The field equations}\label{sec:motionb}

%%%%%%%%%%%%%%%%%%%%%%%%%%%%%%%%%%%%%%%%%%%%%

As we have seen in Section \ref{sec:motion2b} for the case of
two-family gauge fields, recasting the Lagrangian
\eqref{laggenb} in the form \eqref{stuecklaggen} is particularly
convenient when deriving the field equations for $\vf$, the $\Phi_i$
and the $\b_{ijkl}$. Starting form eq.~\eqref{gentraceb}, the $\phi$
variation of \eqref{stuecklaggen} thus yields
\be \label{eqphigen}
\begin{split}
E_{\,\phi} \, & : \ \sum_{p\,=\,0}^{N} \,
\frac{(-1)^{\,p}}{p\,!\,(\,p+1\,)\,!} \ \h^{i_1 j_1}  \ldots \, \h^{i_{\,p} j_{\,p}} \,
\cF^{\,[\,p\,]}(\phi){}_{\,i_1 j_1,\,\ldots\,,\,i_{\,p} j_{\,p}}  \, + \, \12 \ \h^{ij}\,\h^{kl}\, \cB_{\,ijkl} \\
& - \, \frac{1}{4} \, \sum_{p\,=\,0}^{N} \,
\frac{(-1)^{\,p}}{p\,!\,(\,p+1\,)\,!} \ \h^{i_1 j_1}  \ldots \, \h^{i_{\,p} j_{\,p}} \, \pr^{\,k}\pr^{\,l} \, Y_{\{4,2^{p-1}\}} \, T_{kl} \,
\phi^{\,[\,p\,]}{}_{i_1 j_1,\,\ldots\,,\,i_{\,p} j_{\,p}} \, = \, 0 \, ,
\end{split}
\ee
where
\be
\cB_{\,ijkl} \, = \, \g_{\,ijkl} \, - \, \frac{1}{2} \, \sum_{p\,=\,0}^{N-1} \, \frac{(-1)^{\,p}}{p\,!\,(\,p+3\,)\,!} \ \h^{m_1n_1} \ldots\,
\h^{m_{p}n_{p}} \,
\pr_{\,(\,i\,}\pr_{\,j}\,\phi^{\,[\,p+1\,]}{}_{kl\,),\,m_1n_1,\,\ldots\,,\,m_{p}n_{p}}
\, ,
\ee
or more concisely
\be
\cB_{\,ijkl} \, = \, \g_{\,ijkl} \, - \, \frac{1}{12}\ \mathpzc{b}_{\,2}[\, \eta \,
T\,] \, \pr_{\,(\,i\,}\pr_{\,j}\,T_{kl\,)} \, \phi \, .
\ee

The arguments summarized in Appendix
\ref{app:idsb} show that, due to the $\{4,2^{p-1}\}$ projections of the multiple
traces of $\phi$, the second line of eq.~\eqref{eqphigen} is
proportional to the double-trace constraints, and therefore can be
eliminated on-shell. In conclusion, using the approach already
described for two-family fields and thus performing the shifts of
eq.~\eqref{lcuncgen}, the equations of motion of the unconstrained
theory can be finally cast in the form
\begin{alignat}{2}
& E_{\,\vf} & & : \ \sum_{p\,=\,0}^{N} \,
\frac{(-1)^{\,p}}{p\,!\,(\,p+1\,)\,!} \ \h^{i_1 j_1}  \ldots \, \h^{i_{\,p} j_{\,p}} \,
\cA^{\,[\,p\,]}{}_{\,i_1 j_1;\,\ldots\,;\,i_{\,p} j_{\,p}}  \, + \, \12 \ \h^{ij}\,\h^{kl}\,
\cB_{\,ijkl}\, = \, 0
 ,\label{eqgenbvf} \\
& E_{\,\b} & & : \ \frac{1}{8} \ \cC_{ijkl} \, = \, 0
\label{eqbgen}
\, ,
\end{alignat}
where the first can also be written more concisely
\be
E_{\,\vf} \, : \ \mathpzc{b}_{\,0} [\, \eta \, T\,]\, \cA  \, + \,
\12 \ \h^{ij}\,\h^{kl}\, \cB_{\,ijkl} \, = \, 0
\label{eqgenbvfs}
\, ,
\ee
Finally, the equations of motion for the compensators $\Phi_i$ are
as usual the conservation condition for external currents,
\be
\pr_{\,i}\, E_{\,\vf} \, +  \sum_{p\,=\,0}^N \,
\frac{2}{p\,!\,(\,p+2\,)\,!}
\ \h^{m_1n_1} \ldots\, \h^{m_pn_p} \, Y_{\{2^p,1\}}\, T_{m_1n_1} \ldots\, T_{m_pn_p} \,
\pr^{\,j}\pr^{\,k}\pr^{\,l} \, (E_\b)_{\,ijkl}\,=\, 0\, ,
\ee
or more concisely
\be
E_{\,\Phi} \, : \ \pr_{\,i}\, E_{\,\vf} \, + \, \mathpzc{b}_{\,1} [\, \eta \, T\,]\,
\pr^{\,j}\pr^{\,k}\pr^{\,l} \, \cC_{\,ijkl}\,=\,0\, .
\ee
As we stressed in the previous section, these results could have
been derived even more directly starting from the Lagrangian
\eqref{laggen_invmult} and taking into account the self-adjointness
of the portion of its terms that are quadratic in $\phi$.

\vskip 24pt

%%%%%%%%%%%%%%%%%%%%%%%%%%%%%%%%%%%%%%%%%%%%%%%%%%%%%%%%%%%%%%%%%%%%%%%%%%

\scss{Comments on the on-shell reduction to $\cF=0$}\label{sec:redNfam}

%%%%%%%%%%%%%%%%%%%%%%%%%%%%%%%%%%%%%%%%%%%%%%%%%%%%%%%%%%%%%%%%%%%%%%%%%%

The next step would be the reduction of \eqref{eqgenbvf} to $\cA=0$
and thus, after a partial gauge fixing, to the non-Lagrangian
Labastida form $\cF=0$. The operators on which these more general
systems rest are the $gl(N)$ counterparts of those exhibited in
Section \ref{sec:reduction2b}, and the experience developed with
two-family fields suggests the emergence, in low enough dimensions,
of a similar type of phenomena related to Weyl-like symmetries. In
the following we shall discuss their general structure and we shall
also illustrate them in a class of significant, if relatively
simple, examples with three or more index families. On the other
hand, one can argue rather simply that in the $D \to
\infty$ limit $N$-family gauge fields exhibit a universal behavior,
free of special poles, that can be captured ignoring altogether the
intricacies introduced by the $S^{\,i}{}_j$ operators. In this
limit, in fact, the reduction proceeds directly along the lines of
the symmetric case, and the field equations reduce manifestly and
directly to $\cF=0$. Moreover, there are all reasons to expect that,
even for higher values of $N$, special behaviors are confined to
dimensions $D\leq 2\,N+1$, where the corresponding models are at
most dual to other representations characterized by lower values of
$N$, although at the moment we do not have a complete argument to
this effect.

\scsss{Weyl-like symmetries}

In discussing the reduction of the Lagrangian equations for
two-family fields in Section \ref{sec:reduction2b}, we have combined
them with their traces, keeping track explicitly both of $\cA$,
which is the kinetic operator for the physical field $\vf$, and of
the $\cB_{ijkl}$, which simply relate the $\b_{ijkl}$ Lagrange
multipliers to the other fields. However, in the sporadic cases
where the procedure actually proved problematic, new Weyl-like
symmetries surfaced, while keeping track explicitly of the
$\cB_{ijkl}$ played essentially no role in their identification.
This fact can be turned to our own advantage when trying to extend
the analysis to $N$-family fields, since in this more general
setting one is inevitably confronted with higher technical
complications, so that a smooth track is highly preferable. To wit,
the successive traces of the original $\vf$ field equation soon
become unwieldy, due to the proliferation of the $S^{\,i}{}_j$
operators, and thus it is convenient to leave out
from the start some inessential details. For instance, the crucial
option of factoring out the $\cO\,[D-2]$ operator in the first step
of the reduction procedure is not available with three or more
families. Nonetheless, we can illustrate an iterative procedure to
characterize the Weyl-like symmetries that emerge whenever the
Lagrangian field equations leave some of the traces of $\cA$
undetermined.

As we already stressed in our discussion of two-family fields, the
equation of motion \eqref{eqgenbvf} sets to zero the traceless part
of $\cA$, while in special circumstances some of its traces may be
left undetermined. In order to get a handle on this phenomenon for
$N$-family fields, it is useful to study the symmetries of
eq.~\eqref{eqgenbvf} under transformations of the type
\be \label{shift_gen}
\d \, \cA \, = \, \h^{\,ij}\, \O^{\,(1)}{}_{ij} \, ,
\ee
that in general shift the Einstein-like tensor of
eq.~\eqref{einstein_gen} as
\be \label{shift_einstein_start}
\begin{split}
\d \, \cE \, & = \, \sum_{p\,=\,0}^N \, \frac{(-1)^{\,p}}{p\,!\,(\,p+1\,)\,!}\ \h^{i_1j_1}
\ldots\, \h^{i_pj_p}\, [\, Y_{\{2^p\}}\, T_{i_1j_1} \ldots\, T_{i_pj_p} \comma \h^{\,kl} \,] \, \O^{\,(1)}{}_{kl} \\
& + \, \sum_{p\,=\,0}^N \, \frac{(-1)^{\,p}}{p\,!\,(\,p+1\,)\,!}\
\h^{i_1j_1} \ldots\, \h^{i_pj_p}\, \h^{\,kl} \,
\left(\, Y_{\{2^p\}}\, T_{i_1j_1} \ldots\, T_{i_pj_p} \,\right) \O^{\,(1)}{}_{kl} \, .
\end{split}
\ee
Taking into account the $\{2^p\}$ Young projection in the family
indices, the commutator of eq.~\eqref{T^kh} reduces to
\be \label{shift_einstein_comm}
\begin{split}
[\, Y_{\{2^p\}}\, T_{i_1j_1} \ldots\, T_{i_pj_p} \comma \h^{\,kl} \,] \, \O^{\,(1)}{}_{kl} \, & = \,
(\,D-p+1\,)\, Y_{\{2^p\}}\, \sum_{n\,=\,1}^p \, \prod_{r\,\neq\,n}\, T_{i_rj_r}\, \O^{\,(1)}{}_{i_nj_n} \\
& + \, Y_{\{2^p\}}\, \sum_{n\,=\,1}^p \, S^{\,k}{}_{(\,i_n\,|}\,
\prod_{r\,\neq\,n}\, T_{i_rj_r}\, \O^{\,(1)}{}_{|\,j_n\,)\,k} \, ,
\end{split}
\ee
as can be seen resorting to the techniques of Appendix
\ref{app:idsb}. Once the traces in the last term of
eq.~\eqref{shift_einstein_comm} are then moved to the left of the
$S^{\,i}{}_j$ and the two-column projection of the second line of
eq.~\eqref{shift_einstein_start} is combined with the first, one is
left with
\begin{align} \label{shift_einstein}
\d \, \cE & = \sum_{p\,=\,1}^N \, \frac{(-1)^{\,p}}{p\,!\,(\,p+1\,)\,!}\ \h^{i_1j_1} \!\ldots
\h^{i_pj_p} \, Y_{\{2^p\}} \!\sum_{n\,=\,1}^p \, \prod_{r\,\neq\,n} T_{i_rj_r} \left\{\, (\,D-2\,)\,
\O^{\,(1)}{}_{i_nj_n} \!+\, S^{\,k}{}_{(\,i_n} \O^{\,(1)}{}_{j_n\,)\,k} \,\right\} \nn \\
& + \sum_{p\,=\,2}^N \, \frac{(-1)^{\,p}}{(\,p\,!\,)^{\,2}}\
\h^{i_1j_1} \!\ldots \h^{i_pj_p}\, Y_{\{4,2^{p-2}\}}\,
\sum_{n\,=\,1}^p \, \Big(\, Y_{\{2^{p-1}\}}\, \prod_{r\,\neq\,n}\,
T_{i_rj_r} \,\Big)\, \O^{\,(1)}{}_{i_nj_n} \, .
\end{align}
The arguments of Appendix \ref{app:idsb} now guarantee that the
$\{4,2^{\,p-2}\}$ projection present in eq.~\eqref{shift_einstein}
can be recast in a form where the family indices carried by a pair
of $\h$ tensors are \emph{fully} symmetrized. Hence, the terms in
the second line can be compensated by a shift of the $\cB_{ijkl}$,
in sharp contrast with the whole first line, that vanishes if one
tries to symmetrize three or more indices. Therefore, these last
terms can not be canceled redefining the $\cB_{ijkl}$, unless their
traceless parts vanish so that they actually embody at least an
additional $\h$ tensor. As a result, in order to identify a symmetry
of the equation of motion \eqref{eqgenbvf}, one ought to distinguish
two possibilities. The first one is that
\be \label{firsteq_shift}
\cO \, [\,D -
2\,]_{\,ij}{}^{\,kl}\, \O^{\,(1)}{}_{kl}\,\equiv \, (\,D-2\,)\,
\O^{\,(1)}{}_{ij} \!+\, S^{\,k}{}_{(\,i}\,
\O^{\,(1)}{}_{j\,)\,k} \, = \, 0
\, ,
\ee
which recovers the condition already met for two-family gauge fields
in eq.~\eqref{shift_cond}, while a consistent alternative is
\be \label{shift_delta}
(\,D-2\,)\, \O^{\,(1)}{}_{ij} \!+\, S^{\,k}{}_{(\,i}\,
\O^{\,(1)}{}_{j\,)\,k} \, =
\, \h^{\,kl} \, \D^{\,(1)}{}_{ij\,,\,kl} \, ,
\ee
that implies
\be \label{shift_double}
\d \, \cA \, = \, \h^{\,ij}\, \h^{\,kl}\, \O^{\,(2)}{}_{ij \comma kl}\, ,
\ee
provided the parameters $\O^{\,(2)}{}_{ij
\comma kl}$ satisfy some conditions that we are about to specify.
This can be justified recalling that the commutator with an $\eta$
tensor of the $S$ operators, and thus of the whole $\cO\,[\l]$
operators, is still proportional to an $\eta$ tensor. As a result,
inverting eq.~\eqref{shift_delta} forces indeed $\O^{\,(1)}$ to
contain an additional $\eta$.

Before moving to analyze eq.~\eqref{shift_double}, it is worth
exploring further the first alternative, recalling a result that we
already used in Section
\ref{sec:reduction2b}. Namely, the double-trace constraints
\eqref{symmcomp}, or their on-shell analogs for the $\cA$ tensor,
bring about the further condition
\be
T_{(\,ij}\, T_{kl\,)}\, \d\, \cA \, = \, T_{(\,ij\,|}\,
\cO\,[\,D-2\,]_{\,|\,kl\,)}{}^{mn} \, \O^{\,(1)}{}_{mn} \, + \, \h^{\,mn}\,
T_{(\,ij}\, T_{kl\,)}\, \O^{\,(1)}{}_{mn} \, = \, 0 \, .
\ee
The reader should now appreciate that, if eq.~\eqref{firsteq_shift}
holds, this becomes essentially a double-trace constraint for the
$\O^{\,(1)}{}_{mn}$. These remarks are meant to stress that one
really ought to classify shift symmetries of the whole set of
conditions that a kinetic operator $\cA$ must satisfy, not just of
the field equations. In this respect, it is instructive to describe
the effect of the transformation \eqref{shift_gen} on the Bianchi
identities, that must be preserved. Starting from the shift of
eq.~\eqref{shift_gen}, one thus finds
\be\label{Bianchi_simple}
\pr_{\,i} \, \d \, \cA \, - \, \frac{1}{2}\ \pr^{\, j}\, T_{ij}\,
\d\, \cA \, = \, - \, \frac{1}{2}  \ \cO \, [\,D -
2\,]_{\,ij}{}^{\,kl}\, \O^{\,(1)}{}_{kl} \,+\, \h^{\,kl}
\left(\, \pr_{\,i} \, \O^{\,(1)}{}_{kl} \, - \, \frac{1}{2}\ \pr^{\,j} \, T_{ij}\, \O^{\,(1)}{}_{kl} \,
\right)\, .
\ee
Therefore, quite interestingly, if eq.~\eqref{firsteq_shift} holds,
the $\O^{\,(1)}{}_{mn}$ must also satisfy the Bianchi identities, so
that they can be regarded as genuine Fronsdal-Labastida kinetic
operators. Actually, as we already saw for two families, a shift of
the type
\be
\d\, \vf = \h^{\,ij}\, \omega_{\, ij} \, ,
\ee
for the basic field $\vf$ leads generically to the variation
\be
\cF\,(\,\vf\,) \,\to \, \cF\,(\,\vf\,)\,+\, \frac{1}{2}\ \pr^{\,i}\pr^{\,j}\, \cO \, [\,D -
2\,]_{\,ij}{}^{\,kl}\, \omega_{\,kl} \, +\,\h^{\,ij} \,
\cF\,(\,\omega_{\,ij}\,)
\ee
for the Fronsdal-Labastida tensor. As a result, whenever
eq.~\eqref{firsteq_shift} holds, a condition that, we should stress,
depends only on the index structure of the tensors involved, one
recovers precisely a variation like \eqref{shift_gen}, with
\be
\O^{\,(1)}{}_{ij} \, = \, \cF\,(\,\omega_{\,ij} \,)\, .
\ee
In other words, this line of reasoning finally connects the
transformation \eqref{shift_gen} of $\cA$ to an operation effected
acting on the fundamental field $\vf$, as pertains to a symmetry
transformation proper. The end result is a link between the special
types of fields for which $\cA^{\,\pe}$ is not determined by the
Lagrangian field equation and a genuine Weyl-like symmetry of the
action, that generalizes the results of Section
\ref{sec:reduction2b} to fields with an arbitrary number of index
families.

To reiterate, a key lesson of the previous analysis is that the
relevant symmetry operations are to preserve, at the same time, the
Bianchi identities, the double-trace constraints that the $\cA$
tensor satisfies on shell and its equation of motion. Keeping this
in mind, we can now turn our attention to higher-order shifts of the
type
\be\label{shift_ordp}
\d\, \cA \, = \, \h^{k_1l_1}\ldots\,\h^{k_pl_p}\ \O^{\,(p)}{}_{k_1l_1,\, \ldots \, ,\,
k_pl_p}\, ,
\ee
where the parameters contain a priori all available independent
projections in family indices compatible with those admitted by a
product of $\h$ tensors, are thus fully symmetric under interchanges
of pairs of symmetric indices. These transformations generalize
eq.~\eqref{shift_double} and naturally emerge in an iterative
fashion, as we shall see shortly. Indeed, for a shift of this form
the simplest of the three basic conditions that we just stated, the
Bianchi identities, imply the relations
\be\label{Bianchi_ordp}
\begin{split}
\pr_{\,i} \, \d \, \cA \, &- \, \frac{1}{2}\ \pr^{\, j}\, T_{ij}\,
\d\, \cA \, = \, -\, \frac{p}{2}\ \h^{k_1l_1}\ldots\,
\h^{k_{p-1}l_{p-1}}\, \pr^{\,j}\Big[\,(\,D-2\,)\, \O^{\,(p)}{}_{ij\,,\,k_1l_1,\, \ldots \, ,\,
k_{p-1}l_{p-1}}  \\
&+\,\sum_{n\,=\,1}^{p-1}\,\O^{\,(p)}{}_{k_n\,(\,i\,,\,
j\,)\,l_n,\,\ldots\,,\, k_{r\neq n}l_{r \neq n},\, \ldots}\,+\,
S^{\,m}{}_{(\,i}\, \O^{\,(p)}{}_{j\,)\,m\,,\,k_1l_1,\, \ldots \, ,\,
k_{p-1}l_{p-1}} \,\Big]\\
&+\, \h^{k_1l_1}\ldots\,\h^{k_pl_p} \Big[\,\pr_{\,i} \,
\O^{\,(p)}{}_{k_1l_1,\, \ldots \, ,\, k_pl_p} \, - \, \frac{1}{2}\
\pr^{\, j}\, T_{ij}\,
\O^{\,(p)}{}_{k_1l_1,\, \ldots \, ,\,
k_pl_p} \,\Big]\, =\, 0\, ,
\end{split}
\ee
so that they are preserved if
\be\label{cond_ordp}
\begin{split}
&(\,D-2\,)\, \O^{\,(p)}{}_{ij\,,\,k_1l_1,\, \ldots \, ,\,
k_{p-1}l_{p-1}}
\,+\,\sum_{n\,=\,1}^{p-1}\,\O^{\,(p)}{}_{k_n\,(\,i\,,\,
j\,)\,l_n,\,\ldots\,,\, k_{r\neq n}l_{r \neq n},\, \ldots}\\&+\,
S^{\,m}{}_{(\,i}\, \O^{\,(p)}{}_{j\,)\,m\,,\,k_1l_1,\, \ldots \, ,\,
k_{p-1}l_{p-1}}\, = \, 0\, ,
\end{split}
\ee
and if $\O^{\,(p)}$ satisfies them as well. Alternatively, the type
for reasoning presented after eq.~\eqref{shift_delta} leads one to
consider the next class of transformations in the sequence, that
obtains replacing $p$ with $p+1$ in \eqref{shift_ordp}.

As was the case for the order-$\h$ shift determined by $\O^{\,(1)}$,
the non-trivial solutions of eq.~\eqref{cond_ordp} connect naturally
to Weyl-like symmetries, since if eq.~\eqref{cond_ordp} holds
\be
\d\, \vf \, = \, \h^{k_1l_1}\ldots\,\h^{k_pl_p}\ \omega^{\,(p)}{}_{k_1l_1,\, \ldots \, ,\,
k_pl_p}
\ee
translates into
\be
\cF\,(\,\vf\,) \,\to \, \cF\,(\,\vf\,)\,+\, \h^{k_1l_1}\ldots\,\h^{k_pl_p}\ \cF \,(\, \omega^{\,(p)}{}_{k_1l_1,\, \ldots \, ,\,
k_pl_p}\,)\, ,
\ee
and as a result one is led to identify $\O^{\,(p)}$ with
$\cF(\omega^{\,(p)})$. In a similar fashion, if
eq.~\eqref{cond_ordp} admits non-trivial solutions, one can show
that the double-trace constraints on $\vf$ translate essentially
into double-trace constraints on the $\O^{\,(p)}$. Finally, after
some algebra one can also show that an invariance of the equation of
motion obtains corresponding to this Weyl-like symmetry. This can be
actually done most conveniently restricting the attention to
parameters $\O^{(p)}$ that are two-column projected. In fact, the
mere assumption that eq.~\eqref{cond_ordp} admit a non-trivial
solution suffices to eliminate all $S^{\,i}{}_j$ from the variation
of the Einstein tensor, and then all parameters $\O^{(p)}$ that
carry projections with more than two columns can be seen to simply
call for a redefinition the $\cB_{ijkl}$. On the other hand, for
two-column projected $\O^{(p)}$ the invariance of the equation of
motion is less evident, and rests on the precise form that
eq.~\eqref{cond_ordp} takes in this case,
\be\label{modified_eq}
\cO\,[\,D-p+1\,]_{\,ij}{}^{\,mn}\,\O^{\,(p)}{}_{mn\,,\,k_1l_1,\, \ldots \, ,\,
k_{p-1}l_{p-1}} \, =\,0 \, .
\ee

Let us also point out that the order-$p$ equation \eqref{cond_ordp}
clearly admits solutions of the type $\eta
\, \O^{\,(p+1)}$, simply because its structure only feels the
space-time dimension $D$ and the $gl(D)$ tensorial properties of the
quantities involved. However, we would like to emphasize that the
eigenvalue equation at level $p+1$ is really different, and thus
yields genuinely new solutions. We saw this fact explicitly for two
families when we solved in Section \ref{sec:reduction2b} the
reducibility conditions for the double trace of $\cA$ for a
$\{2,2\}$ field in three dimensions.

We defer to a future work a more extensive analysis of these
systems, since here we have only confined ourselves to
characterizing the relations that signal the onset of the type of
phenomena that we investigated in detail in Section
\ref{sec:reduction2b}. It would be interesting to solve explicitly
eqs.~\eqref{cond_ordp}, or their simpler counterparts
\eqref{modified_eq} for two-column parameters, extending to
the more intricate $sl(N)$ weight lattices the results obtained with
the $sl(2)$ analysis. However, we can not refrain from displaying a
simple class of explicit solutions. To this end, let us consider
two-column gauge fields, that as we saw in Section
\ref{sec:reduction2b} provide an interesting
playground to exhibit some of the properties of multi-family
tensors. For instance, for $\{2^N\}$-projected irreducible fields of
the type $\vf_{\,\m^1_1\m^1_2;\,
\ldots\, ;\,\m^N_{\,1}\m^N_{\,2}}$, the previous equations greatly
simplify, to the extent that only a single independent trace emerges
at any order. In Section
\ref{sec:irreducible} we shall describe in detail how to deal with
this type of irreducible fields, but for the time being suffice it
to say that the relevant conditions are
\be
S^{\,k}{}_1\, \O_{\,1k} \, = \, - \ \O_{\,11} \qquad (\,
\textrm{for}\ k
\ {\rm fixed\ and} \,k\, \neq\,1\,) \, ,
\ee
that generalize to
\be
S^{\,k}{}_1\, \O_{\,1k\,,\,22\,,\, \ldots \,,\,pp} \, = \, - \
\O_{\,11\,,\,22\,,\, \ldots \,,\,pp} \qquad (\,
\textrm{for}\ k
\ {\rm fixed\ and} \,k\, > \,p\,)
\ee
for the higher shifts of eq.~\eqref{shift_ordp}. As a result, the
$\frac{N(N+1)}{2}$ conditions of eq.~\eqref{firsteq_shift}
effectively reduce to the single equation
\be
(\,D-2\,)\, \O_{\,11} \, + \, 2
\, \sum_{k\,=\,2}^N S^{\,k}{}_{1}\, \O_{\,1k} \, = \, 0  \, ,
\ee
and finally to
\be
(\,D-2\,N\,)\ \O_{\,11} \, = \, 0 \, ,
\ee
that clearly admits a non-trivial solution for $D = 2N$. In a
similar fashion, one can recognize that all equations of the form
\eqref{modified_eq} are equivalent to the conditions
\be
(\,D-2\,N+p-1\,)\ \O_{\,11\,,\,22\,,\, \ldots \,,\,pp} \, = \, 0 \,
,
\ee
so that the $p$-th trace of a $\{2^{\,N}\}$-projected field
$\vf_{\,\m^1_1\m^1_2;\, \ldots\, ;\,\m^N_{\,1}\m^N_{\,2}}$ is not
determined by the Lagrangian field equation in $D =
2\,N\,-\,p\,+\,1$.

\vskip 24pt

%%%%%%%%%%%%%%%%%%%%%%%%%%%%%%%%%%%%%%%%%%%%%%%%%%%%%%%%%%%%%%%%%%%%%%%%%%

\scs{Irreducible Bose fields of mixed symmetry}\label{sec:irreducible}

%%%%%%%%%%%%%%%%%%%%%%%%%%%%%%%%%%%%%%%%%%%%%%%%%%%%%%%%%%%%%%%%%%%%%%%%%%

The theory of higher-spin fields of mixed symmetry discussed in the
previous sections is essentially all one needs to compare with
String Theory. This is directly true for the bosonic string, whose
fields, as we have stressed, accompany products of mutually
commuting string oscillators, so that they are naturally reducible
symmetric tensors, while a convenient discussion of the bosonic
excitations of superstrings would also require the extension to the
case of multi-forms. This will actually be described in Section
\ref{sec:multiforms}, and as we shall see it is quite simple.
For definiteness, we thus continue momentarily to refer to
multi-symmetric tensors, and describe how the previous results can
be adapted to irreducible tensor fields, that are more along the
lines of what one is used to for low spins.

The condition of irreducibility basically states that the
symmetrization of a given line of the corresponding Young tableau
with any index belonging to one of the lower lines gives a vanishing
result. This type of operation can be simply described via the
$S^{\,i}{}_j$ operators defined in eq.~\eqref{flip} of Appendix
\ref{app:MIX}: an irreducible higher-spin tensor field thus
satisfies the conditions
\be
S^{\,i}{}_j \, \vf \, = \, 0  \quad (\,i<j\,)\, .
\label{irredcond}
\ee

The Fronsdal-Labastida operator $\cF$ commutes with the
$S^{\,i}{}_j$ operators, a property that can be proved directly making
use of the commutation relations listed in Appendix \ref{app:MIX}. As a result,
the constrained dynamical equations, and actually the whole
constrained Einstein-like tensor, take exactly the same form for
irreducible fields. On the other hand, both the compensator fields
and the Lagrange multipliers adapt to the given irreducible
components in a way that we now describe. To begin with, let us
consider the gauge transformations, that for a reducible field are
given in eq.~\eqref{gauge} so that, as we have seen in detail so
far, the theory relies on one independent gauge parameter for any
index family. In the irreducible case they should rather read
\be
\delta \, \vf \, = \, Y\, \partial^{\,i}\, \Lambda_{\,i} \, , \label{irrgaugeb}
\ee
where $Y$ is the Young projector corresponding to the irreducible
symmetry of $\vf$, that commutes with $\cF$ on account of the
previous discussion. Indeed, $Y$ projects onto the kernel of the
relevant $S^{\,i}{}_j$, to which $\vf$ belongs. In general,
irreducible fields involve fewer independent gauge parameters than
their reducible counterparts, and how this number is reduced was
already stated in
\cite{labastida1}: the relevant \emph{irreducible} gauge parameters
can be associated to all admissible Young diagrams obtained
stripping one box from the diagram corresponding to the gauge field.
This result reflects the structure of eq.~\eqref{irrgaugeb}, whose
right-hand side determines the variation of the field via a tensor
product with gradients carrying additional types of Lorentz indices,
belonging to the different families, up to a proper projection. The
differences with respect to the reducible case are sizable. For
instance, a reducible rank-$(4,2)$ bosonic field of the type
$\vf_{\m_1\m_2\m_3\m_4,\n_1\n_2}$ admits two independent gauge
parameters \footnote{In the following we shall continue to abide to
this notation, introduced in Section
\ref{sec:examples2b}, adapting it to the
irreducible case.}, $\L^{(1)}_{\m_1\m_2\m_3,\n_1\n_2}$ and
$\L^{(2)}_{\m_1\m_2\m_3\m_4,\n_1}$, as many as those admitted by an
irreducible $\{4,2\}$ field, but these are themselves irreducible.
Another significant example is provided by reducible fields of rank
$(s,s)$, that admit two independent reducible gauge parameters while
their irreducible counterparts admit a single irreducible gauge
parameter of rank $(s,s-1)$. More generally, if an irreducible gauge
field is characterized by a Young diagram containing a number of
identical rows, only a single gauge parameter is associated to all
of them.

It is perhaps instructive to recover these results in a slightly
different way. They follow in fact rather directly from the
structure of the $S^{\,i}{}_j$ operators that implement the
irreducibility condition, and from the requirement that
eq.~\eqref{irredcond} be invariant under gauge transformations, that
can be cast in the form
\be
\partial^{\,k} \left(\, S^{\,i}{}_j \, \L_{\,k} + \d^{\,i}{}_k \, \L_{\,j} \,\right) = \, 0 \quad (\,i <j\,) \,
,
\ee
which forces the entry to define a gauge-for-gauge transformation:
\be
S^{\,i}{}_j \, \L_{\,k} \, + \,  \d^{\,i}{}_k \, \L_{\,j} \, = \,
\partial^{\,l} \, \L^{i}{}_{[kl]j}  \quad (\,i <j\,) \, .
\label{gaugeparcondirr}
\ee
Up to the last irrelevant term, one thus obtains a set of
constraints that are conveniently analyzed starting from the highest
available value of $k$. In this case, in fact, $i$ is necessarily
less than $k$, being constrained to be less than $j$, so that
eq.~\eqref{gaugeparcondirr} reduces to the more familiar condition
that the last gauge parameter in the chain, $\L_k$, be irreducible,
\be
S^{\,i}{}_j \, \L_{\,k} \,=\, 0     \, .
\label{lastcondirr}
\ee
The remaining contributions can then be used to determine the other
parameters corresponding to lower values of $k$ in terms of this
solution and of additional independent ones that emerge, one for
each step, from the homogeneous parts of eqs.~\eqref{lastcondirr}.
Let us stress that the independent parameters solve further
irreducibility conditions, and therefore can only exist if the
corresponding Young diagrams are admissible. In conclusion, one thus
obtains an irreducible gauge parameter for each admissible Young
diagram built stripping one box from the original diagram for the
gauge field, precisely as stated in
\cite{labastida1}. The independent parameters lead by construction to
irreducible gauge transformations where no explicit Young projectors
as that present in eq.~\eqref{irrgaugeb} are needed. Let us see how
this works, referring to an irreducible $\{4,2\}$ gauge field, for
which the conditions \eqref{gaugeparcondirr} read
\begin{align}
& S^{\,1}{}_2\, \L_{\,1} \, + \, \L_{\,2} \, = \, 0 \, ,  \nonumber\\
& S^{\,1}{}_2\, \L_{\,2} \, = \, 0 \, . \label{irredgauge42}
\end{align}
To begin with, $\L_1$ could have three irreducible components, a
$\{3,2\}$, a $\{4,1\}$ and a $\{5\}$, while $\L_2$ could have a
$\{4,1\}$ and a $\{5\}$. Now the second of eqs.~\eqref{irredgauge42}
eliminates directly the $\{5\}$ component of $\L_2$, and then the
first eliminates the $\{5\}$ component of $\L_1$. The two $\{4,1\}$
components are then connected by the first equation, while the
$\{3,2\}$ component is a zero mode of the first equation, and as
such is not constrained. In space-time notation the conclusion is
that, when adapted to an irreducible $\{4,2\}$ field, the original
reducible gauge transformation \eqref{redgauge42} can be cast in the
form
\be
\begin{split}
\d\, \vf_{\m_1\m_2\m_3\m_4;\, \n_1\n_2} \, & = \,
\pr_{\,(\m_1}\L^{(1)\{3,2\}}{}_{\m_2\m_3\m_4);\, \n_1\n_2}\,\\ &+\,
\left( \pr_{\,(\m_1}\L^{(1)\{4,1\}}{}_{\m_2\m_3\m_4);\, \n_1\n_2}\,-\, \pr_{\,(\n_1|}\L^{(1)\{4,1\}}{}_
{(\, \m_1\m_2\m_3;\, \m_4)\,|\n_2)}\right)\, .
\end{split}
\ee
Let us stress again that no explicit Young projector is needed here,
since both combinations are already $\{4,2\}$ projected.

In a similar fashion, the unconstrained counterpart $\cA$ of $\cF$
is annihilated by the $S^{\,i}{}_j$ operators for $i<j$ \emph{only}
if the compensators satisfy a nested set of conditions that are the
direct counterpart of those in eq.~\eqref{gaugeparcondirr},
\be
\pr^{\,i}\pr^{\,j}\pr^{\,k} \left(\, S^{\,m}{}_n\, \a_{\,ijk}\,+\, \d^{\,m}{}_{(\, i}\, \a_{\,jk\,)n}  \,\right) \, = \,
0\label{irredalpha}
\quad (\,m <n\,) \, .
\ee
These, in their turn, reflect the fact that the independent
compensators $\Phi_i$ satisfy the same constraints of
eq.~\eqref{gaugeparcondirr} satisfied by the $\L_i$, consistently
with their gauge transformation \eqref{gaugephii}. Solving
eqs.~\eqref{irredalpha} shows that there is a single independent
compensator $\a_{ijk}$ for any allowed irreducible projection of the
original reducible compensators, so that the irreducible $\{4,2\}$
$\cA$ tensor can be cast in the form
\be
\begin{split}
\cA^{\{4,2\}}{}_{\,\m_1 \ldots\, \m_4;\,\n_1\n_2}\, & = \,
\cF^{\{4,2\}}{}_{\,\m_1 \ldots\,
\m_4;\,\n_1\n_2} - \, 3 \, \pr_{\,(\m_1\,}\pr_{\,\m_2\,}\pr_{\,\m_3}\, \a^{(1)\{2,1\}}{}_{\m_4);\,\n_1\n_2} \\
& + \, \pr_{\,(\n_1|\,}\pr_{\,(\m_1\,}\pr_{\,\m_2}\,
\a^{(1)\{2,1\}}{}_{\m_3;\,\m_4)\,|\n_2)} - \, 3 \, \pr_{\,(\m_1\,}\pr_{\,\m_2\,}\pr_{\,\m_3}\,
\a^{(1)\{3\}}{}_{\m_4)\,\n_1\n_2} \\
& + \, 2 \, \pr_{\,(\n_1|\,}\pr_{\,(\m_1\,}\pr_{\,\m_2}\,
\a^{(1)\{3\}}{}_{\m_3\m_4)\,|\n_2)} - \, 3 \,
\pr_{\,\n_1\,}\pr_{\,\n_2\,}\pr_{\,(\m_1}\,
\a^{(1)\{3\}}{}_{\m_2\m_3\m_4)}\, .
\end{split} \label{a42irred}
\ee

The other key ingredient of the construction are the Labastida
double-trace constraints or, in our unconstrained setting, the
corresponding $\cC_{ijkl}$ tensors. The key irreducibility
constraints on these quantities are
\be
S^{\,m}{}_n \, \cC_{\,ijkl} \, + \, \d^{\,m}{}_{(\,i}\,\cC_{\,jkl\,)\,n} \,
=
\, 0 \quad (\,m <n\,) \, ,
\ee
as can be seen either directly from eq.~\eqref{C2} or from the
Bianchi identities. Solving them shows that in the irreducible
$\{4,2\}$ case there is a single independent quantity of this type,
say $\cC^{(1)}{}_{\n_1\n_2}$ of eq.~\eqref{l42expl}, and thus also a
single Lagrange multiplier. In fact, in space-time notation the
multiplier terms of the Lagrangian for an irreducible $\{4,2\}$
field are simply
\be
3\left( \b^{(1)\{2\}\, \n_1\n_2} \, - \, 2\, \b^{(2)\{2\}\,
\n_1\n_2} \,+\,
\b^{(3)\{2\}\, \n_1\n_2}\right)\, \cC^{\,(1)}{}_{\n_1\n_2} \, ,
\ee
so that the single surviving Lagrange multiplier $\widetilde{\b}$ is the particular
combination of the three $ \b^{(i)\{2\}}$ above.

Similar steps allow to express the Lagrangian in terms of the
independent compensators $\a^{(1)\{2,1\}}{}_{\m;\,\n_1\n_2}$ and
$\a^{(1)\{3\}}{}_{\m_1\m_2\m_3}$ of eq.~\eqref{a42irred}, and
determine the independent irreducible components of the multiple
traces of $\cA$ and of their divergences. In particular, for two
families the relevant conditions are
\begin{align}
& S^{\,m}{}_n \, T_{ij}\, \cA \, +\, \d^{\,m}{}_{(\,i}\,T_{\,j\,)\,n} \,\cA
\, = \, 0 \, , \nonumber \\[2pt]
& S^{\,m}{}_n \, T_{ij} \, T_{kl}\, \cA \, +\,
\d^{\,m}{}_{(\,i}\,T_{\,j\,)\,n}\, T_{kl}\, \cA \, +\,
\d^{\,m}{}_{(\,k}\,T_{\,l\,)\,n}\,T_{ij}\, \cA
\, = \, 0 \, , \nonumber \\[2pt]
& S^m{}_n \, \pr_{\,i} \, T_{jk}\, \cA \, +\,\d^{\,m}{}_{i}\,
\pr_{\,n} \,T_{jk}
\,\cA \,+\,
\d^{\,m}{}_{(\,j}\,T_{\,k\,)\,n} \, \pr_{\,i}
\,\cA
\, = \, 0 \, .
\end{align}
In conclusion, the Lagrangian for an irreducible $\{4,2\}$ gauge
field, when expressed in terms of independent quantities, can be
presented in the form
\be
\begin{split}
\cL \, & = \, \frac{1}{2}\ \vf^{\,\{4,2\}\,
\m_1\ldots\, \m_4;\,\n_1\n_2}\Big\{\, \cA^{\{4,2\}}{}_{\m_1\ldots\, \m_4;\,\n_1\n_2}
\, - \, \frac{1}{2} \ \h_{\,(\,\m_1\m_2\,} \cA^{\,\pe_\m \{2,2\}}{}_{\m_3\m_4);\,\n_1\n_2} \\
& - \, \12 \ \h_{\,(\,\m_1\m_2\,} \cA^{\,\pe_\m
\{3,1\}}{}_{\m_3\m_4);\,\n_1\n_2} \, + \, \frac{1}{4} \ \h_{\,(\,\n_1|\,(\,\m_1\,} \cA^{\,\pe_\m \{3,1\}}{}_{\m_2\m_3;\,\m_4)\,|\,\n_2)} \, \\
& - \, \12 \ \h_{\,(\,\m_1\m_2\,} \cA^{\,\pe_\m
\{4\}}{}_{\m_3\m_4)\,\n_1\n_2} \, + \, \frac{3}{4} \ \h_{\,(\,\n_1|\,(\,\m_1\,} \cA^{\,\pe_\m \{4\}}{}_{\m_2\m_3\m_4)\,|\,\n_2)} \, - \, 3 \ \h_{\n_1\n_2\,} \cA^{\,\pe_\m \{4\}}{}_{\m_1 \ldots\, \m_4} \\
& + \, \fr{1}{18} \left(\, 2\ \h_{\, \n_1 \n_2} \,
\h_{\,(\,\m_1\m_2\,|} \, - \, \h_{\, \n_1 \,(\, \m_1}
\h_{\,\m_2\,|\,\n_2} \,\right) \left(\, \cA^{\, \pe_{\m} \pe_{\n}} -
\, \cA^{\, \hpe\, \hpe} \,\right){}_{\!|\,\m_3\m_4\,)} \,\Big\} \\
& - \, 4 \ \a^{(1)\{2,1\}\,\m\,;\,\n_1\n_2} \, \pr^{\,\l} \Big\{\, \cA^{\,\pe_\m \{2,1\}}{}_{\l\,\m\,;\,\n_1\n_2} \, - \, \frac{1}{18} \ Y_{\{2,1\}} \, \h_{\,\n_1\n_2} \left(\, \cA^{\, \pe_{\m} \pe_{\n}} -
\, \cA^{\, \hpe\, \hpe} \,\right){}_{\m\l}  \,\Big\} \\
& - \, 10 \ \a^{(1)\{3\}\,\m_1\m_2\m_3} \, \pr^{\,\l} \Big\{\, \cA^{\,\pe_\m \{3\}}{}_{\l\,\m_1;\,\m_2\m_3} \, - \, \frac{1}{18} \ \h_{\,(\,\m_1\m_2}  \left(\, \cA^{\, \pe_{\m} \pe_{\n}} -
\, \cA^{\, \hpe\, \hpe} \,\right){}_{|\,\m_3)\,\l} \,\Big\} \\
& + \, 3 \ \widetilde{\b}^{\,\n_1\n_2} \, \cC^{\,(1)}{}_{\n_1\n_2}\,
.
\end{split}
\ee
Alternatively, one could work with the original Einstein-like tensor
of eq.~\eqref{einst4,2}, that would maintain the same form for
projected gauge fields, as we stressed above, but whose traces would
not be independent. The procedure that we have illustrated with this
example is of general import, since its basic ingredients, the
irreducibility conditions and their natural generalizations to
multiple traces, clearly hold for generic multi-family gauge fields.

\vskip 24pt

%%%%%%%%%%%%%%%%%%%%%%%%%%%%%%%%%%%%%%%%%%%%%%%%%%%%%%%%%%%%%

\scs{Elimination of higher derivatives}\label{sec:lowerder}

%%%%%%%%%%%%%%%%%%%%%%%%%%%%%%%%%%%%%%%%%%%%%%%%%%%%%%%%%%%%%

In this section we would like to describe a systematic way of
eliminating the higher-derivative terms appearing in our
unconstrained Lagrangians while preserving, insofar as possible, the
simplicity of the construction.

\vskip 24pt

\scss{The problem}\label{sec:lowerintro}

Let us begin by stressing once more that all terms involving more
than two derivatives enter our Lagrangians in connection with
compensator fields, that can be eliminated by a suitable partial
gauge fixing. And indeed all tests available in the free theory, and
in particular the current-exchange amplitudes for symmetric tensors
of
\cite{fms,fms2}, confirm the obvious feeling that their presence
does not bring about any complications.
Still, it is both interesting and natural to try and bring these
systems closer to their lower-spin counterparts.

An unconstrained formulation of free symmetric higher spins without
higher-derivative terms and with a fixed number of extra fields was
first attained in \cite{buchnew}. It is an interesting off-shell
variant of the on-shell truncation of the ``triplets''
\cite{triplets} of String Field Theory \cite{sftheory} obtained in
\cite{fs2,st} (see also \cite{sorokvas} for some recent
developments in the ``frame-like'' formalism). Our aim in this
section is to try and extend these types of results to
mixed-symmetry fields, starting however from the alternative
construction of \cite{dario07}, that was tailored to the compensator
constructions of \cite{fs3,fms}.

The idea underlying our procedure is the replacement of the $\a_{\,
i j k}$ of (\ref{compensator}) with other fields whose dimensions
are at least as high as that of the gauge potentials $\vf$. Let us
stress that such fields are not pure gauge, but for instance in the
solution for the one-family case given in \cite{dario07} they vanish
on shell.

It is simple to construct kinetic tensors similar to $\cA$ but free
of higher derivatives, and indeed several options are available, but
it is less straightforward to make sure that the resulting
Lagrangians do not propagate additional degrees of freedom. Let us
begin with a brief analysis of symmetric tensors, aimed at
clarifying the origin of some potential difficulties. A close
scrutiny of this simpler case will provide clues for the more
general approach that will be presented in the next section. For
fully symmetric bosonic fields, there are in principle several
options to compensate the gauge transformation of the Fronsdal
tensor
\footnote{A reader familiar with some previous papers, such as
\cite{fs3,fms}, will recognize that there we found it more
convenient to define products of mutually commuting objects, and of
derivatives in particular, with different normalizations. For
example, in this case we would have introduced the convenient symbol
$\pr^{\,3}$ to denote the product of three $\pr$'s, up to an overall
factor 6. For the sake of clarity and for an easier comparison with
the multi-family case, however, here we prefer to conform, insofar
as possible, to the notation of the preceding sections. }
\be
\d \, \cF \, = \, \frac{1}{2} \, \pr \,\pr \,\pr \, \L^{\, \pe} \, .
\ee
In the first part of this section, for brevity, we are hiding the
single-valued family indices, so that here gradients are denoted by
$\pr$ rather than $\pr^{\,1}$, divergences are denoted by $\partial
\, \cdot$ rather than $\partial_{\, 1}$ and
the only available trace, $T_{11}$, is denoted by a ``prime''. The
choice made in
\cite{fs2, st, fs3}\footnote{The first occurrence of this type of
field for spin 3 actually dates back the work of Schwinger
\cite{schwinger}, as was pointed out to us by G. Savvidy.} was to
resort to a compensator $\a$, the one-family analogue of the
$\a_{ijk}$ but treated as an independent field, with gauge
transformation
\be
\d \, \a \, = \, \L^{\, \pe} \, .
\ee

On the other hand, three other choices could provide in principle
viable alternatives avoiding the introduction of higher derivatives:
\begin{align} \label{compens}
& \th_{(1)} \, : & &\d \, \th_{(1)} \, = \, \pr \,  \L^{\, \pe} &&
\,
\ra
\, & &
\cA_{\, \th_{(1)}} \, = \, \cF \, - \, \frac{1}{2}\ \pr \, \pr \, \th_{(1)} \, ,& \nonumber \\
& \th_{(2)} \, : & &\d \, \th_{(2)} \, = \, \pr \, \pr
\,
\L^{\,
\pe} &&
\,
\ra \, & &
\cA_{\, \th_{(2)}} \, = \, \cF \, - \, \frac{1}{2}\ \pr \, \th_{(2)} \, ,& \\
& \th_{(3)} \, : & &\d \, \th_{(3)} \, = \,
\pr\,\pr\,\pr
\,
\L^{\,
\pe} &&
\,
\ra \, & &
\cA_{\, \th_{(3)}} \, = \, \cF \, - \, \frac{1}{2}\ \th_{(3)} \, .& \nonumber
\end{align}
Notice that in all these cases the additional field is \emph{not}
manifestly pure gauge, but one can aim nonetheless at constructions
where this condition is enforced on-shell, for instance, via a term
of the form
\be
\bra \c \comma \th_{(1)} \,-\, \pr \, \a \ket\, , \label{theta1_a}
\ee
where for definiteness we are focusing on the case of $\th_{(1)}$,
and where $\c$ is a gauge-invariant Lagrange multiplier.

The three kinetic tensors $\cA_{\, \th_{(i)}}$ indeed do not contain higher
derivatives and satisfy the Bianchi identities
\begin{align} \label{bianchis}
& \th_{(1)} : & &\!\prd \cA_{\, \th_{(1)}} \, - \, \12 \, \pr \,
\cA_{\,
\th_{(1)}}^{\, \pe} = \, - \, \fr{1}{4}\, \pr \, \{\pr\, \pr \,
\vf^{\, \pe \pe} + \, 2 \, \Box \, \th_{(1)} \, - \, 2 \,
\pr \, \prd \th_{(1)} \, -  \, \pr\, \pr \, \th^{\, \pe}{}_{\!(1)}\} \,
\eq - \, \12 \, \pr \, \cC_{(1)} \, ,  \nonumber\\
& \th_{(2)} : & &\!\prd \cA_{\, \th_{(2)}} \, - \, \12 \, \pr \,
\cA_{\,
\th_{(2)}}^{\, \pe} = \, - \, \fr{1}{4}\, \{\pr \, \pr \, \pr \, \vf^{\, \pe
\pe} - \, 2 \, \Box \, \th_{(2)} \, + \,
 \pr\, \pr \, \th^{\, \pe}{}_{\!(2)}\} \,  \eq \, - \, \cC_{(2)} \, , \nonumber \\
& \th_{(3)}: & & \!\prd \cA_{\, \th_{(3)}} \, - \, \12 \, \pr \,
\cA_{\,
\th_{(3)}}^{\, \pe} = \, - \, \fr{1}{4}\, \{\pr\, \pr\, \pr \, \vf^{\, \pe
\pe} + \, 2 \, \prd \th_{(3)} \, - \,
 \pr \, \th^{\, \pe}{}_{\!(3)}\} \, \eq \, - \, \fr{3}{2} \, \cC_{(3)} \, .
\end{align}
These relations already display a delicate feature of this approach,
related to the request that the resulting Lagrangians only describe
the propagation of massless, irreducible spin-$s$ degrees of
freedom. For the sake of comparison, let us recall that in the
theory described by the minimal Lagrangian (\ref{lagone})
\begin{itemize}
 \item the equation for $\vf$ can be reduced to the Fronsdal form;
 \item the compensator $\a$ can be set to zero with a gauge choice;
 \item on-shell the Lagrange multiplier $\b$ can be expressed
 in terms of the field $\vf$, which guarantees that it does not carry
 additional degrees of freedom.
\end{itemize}
If one were to retrace our usual procedure, introducing in the
Lagrangian quantities of the form $\bra \b_{(i)} \comma
\cC_{(i)}\ket$, $i= 1, 2, 3$, with $\b_{(i)}$ independent Lagrange
multipliers, these fields would enter the equation for $\vf$ in
combinations involving their divergences, in a way that depends on
the corresponding choice for $\th_{(i)}$. Thus, differently from the
usual case, the equation for $\vf$ would not fix directly the
multipliers in terms of the gauge potential, simply because not all
gradients can be factored in eq.~\eqref{bianchis}. In addition, the
presence of D'Alembertian operators in $\cC_{(1)}$ and $\cC_{(2)}$
would have the unpleasant consequence of introducing in the
equations for the compensators $\th_{(1)}$ and $\th_{(2)}$ wave
operators acting on the multipliers $\b_{(1)}$ and $\b_{(2)}$.

An alternative approach that is devoid of these difficulties was
proposed in \cite{dario07}, and is based on the following two
observations:
\begin{itemize}
 \item rather than introducing potentially propagating  multipliers,
one can look for combinations of fields already present in the
Lagrangian and possessing the required gauge transformations;
\item in order to guarantee that no spurious degrees
of freedom  propagate, one can add an independent Lagrange
multiplier $\b$ to enforce the usual double trace constraint $\bra
\b \comma \cC\ket$ of \cite{fs3,fms}, together with an additional
one for the constraint relating the $\th_{(i)}$ to $\a$, as in
eq.~\eqref{theta1_a}. In this construction, these additional
multipliers would be gauge invariant, while their equations of
motion would provide the conditions needed to recover the Fronsdal
form.
\end{itemize}
Actually, while the second step can always be performed, only
$\th_{(1)}$ can build a combination transforming as an effective
multiplier. In this case the complete Lagrangian without higher
derivatives is then
\be
\begin{split} \label{basiclagr}
\cL \, = & \, \12 \, \bra \vf \, , \, \cA_{\, \th_{(1)}} \, - \, \12 \, \h \, \cA^{\, \pe}{}_{\th_{(1)}}\ket
\, + \, \fr{1}{4} \, \bra\th_{(1)} \, , \, \cA^{\, \pe}{}_{\th_{(1)}} \ket
+ \, \fr{1}{4} \, \bra \vf^{\, \pe} \, - \, \th_{(1)} \, , \,  \cC_{(1)} \ket \\
&  +  \, \bra \c \, , \, \th_{(1)} \, - \, \pr \, \a\ket \, + \, 3
\,  \bra \b \, , \, \vf^{\, \pe \pe} \, - \, 4 \, \prd \, \a
\, - \, \pr \, \a^{\,\pe} \ket \, ,
\end{split}
\ee
where $(\,\vf^{\, \pe} \, - \, \th_{(1)}\,)$ is precisely the
combination of the original fields playing the role of an effective
Lagrange multiplier for $\cC_{(1)}$. Notice that the terms of this
Lagrangian involving $\cA_{\, \th_{(1)}}$ have the same form as in
eq.~\eqref{lagone}, since they are still driven by the Bianchi
identity. Furthermore, out of the three fields of
eq.~\eqref{compens} only $\theta_{(1)}$ can build a direct local
coupling to $\cA^{\,
\pe}{}_{\th_{(i)}}$, that already contains two derivatives,
precisely because it has dimension one. Indeed, in
\cite{dario07} it was shown that the resulting equations of motion
set to zero on-shell some of the additional fields and relate the
others to $\vf$, so that the system reduces correctly to the
Fronsdal theory after a partial gauge fixing. The field content of
\eqref{basiclagr} is apparently the minimal one allowing a local description
of unconstrained symmetric bosons of all spins without higher
derivatives \ft{For instance, the Lagrangian of \cite{buchnew}
provides an elegant off-shell truncation of the ``triplets'' of
\cite{triplets}, and contains the higher-spin field $\vf$, the compensator $\a$, two auxiliary
fields $C$ and $D$ and two Lagrange multipliers $\l_1$ and $\l_2$.
It can actually be reduced to
\eqref{basiclagr} in two steps. First, solving the equations for $C$
and $\l_1$ and substituting back, so as to eliminate $C$, $D$ and
$\l_1$, and finally making the identifications $\pr
\, \a =
\th_{\, (1)}$, $\l_2 = \b$ and adding the constraint $\bra
\c \, , \, \th_{(1)} - \pr \, \a\ket$.}.

For fields of mixed symmetry no simple solution of this type is
apparently available, so that one is led to consider more general
possibilities for the structure of the compensators. In order to
appreciate the difficulties that one is confronted with in the
general case, let us try a straightforward extension of the approach
of \cite{dario07}, defining the gauge invariant combination
\be
\cA_{\, \th} \, = \, \cF \, - \, \12 \, \pr^{\, i} \pr^{\, j} \, \th_{\, i j} \, ,
\ee
where
\be
\d \, \th_{\, i j} \, = \, \fr{1}{3} \, \pr^{\, k} \, T_{\, (\,i j}\,\L_{\, k\,)} \,
.
\ee
As for symmetric fields, the subtleties in the construction of the
full Lagrangian can be traced to the structure of the Bianchi
identity satisfied by $\cA_{\, \th}$,
\be
\pr_{\, i} \, \cA_{\, \th} \, - \, \12 \, \pr^{\, j} \, T_{\, ij} \, \cA_{\, \th} \, =
 \, \12 \, \pr^{\, j} \, \cC^{\,\th}{}_{ij} \, .
\ee
The gauge invariant constraint tensors $\cC^{\, \th}{}_{ij}$
generalize the $\cC_{(1)}$ of the previous example, and are given by
\be
\begin{split}
\cC^{\, \th}{}_{ij} \, & = \, \fr{1}{6} \, \pr^{\, k} \pr^{\, l} \, T_{(\,ij}\,T_{kl\,)}\,
\vf \, + \, \Box \, \th_{\, i j} \, - \, \pr^{\, k}\pr_{\, k} \, \th_{\, i j} \\
& - \, \fr{1}{8} \, \pr^{\, k}\pr^{\, l} \, \left(\, 2 \,  T_{\,
ij}
\,
\th_{\, kl} \, - \, 2 \, T_{\, kl} \, \th_{\, ij} \, + \, T_{\, i\,(\,k}
\, \th_{\, l\,)\,j} \, + \, T_{\, j\,(\,k} \, \th_{\, l\,)\, i} \,\right) \,
,
\end{split}
\ee
so that, as expected, a D'Alembertian operator is again present in
the constraint tensor associated to the ``lower derivative''
compensators $\th_{\, ij}$.

On account of the results displayed here and in the previous
sections, in the general case of $N$-family fields, one can consider
the trial Lagrangian
\be \label{trialmix}
\begin{split}
\cL_0 \, & = \, \12 \, \bra\, \vf \,\comma\, \sum_{p\,=\,0}^{N} \,
\frac{(-1)^{\,p}}{p\,!\,(\,p+1\,)\,!} \ \h^{i_1 j_1} \ldots \, \h^{i_{\,p} j_{\,p}} \,
\cA^{\,[\,p\,]}{}_{i_1 j_1,\,\ldots\,,\,i_{p} j_{p}} \,\ket \\
& + \frac{3}{4} \, \bra\, \theta_{\,ij} \,\comma\,
\sum_{p\,=\,0}^{N-1}
\,
\frac{(-1)^{\,p}}{p\,!\,(\,p+3\,)\,!} \ \h^{i_1 j_1} \ldots \, \h^{i_{p} j_{p}} \,
\cA^{\,[\,p+1\,]}{}_{ij\,,\,i_1 j_1,\,\ldots\,,\,i_{p} j_{p}}
\,\ket \\
& + \, \bra \c_{\, ij} \comma \th_{\, i j} \, - \, \pr^{\, k}
\, \a_{\, i j k}(\Phi) \ket \, ,
\end{split}
\ee
where we already included the constraints on the $\th_{\, i j}$
compensators. The gauge variation of eq.~ (\ref{trialmix}) generates
the remainder
\be \label{gaugelower}
\d \, \cL_0 \, = \, \frac{1}{4} \, \bra \sum_{p\,=\,0}^{N} \, \frac{(-1)^{\,p}}{p\,!\,(\,p+2\,)\,!} \
\pr_{\,(\,i\,|} \, \h^{m_1n_1} \ldots\, \h^{m_pn_p} \, Y_{\{2^p,1\}}
\, T_{m_1n_1} \ldots\, T_{m_pn_p} \, \L_{\,|\,j\,)} \comma \cC^{\,\th}{}_{ij} \ket \,
,
\ee
and again, in strict analogy with what was already seen in the
one-family case, one can try to look for combinations of the fields
already present in the Lagrangian possessing the gauge
transformation in (\ref{gaugelower}). However, in sharp contrast
with the previous example, with two or more families the
generalization of the composite multipliers of the symmetric case
transform as
\be
\d \left(\, T_{ij} \, \vf \, -\, \th_{\,ij}\, \right)\, = \,
\pr_{\,(\,i} \, \L_{\,j\, )} + \, \frac{1}{3} \left(\, 2\, T_{ij}\, \L_k \, - \, T_{k\,(\, i}\, \L_{j\,)} \,\right)\, ,
\ee
with a relative factor between the two terms that does not allow to
cancel the first term in \eqref{gaugelower}. In order to recover a
formulation of mixed-symmetry gauge fields that is free of
higher-derivative terms, one is therefore led to consider more
general possibilities for the compensators $\th_{ij}$. A systematic
procedure leading to a successful choice for general multi-family
fields is outlined in the next section.

\scss{A general solution}\label{lowermixgeneral}

As we have seen, the remainder or ``classical anomaly'' in the
Bianchi identities is the crucial ingredient when one tries to build
unconstrained Lagrangians, while the main virtue of our basic choice
(\ref{newcomp}) is that in the Bianchi identity (\ref{bianchiA}) the
remainder does not contain the D'Alembertian operator. This grants
from the very beginning that no pathologies related to the behavior
of the multipliers $\b_{\, i j k l}$ defined in (\ref{mult}) can
present themselves. In the search for unconstrained Lagrangians
without higher-derivative terms, one should thus be ready to
consider more complicated gauge invariant completions of the
Fronsdal-Labastida tensor, if their Bianchi identities are free of
D'Alembertian operators acting on the compensators. On the other
hand, since the standard form (\ref{gauge_fronsdal}) of the
variation of $\cF$ does not seem to provide a good guidance, it
might be helpful to try and re-express it in alternative ways, so as
to suggest which additional fields are actually needed.

A conceptually simple possibility is to separate in the gauge
parameters $\L_i$ two contributions, letting
\be \label{gaugesep}
\L_{\, i} \, = \, \L^{(t)}{}_{i} \, + \, \h^{\, jk} \, \L^{(p)}{}_{ijk} \, ,
\ee
where the $\L^{(t)}{}_{i}$ satisfy the conditions
\be
T_{(\,ij} \, \L^{(t)}{}_{k\,)} \, \equiv \, 0 \, ,
\ee
so that they are effectively the gauge parameters of the Labastida
theory, while the $\L^{(p)}{}_{ijk}$ carry the full amount of gauge
symmetry that one would like to add, and are such that
\be \label{deflambdap}
T_{\, (\,ij } \ \h^{\, lm} \, \L^{(p)}{}_{\, k\,)\, lm} \, \equiv \, T_{\, (\,ij} \, \L_{\, k\,)} \, .
\ee
Working in terms of $\L^{(t)}{}_{i}$ and $\L^{(p)}{}_{ijk}$, the
gauge variation of the Labastida tensor reads
\be
\begin{split}
\d \, \cF \, & = \, \frac{1}{6} \, \pr^{\,i} \pr^{\,j} \pr^{\, k} \left\{\, 3\, D\, \L^{(p)}{}_{ijk} \, +
\, 2\, S^{\,l}{}_{(\,i}\, \L^{(p)}{}_{jk\,)\,l} \, + \, \h^{\, lm}\,T_{(\,ij}\, \L^{(p)}{}_{k\,)\,lm} \,\right\} \, ,
\end{split}
\ee
a result that calls for the introduction of compensator fields
$\th_{\, ij}$ such that
\be \label{lowercomp}
\d \, \th_{\, ij}  \, = \, \pr^{\, k} \, \L^{(p)}{}_{ijk} \, .
\ee
One is thus led to define the rather unconventional gauge invariant
kinetic tensor
\be \label{kinlower}
\cA_{\,\th} \, = \, \cF \, - \, \12 \, \left[\, (\,D - 2\,)\,
\pr^{\,i} \pr^{\, j} \, + \, \pr^{\, k} \pr^{\, (\,i\,} S^{\, j\,)}{}_k \,\right] \, \th_{\, ij}
\, - \, \h^{\, ij} \, \cF_{ij\,}(\theta) \, ,
\ee
where in particular the $\cF_{ij\,}(\theta)$ are Labastida tensors
for the collection of $\th_{\, ij}$ fields. According to our
previous discussion, despite the rather involved form of
$\cA_{\,\th}$, it is the structure of its Bianchi identity that
should tell us whether such a choice for the compensators might
prove useful in the construction of a suitable gauge invariant
Lagrangian. The explicit computation gives the gratifying result
\be
\begin{split}
\pr_{\, i} \, \cA_{\,\th} \, - \, \12 \, \pr^{\, j} \, T_{\, i j} \, \cA_{\,\th} \, = &
- \, \frac{1}{12} \, \pr^{\,j} \pr^{\,k} \pr^{\,l}
\, T_{(\,ij}\,T_{kl\,)}
\left(\,
\vf - \h^{\, mn} \, \th_{\,mn} \,\right)\, ,
\end{split}
\ee
where all contributions involving a D'Alembertian operator acting on
$\th_{\, ij}$ have disappeared thanks to the presence of the
Labastida tensors $\cF_{ij\,}(\theta)$.

The previous result is clearly a strong hint that the definition
(\ref{kinlower}) can lead to the desired solution. That this is
actually the case is confirmed by the following observations, that
also suggest a simpler route leading to complete Lagrangians. Let us
in fact recall that the completion of the gauge symmetry of the
theory, leading from the Fronsdal-Labastida tensor $\cF$ to the
basic unconstrained tensor $\cA$ defined in (\ref{A}), could be
attained working directly at the level of the gauge field $\vf$. As
observed in Section \ref{sec:lagrangian2b}, the tensor $\cA$ can
indeed be regarded as the result of a Stueckelberg-like substitution
performed in $\cF$
\be \label{shift}
 \vf \, \ra \, \vf - \, \pr^{\, i} \, \Phi_{\, i} \, ,
\ee
where the fields $\vf$ and $\Phi_{\, i}$ possess the gauge
transformations
\begin{alignat}{2}
&\d \, \vf & & = \, \pr^{\,i} \, \L_{\, i} \, , \nonumber \\
&\d \, \Phi_{\, i} & & = \, \L_{\,i} \, .
\end{alignat}
Now, in order to better understand the meaning of the $\th_{\, ij}$
and of their gauge transformations (\ref{lowercomp}), it is worth
stressing the otherwise obvious fact that under the substitution
(\ref{shift}) \emph{any} function of $\vf$ would be gauge invariant.
One might thus wonder whether the choice (\ref{shift}) is really the
conceptually simplest option for our purposes, since when performing
this shift in the Fronsdal-Labastida tensor one overlooks the fact
that the  theory already possesses a constrained gauge invariance,
that is somehow built anew in terms of the additional fields.

In other words, the Stueckelberg-like shift (\ref{stueck}) can make
\emph{any} theory gauge invariant, even one that does not possess,
to begin with, a gauge symmetry. Our aim here is rather
\emph{to enlarge to the unconstrained level a constrained gauge
symmetry that is already present}. In this sense, the substitution
(\ref{shift}) is really somewhat unnatural, while it looks more
logical, if technically more involved, to decompose the gauge
variation of $\vf$, following eq.~(\ref{gaugesep}), as
\be
\d \, \vf \, = \, \pr^{\, i} \, \L^{(t)}{}_{i} \, + \, \h^{\, ij} \, \pr^{\, k} \,  \L^{(p)}{}_{ijk} \, ,
\ee
and then to exploit the transformation properties (\ref{lowercomp})
of the $\th_{\, ij}$ in order to define an improved form of the
substitution (\ref{stueck}),
\be \label{stueckimpr}
\vf \, \rightarrow \, \vf \, - \, \h^{\, ij} \, \th_{\, ij} \, ,
\ee
that indeed turns the Fronsdal-Labastida tensor (\ref{fronsdal})
into the kinetic tensor $\cA_{\,\th}$ defined in (\ref{kinlower}).
The reader will not fail to notice that this type of shift was
already encountered when we discussed Weyl-like symmetries. Here,
however, we do not require that the operator $\cO[D-2]$ of
eq.~\eqref{operatorO} annihilate $\th_{ij}$, but we simply assign to
these fields a specific gauge transformation.

Indeed, the key difference between (\ref{stueckimpr}) and the naive
Stueckelberg shift (\ref{stueck}) is that the combination $\vf
\, - \, \h^{\, ij} \, \th_{\, ij}$ transforms precisely
as the Fronsdal-Labastida field, even in the presence of
unconstrained gauge parameters $\L_i$. The substitution can now be
effected in any of the two Labastida-like Lagrangians that we have
presented in Section \ref{sec:generalb} in eqs.
\eqref{stuecklaggen} and \eqref{laggen_invmult}. For instance, referring
to eq.~\eqref{stuecklaggen} and adding a further constraint forcing
the $\th_{\, ij}$ to be pure gauge, in the same spirit as in
(\ref{basiclagr}), leads to
\be \label{lowerlagr}
\begin{split}
\cL \, & = \, \12 \, \bra \vf  \, - \, \h^{\, mn} \, \th_{\, mn} \comma \cA_{\,\th} \, + \,
\sum_{p\,=\,1}^{N} \, \frac{(-1)^{\,p}}{p\,!\,(\,p+1\,)\,!}
 \ \h^{\,p} \, \cA_{\,\th} ^{\,[\,p\,]} \ket \\
& + \, \fr{1}{8} \, \bra \b_{\, i j k l} \comma T_{(\,ij}\,T_{kl\,)}
\, (\vf \, - \, \h^{\, mn} \, \th_{\, mn})\ket
\, + \, \bra \c_{\, ij} \comma \th_{\, ij} \, - \, \pr^{\, k} \, \P_{\, i j k}^{\, l m n} \, \a_{\, l m n} (\Phi)\ket\, .
\end{split}
\ee
This Lagrangian is invariant under the gauge transformations that we
already defined, and that we collect here:
\be
\begin{split}
& \d \, \vf \, = \, \pr^{\, i} \, \L_{\, i} \, ,\\
& \d \, \th_{\, ij}  \, = \, \pr^{\, k} \, \L^{(p)}{}_{ijk} \, , \\
& \d \, \a_{\, ijk} \, = \, \frac{1}{3} \, T_{(\,ij} \, \L_{\,k)} \, , \\
& \d \, \b_{\, ijkl} \,  = \, \frac{1}{2} \, \sum_{p\,=\,0}^{N} \, \frac{(-1)^{\,p}}{p\,!\,(\,p+2\,)\,!} \
\pr_{\,(\,i\,}\pr_{\,j\,}\pr_{\,k\,|} \, \h^{m_1n_1} \!\ldots\, \h^{m_pn_p} \, Y_{\{2^p,1\}}
\, T_{m_1n_1} \ldots\, T_{m_pn_p} \, \L^{(t)}{}_{|\,l\,)} \, , \\
& \d \, \c_{\,ij} \, = \, 0 \, .
\end{split}
\ee
In (\ref{lowerlagr}), $\P_{\, i j k}^{\, l m n}$ is the projector
defining the solution of eq.~(\ref{deflambdap}) for $\L^{(p)}{}_{i j
k}$ in the form
\be \label{sollambdap}
\L^{(p)}{}_{i j k }\, = \, \P_{\, i j k}^{\, l m n} \, T_{\, (\,l m } \, \L_{\, n\,)} \, .
\ee
Computing the projector $\P_{\, i j k}^{\, l m n}$ represents the
main technical difficulty of this construction, and indeed we were
not able to obtain for it an explicit closed form in the general
case, although it is rather straightforward, if lengthy, to compute
it explicitly in specific cases of interest. Thus, for instance, in
the one-family case of symmetric tensors the explicit relation
between $\L^{\, (p)}$ and $\L$ is
\be \label{explambdp}
\L^{(p)} \, = \, \sum_{k = 0}^{[\fr{s - 3}{2}]} \, \fr{(-1)^{\, k} \, k\, !}{\prod_{i =
0}^{k}
\, \sum_{j = 0}^{i} \, \left[D \, + \, 2 \, (s \, - 2 \, j \, - \, 3)\right]} \ \h^{\, k} \, \L^{\, [k+1]} \,
,
\ee
where $\L^{\, [k+1]}$ denotes the $(k+1)$-th trace of $\L$ and
$\h^{\, k}$ is a product of $k$ Minkowski metric tensors written
with unit overall normalization and with the minimal number of terms
needed to be totally symmetric.

Finally, the reduction of the equations of motion to the
Fronsdal-Labastida form $\cF = 0$ would follow steps similar to
those illustrated for the minimal higher-derivative Lagrangians.

\vskip 24pt

%%%%%%%%%%%%%%%%%%%%%%%%%%%%%%%%%%%%%%%%%%%%%%%%%%%%%%%%%%%%%%%%%%%%%%%%%%%%%%

\scs{Multi-form gauge fields} \label{sec:multiforms}

%%%%%%%%%%%%%%%%%%%%%%%%%%%%%%%%%%%%%%%%%%%%%%%%%%%%%%%%%%%%%%%%%%%%%%%%%%%%%%

So far we have discussed in some detail the properties of
multi-symmetric gauge fields of the type $\vf_{\m_1 \ldots
\m_{s_1}\,;\,\n_1 \ldots
\n_{s_2}\,;\, \ldots}$, with arbitrary numbers of ``families'' of
fully symmetric index sets. As we have stressed, the interest of a
general theory for these higher-spin fields lies to a large extent
in their direct link with the massive excitations of the bosonic
string. Fields of this type are in fact natural partners of generic
products of bosonic string oscillators of the type $\a_{-1}^{\m_1}
\ldots \a_{-1}^{\m_{s_1}}\,\a_{-2}^{\n_1}
\ldots
\a_{-2}^{\n_{s_2}}\,\ldots$. The resulting theory, however, is not
fully conventional for two reasons. First, as we have stressed,
these types of fields carry \emph{reducible} representations of the
Lorentz group. Moreover, they are vastly redundant, since fields of
this type carrying a single index for each family would suffice to
build arbitrary irreducible representations of the Lorentz group.
Nonetheless, in comparing with the superstring it is worthwhile to
study yet another class of fields, multi-forms of the type
$\vf_{[\m_1
\ldots \m_{s_1}]\,;\,[\n_1 \ldots \n_{s_2}]\,;\, \ldots}$, where
now the sets of Lorentz indices belonging to a given family are
totally antisymmetric, rather than totally symmetric as before.
While this is clearly another step in the direction of redundancy,
it is a fact that fields of this type accompany arbitrary products
of fermionic oscillators in massive superstring excitations. It is
therefore both interesting and useful to also have at one's disposal
a general theory of reducible multi forms, and even for more general
types of fields combining some fully symmetric and some fully
antisymmetric index families, that exhaust all possible types of
massive superstring excitations.

In this section we thus describe briefly how the present theory can
be adapted to the case of multi-form gauge fields. The mixed case
can then be treated combining the prescriptions we are about to
spell out with the results of the preceding sections. It was
originally pointed out in \cite{bb2} that this variant of the
Labastida theory can be built in a remarkably simple fashion. With
some changes of notation and conventions, we shall soon confirm this
conclusion within our formalism.

The key step in the transition to multi-forms is to modify the
definition of the few ingredients of the construction, the gradient
$\partial^{\,i}$, the divergence $\partial_{\,i}$, the trace
$T_{ij}$ and the metric tensor in family space $\h^{ij}$, making the
first two anticommuting and the last two antisymmetric. Once this is
done, the previous results change as follows.

To begin with, the gauge transformations maintain the same form of
eq.~\eqref{gauge}, but now the gauge-for-gauge transformations are
associated to \emph{symmetric} parameters,
\be
\delta \, \L_{\,i} = \pr^{\,j} \, \L_{\,(\,ij\,)}\,
,\label{gaugexgaugeform}
\ee
consistently with the fact that they exist already for a single
family,
\emph{i.e.} for the usual form fields.

In addition, the Fronsdal-Labastida operator maintains the form of
eq.~\eqref{fronsdal}, but its gauge variation becomes
\be \label{gauge_fronsdalform} \d \, \cF \, =  \, \frac{1}{6} \, \pr^{\,i} \pr^{\,j} \pr^{\,k} \,
T_{[\,ij} \, \L_{\,k\,]} \, ,\ee
since the derivatives are now anticommuting. As a result, the trace
constraints on the gauge parameters now become
\be
T_{[\,ij} \, \L_{\,k\,]} \, = \, 0 \, , \label{labacgaugebform}
\ee
so that the $\a_{ijk}$ are now fully \emph{antisymmetric} in their
family indices. Furthermore, the Bianchi identity now becomes
\be \label{bianchiFform}
\pr_{\,i} \, \cF \, - \, \12 \, \pr^{\,j} \, T_{ij} \, \cF \, = \, - \, \frac{1}{12} \, \pr^{\,j} \pr^{\,k} \pr^{\,l} \,
T_{[\,ij}\,T_{kl\,]} \, \vf \, , \ee
so that one is now led to constrain the fully antisymmetrized double
trace of $\vf$. In deriving these results, we have made use of the
multi-form variants of eqs.~\eqref{didj} and
\eqref{tijdk}, that read
\bea
&&\{ \, \pr_{\,i} \comma \pr^{\,j} \, \} \ = \ \Box \, \d_{\,i}{}^{\,j} \, ,\\[0pt]
&&[\, T_{ij} \comma \pr^{\,k} \,] \ = \ -\,
\pr_{\,[\,i}\,\d_{\,j\,]}{}^k
\, .
\eea

Proceeding as in the previous sections, one can now introduce the
compensators, or better the corresponding $\Phi_i$, and define the
$\cA$ tensor, which is the starting point to build unconstrained
Lagrangians. The last ingredients are the Lagrange multipliers,
$\b_{ijkl}$, that are now fully antisymmetric in their family
indices. As a result, the $\a_{ijk}$ and the $\b_{ijkl}$ start to
play a role at three and four families, respectively. A further key
ingredient are the traces of the Bianchi identity, the first of
which now reads
\be \label{first_bianchiform} \pr_{\,i} \, T_{jk} \, \cA \, - \, \12 \
\pr_{\,[\,j}\, T_{k\,]\,i} \, \cA \, = \, \12 \ \pr^{\,l} \, T_{il} \,
T_{jk} \, \cA \, - \, \frac{1}{4} \, T_{jk} \,
\pr^{\,l}\pr^{\,m}\pr^{\,n}\, \cC_{\,ilmn}  \, , \ee
so that its content is precisely as before. There is again a
dynamical portion relating divergences and gradients, that is still
associated to the $\{2,1\}$ projection in family space. However, the
projection is now effected in the antisymmetric basis, so that in
the present example there is manifest antisymmetry in $j$ and $k$,
rather than manifest symmetry as in eq.~\eqref{first_bianchi}. As a
technical note, we can add that the Young tableau is now filled in a
different fashion, with $j$ and $k$ along the vertical, so that the
diagram is effectively ``flipped'' about its diagonal with respect
to eq.~\eqref{first_bianchi}.

For higher traces, one needs the analogue of eq.~\eqref{chain}, that
reads
\be \label{chainform}
\begin{split}
& (\,p+2\,) \ Y_{\{p+1,p\}} \, \pr_{\,k} \, \cA^{\,[\,p\,]}{}_{i_1
j_1;\,\ldots\,;\,i_p j_p}  - \, \pr^{\,l} \,
\cA^{\,[\,p+1\,]}{}_{i_1
j_1;\,\ldots\,;\,i_p j_p\,;\,k\,l} \, = \\
& = \, - \, \12 \ Y_{\{p+1,p\}} \, T_{i_1 j_1} \ldots \, T_{i_p j_p}
\, \pr^{\,l}\pr^{\,m}\pr^{\,n}\, \cC_{\,klmn} \, .
\end{split}
\ee
Now semi-colons separate antisymmetric index pairs, since the
previous $\{2^p\}$ projection is replaced by a flipped diagram, that
in our notation is of $\{p,p\}$ type and is defined in the
antisymmetric basis. These results translate into corresponding ones
for the Lagrangians, that now read
\be  \label{laggenbform}
\begin{split}
\cL \, & = \, \12 \, \bra\, \vf \,\comma\, \sum_{p\,=\,0}^{N} \,
\frac{(-1)^{\,p}}{p\,!\,(\,p+1\,)\,!} \ \h^{i_1 j_1} \ldots \, \h^{i_{\,p} j_{\,p}} \,
\cA^{\,[\,p\,]}{}_{i_1 j_1;\,\ldots\,;\,i_{p} j_{p}} \,\ket \\
& - \frac{1}{4} \, \bra\, \a_{\,ijk} \,\comma\, \sum_{p\,=\,0}^{N-1}
\,
\frac{(-1)^{\,p}}{p\,!\,(\,p+3\,)\,!} \ \h^{i_1 j_1} \ldots \, \h^{i_{p} j_{p}} \, \pr_{\,[\,i}\,
\cA^{\,[\,p+1\,]}{}_{jk\,];\,i_1 j_1;\,\ldots\,;\,i_{p} j_{p}}
\,\ket\\
 &+ \, \frac{1}{8} \, \bra\, \b_{\,ijkl} \,\comma\, \cC_{\,ijkl}
 \,\ket\, .
\end{split}
\ee
Finally, the $S^{\,i}{}_j$ operators maintain the same algebra,
while the irreducibility conditions discussed in Section
\ref{sec:irreducible} are now to be defined via the columns of the
Young diagrams, and amount to the conditions the any
antisymmetrization beyond a given column vanishes.

This presentation of the formalism is particularly effective for
two-column fields, that as we have stressed repeatedly are somehow
close analogs of the spin-two metric fluctuation, since they need
neither compensators nor Lagrange multipliers. Interesting,  in the
multi-antisymmetric description their Fronsdal-Labastida operators
take the particularly simple form
\be
\cF \, = \, \frac{1}{2} \ T_{ij}\, \partial^{\,i} \partial^{\,j} \, \vf \, ,
\ee
so that they are manifestly related to the corresponding curvatures
via a single trace.

\vskip 24pt

%%%%%%%%%%%%%%%%%%%%%%%%%%%%%%%%%%%%%%%%%

\scs{Conclusions}\label{sec:conclusions}

%%%%%%%%%%%%%%%%%%%%%%%%%%%%%%%%%%%%%%%%%

In this paper we have described in some detail the general
properties of free mixed-symmetry bosonic gauge fields described by
multi-symmetric tensors of the type $\vf_{\m_1 \ldots \, \m_{s_1} ;
\, \n_1 \ldots \, \n_{s_2}; \, \ldots}$ or by corresponding
multi-forms, or in fact by fields combining arbitrary numbers of
families of symmetric or antisymmetric indices. Here we have
discussed massless fields in a Minkowski background, but as for
lower spins the treatment can be directly extended to the massive
case via the harmonics of Kaluza-Klein circle reductions. The
extension to $(A)dS$ has not been carried out here, but is expected
to be possible, proceeding along the lines of
\cite{fronads,fms,fms2}. These general mixed-symmetry fields are
needed to describe all representations of the Poincar\'e group for
$D>5$, and in particular are key ingredients of all massive string
spectra. Whereas their dynamics is still poorly understood, it is
difficult to escape the feeling that they are directly responsible
for the most spectacular properties of String Theory. And, we should
add, that they might even pave the way to possible generalizations
of the string framework.

The theory of mixed-symmetry higher-spin fields was touched upon by
a number of authors in the Seventies and Eighties \cite{curt,mixed},
and most notably by Labastida \cite{labastida1,labastida}, who
managed to generalize the Fronsdal construction of \cite{fronsdal}
to tensors of this type, after identifying the proper constraints on
gauge parameters and gauge fields. His work was then pursued further
in \cite{mixed2}. In this paper we have extended the Labastida
construction to an unconstrained formulation, along the lines of
what was previously done for Fronsdal's case in
\cite{fs2,st,fs3,fms} via a single compensator $\a$ and a single Lagrange
multiplier $\b$. We have followed as closely as possible the
index-free notation that proved so powerful in the symmetric case,
but for the introduction of ``family indices''. These are the
counterpart, in our language, of the non-Abelian oscillator algebra
of \cite{labastida1,labastida}, and in fact the two notations can be
turned into one another almost verbatim. Still, in our opinion the
present notation has the advantage of bringing these systems,
despite the complications introduced by their general nature, closer
to more conventional field theories. For pedagogical reasons, we
have started from two-family gauge fields, that follow most closely
the pattern that emerged in the symmetric case. The Bianchi
identities were again the key ingredients of our construction that,
in its minimal form, together with the gauge fields $\vf$, also
involves compensator fields $\a_{ijk}$ and Lagrange multipliers
$\b_{ijkl}$. The result was a rather streamlined and compact
derivation of the general Lagrangians, from which the constrained
Labastida construction can be recovered almost by inspection. Still,
the unconstrained theory brings about a number of surprises when
compared to the symmetric or single-family case. These have to do,
one way or another, with the lack of mutual independence of the
Labastida constraints, which makes the $\a_{ijk}$ not independent as
well, and forces one to relate them to other more fundamental
compensator fields here called $\Phi_i$, that only enter via the
combinations $T_{(\,ij}\Phi_{k\,)}$ due precisely to the constrained
Labastida symmetry. In addition, the Lagrangians enjoy a local
symmetry under shifts of the Lagrange multipliers that, as a
consequence, are not fully determined by the field equations.
Another key consequence of the non-Abelian structure underlying
these free theories is a rich pattern of sporadic cases where
Weyl-like symmetries emerge, generalizing the well-known property of
two-dimensional gravity even to cases where the Lagrangians are not
topological. Despite these subtleties, however, we have described in
full generality how the field equations of two-family fields can be
reduced on shell and we have exemplified these results in a number
of cases, with or without external currents. Moreover, we have
illustrated the key steps of the general reduction procedure for
$N$-family fields. While the bulk of this paper was devoted to
multi-symmetric \emph{reducible} gauge fields, we have also
described how to adapt the formalism to the cases of irreducible
fields or multi-forms. Finally, we have shown in full generality how
the higher-derivative terms involving the compensators that are
present in our unconstrained formulation can be eliminated at the
expense of a mild enlargement of the field content, generalizing the
construction presented for symmetric bosons in \cite{dario07}.

The companion paper \cite{cfms2} will contain a similar discussion
of Fermi fields. There the corresponding Lagrangians will appear for
the first time in their general ``metric-like'' form, since
Labastida only obtained in \cite{labferm} field equations
generalizing those of Fang and Fronsdal \cite{fangfronsdal}, while
the subsequent literature \cite{mixed2} only contains partial
results in this respect. A key issue for future research is clearly
to attain a better understanding of higher-spin interactions, and
above all of their systematics. Much was recently done in this
respect by a number of authors \cite{othernew}, but a decisive
progress in this respect is clearly expected to be far more
difficult. We hope that this ``metric-like form'' of the theory of
free higher-spin fields will provide useful insights in this
respect.

\vskip 24pt

%%%%%%%%%%%%%%%%%%%%%%%%%%%%%%%%%%%%%%%%%%%%%%%%%%%%%%%%%%%%%%

\section*{Acknowledgments}

%%%%%%%%%%%%%%%%%%%%%%%%%%%%%%%%%%%%%%%%%%%%%%%%%%%%%%%%%%%%%%

We are very grateful to X.~Bekaert and N.~Boulanger for extended
discussions at the beginning of this project. A.C. and D.F. would
also like to thank B. Nilsson for helpful conversations. We are
grateful to the APC-Paris VII, to the Chalmers University of
Technology and to the Scuola Normale Superiore di Pisa for the kind
hospitality extended to one or more of us at various stages while
this work was in progress. The present research was supported in
part by APC-Paris VII, by Scuola Normale Superiore, by INFN, by
CNRS, in particular through the P2I program, by the MIUR-PRIN
contract 2007-5ATT78, by the EU contracts MRTN-CT-2004-503369 and
MRTN-CT-2004-512194 and by the NATO grant PST.CLG.978785.

\newpage

\begin{appendix}

%%%%%%%%%%%%%%%%%%%%%%%%%%%%%%%%%%%%%%%%%%%%%

\scs{Notation and conventions}\label{app:MIX}

%%%%%%%%%%%%%%%%%%%%%%%%%%%%%%%%%%%%%%%%%%%%%

In this paper we use the ``mostly plus'' convention for the
space-time signature and resort to a compact notation eliminating
all space-time indices from tensor relations. The fields of interest
are here multi-symmetric tensors $\varphi_{\mu_1 \ldots\,
\mu_{s_1};\,\nu_1
\ldots\, \nu_{s_2};\,\ldots}$,
whose sets of symmetric indices are here referred to as
``families'', or corresponding multi-forms. There are a number of
sign differences between the cases of multi-symmetric and
multi-antisymmetric fields, but for definiteness in this Appendix,
as in most of the present paper, we refer explicitly to
\emph{multi-symmetric} fields. The key modifications needed in the case
of multi-forms are spelled out in Section
\ref{sec:multiforms}.

While fully symmetric under the interchange of pairs of indices
belonging to the same set, a field like $\varphi_{\mu_1 \ldots\,
\mu_{s_1};\,\nu_1
\ldots\, \nu_{s_2};\,\ldots}$ has no prescribed symmetry
relating different sets, and is thus a reducible $gl(D)$ tensor. As
a result, it is perhaps less familiar than Young projected tensors,
but is a most convenient object to study and plays also a natural
role in String Theory. For instance, in the bosonic string
multi-symmetric tensors of this type accompany generic products of
string oscillators $\alpha_{-1}^{\mu_1}  \ldots
\alpha_{-1}^{\mu_{s_1}}\alpha_{-2}^{\nu_1} \ldots \alpha_{-2}^{\nu_{s_2}}
\ldots$, that are only symmetric under
interchanges of pairs of identical oscillators, just as multi-forms
accompany in superstring similar products of fermionic oscillators.
Recovering more conventional field theories from the present
formulation requires suitable projections, discussed in Section
\ref{sec:irreducible}: the simplest example to this effect, as we
anticipated in the Introduction, is a field $\varphi_{\mu\,;\,\nu}$,
which combines a spin-2 field $\varphi_{(\mu\,;\,\nu)}$ and a
Kalb-Ramond field $\varphi_{[\mu\,;\,\nu]}$. In this paper a generic
multi-(anti)symmetric gauge field of this type is simply denoted by
$\varphi$.

The Lagrangians and field equations of multi-symmetric fields
involve traces, gradients and divergences, as well as Minkowski
metric tensors related to one or two of the previous index sets. As
a result, ``family indices" are needed in order to specify the sets
to which some tensor indices belong. These family indices are here
denoted by small-case Latin letters, and the Einstein convention for
summing over pairs of them is used throughout. It actually proves
helpful to be slightly more precise:
\emph{upper} family indices are thus reserved for operators, like a
gradient, that \emph{add} space-time indices, while \emph{lower}
family indices are used for operators, like a divergence, that
\emph{remove} them. As a result gradients, divergences and traces of a
field $\vf$ are denoted concisely by $\pr^{\, i} \,
\vf$, $\pr_{\, i}\, \vf$ and $T_{ij} \, \vf$. This shorthand
notation suffices to identify the detailed meaning of these symbols,
so that for instance
\begin{eqnarray} \label{operations}
\pr^{\, i} \, \vf & \equiv & \pr_{\,(\,\m^i_1|} \, \vf_{\ldots \,;\, | \, \m^i_2 \, \ldots \, \m^i_{s_i+1} ) \,;\, \ldots} \ , \nonumber \\
\pr_{\, i}\, \vf & \equiv & \pr^{\,\l} \, \vf_{\ldots \,;\, \l \, \m^i_2 \, \ldots \, \m^i_{s_i} \,;\, \ldots} \ , \nonumber \\
T_{ij} \, \vf & \equiv & \vf_{\ldots
\,;\phantom{\m^i_1}}{}^{\!\!\!\l}{}_{\m^i_2
\,
\ldots \, \m^i_{s_i} \,;\, \ldots \,;\, \l \,
\m^j_2 \ldots \, \m^j_{s_j} \,;\, \ldots} \ .
\end{eqnarray}
In addition, as in \cite{fs1,fs2,st,fs3,fms,dario07}, we work with
symmetrizations that are \emph{not} of unit strength, but involve
the minimum possible number of terms, and we use round brackets to
denote them. Thus, for instance, the product $T_{(ij} T_{kl)}$ here
stands for $T_{ij}T_{kl}+T_{ik}T_{jl}+T_{il}T_{jk}$. In addition, we
use square brackets to denote antisymmetrizations. In the
mixed-symmetry case, it is also necessary to introduce a mixed
metric tensor
\be \label{metric} \h^{\,ij} \, \vf \, \equiv \, \12 \,
\sum_{n\,=\,1}^{s_i+1}  \, \h_{\,\m^i_n \, (\,\m^j_1\,|}\, \vf_{\,\ldots \,;\, \ldots \, \m^i_{n-1} \, \m^i_{n+1}\, \ldots  \,;\, \ldots
\,;\, |\,\m^j_2 \ldots\, \m^j_{s_j+1}) \,;\, \ldots} \ , \ee
that here is suitably rescaled in order that its diagonal terms
retain the conventional normalization.

In order to further simplify the combinatorics, it proves very
convenient to introduce the scalar product
\be \label{scalar} \bra \vf \comma \c \ket \, \equiv \, \frac{1}{s_1! \ldots s_n!} \ \vf_{\m^1_1 \ldots \, \m^1_{s_1} \, ; \, \ldots \, ; \,
\m^n_1 \ldots \, \m^n_{s_n}} \, \c^{\m^1_1 \ldots \, \m^1_{s_1} \, ; \, \ldots \, ; \, \, \m^n_1 \ldots \, \m^n_{s_n}} \, \equiv \,
\frac{1}{s_1! \, \ldots s_n!} \ \vf \, \c \, . \ee
Inside the brackets it is then possible to integrate by parts and to
turn $\h$'s into traces without introducing any $s_i$-dependent
combinatoric factors, since
\begin{eqnarray} \label{deriv}
\bra \vf \comma \pr^{\,i} \, \c \ket & \equiv & \frac{1}{s_1! \ldots s_n!} \ \vf \, \pr^{\,i} \, \c \, = \, - \, \frac{s_i}{s_1! \ldots s_n!} \
\pr_{\,i}\, \vf \, \c \, \equiv \, - \, \bra \pr_{\,i}\, \vf \comma \c \ket \, , \\
\bra \vf \comma \h^{ij} \, \c \ket & \equiv & \frac{1}{s_1! \ldots s_n!} \ \vf \, \h^{ij} \, \c \, = \, \12 \, \frac{s_i \, s_j}{s_1! \ldots
s_n!} \ (T_{ij} \, \vf) \, \c \, \equiv \, \12 \, \bra T_{ij} \, \vf
\comma \c \ket \, ,
\end{eqnarray}
where the reader should keep track of the somewhat unusual factor
$\frac{1}{2}$, originating again from our choice of normalization
for $\eta_{ij}$ in eq.~(\ref{metric}). However, in the main body of
this paper we are ignoring, for simplicity, the overall factor
$\prod_{i=1}^N s_i!$, that should accompany the Lagrangian of a
multi-symmetric tensor $\varphi_{\mu^1_1 \ldots\,
\mu^1_{s_1};\,\ldots\, ; \, \mu^N_1 \ldots\, \mu^N_{s_N}}$ to grant it the conventional normalization.

These matrix elements are quite convenient to derive Lagrangians and
field equations for mixed-symmetry fields, but they would be as
convenient for symmetric tensors. In this notation the fully
symmetric, or one-family, Lagrangian would simply read
\be {\cal L} \, = \, \frac{1}{2} \, \bra \varphi \comma \cA - \frac{1}{2} \, \eta \, \cA^{\; \prime} \ket \, - \, \frac{1}{8}\, \bra \alpha \comma
\partial \cdot \cA^{\; \prime} \ket \, + \, \frac{1}{8} \, \bra \beta \comma \cC \ket \ ,  \ee
up to an overall $s!$, to be compared with the corresponding
expression of
\cite{fms}
\be {\cal L} \, = \, \frac{1}{2} \,  \varphi \left( \cA - \frac{1}{2} \, \eta \, \cA^{\; \prime} \right) \, - \, \frac{3}{4} \left( s \atop 3 \right)
\, \alpha \,
\partial \cdot \cA^{\; \prime} \, + \, 3 \left( s \atop 4 \right) \, \beta \, \cC \ , \ee
that contains explicit combinatoric factors.

As in the symmetric case, in order to take full advantage of the
compact notation, it is convenient to collect a number of identities
that are used recurrently in this type of analysis. These, however,
have a more complicated structure than their symmetric counterparts
of \cite{fms}, since they involve a genuinely new type of operation.
This turns a tensor with $s_i$ indices in the $i$-th group into
others with $s_i+1$ indices in the $i$-th group and $s_j-1$ indices
in the $j$-th group, according to
\be \label{flip} S^{\,i}_{~j} \, \vf \, \equiv \, \vf_{\ldots \,;\, ( \m^i_1 \ldots \m^i_{s_i} | \,;\, \ldots \,;\, | \m^i_{s_i+1} ) \,
\m^j_2
\ldots \m^j_{s_j} \,;\, \ldots}  \ . \ee
The new rules one needs follow from the algebra of the various
operators, and can be also derived from a realization in terms of
bosonic oscillators, along the lines of bosonic String Field Theory
and of \cite{labastida1,labastida}:
\begin{align}
&[ \, \pr_{\,i} \comma \pr^{\,j} \, ] \ = \ \Box \, \d_{\,i}{}^{\,j} \, , \label{didj}\\[0pt]
&[\, T_{ij} \comma \pr^{\,k} \,] \ = \ \pr_{\,(\,i}\,\d_{\,j\,)}{}^k \, , \label{tijdk}\\[0pt]
&[\, \pr_{\,k} \comma \h^{\,ij} \,] \ = \ \12 \ \pr^{\,(\,i\,}\d^{\,j\,)}{}_k \, , \\[0pt]
&[\, T_{ij} \comma \h^{\,kl} \,] \ = \ \frac{D}{2} \ \d_{\,i}{}^{\,(\,k\,}\d^{\,l\,)}{}_j \, + \,
\12 \left( \, \d_{\,i}{}^{\,(\,k}S^{\,l\,)}{}_j \, + \, \d_{\,j}{}^{\,(\,k}S^{\,l\,)}{}_i \, \right) , \label{ThS} \\[0pt]
&[\, S^{\,i}{}_j \comma \h^{\,kl} \,] \ = \ \h^{\,i\,(\,k\,} \d^{\,l\,)}{}_j \, , \label{commhS} \\[0pt]
&[\, T_{ij} \comma S^{\,k}{}_l \,] \ = \ T_{l\,(\,i}\, \d_{\,j\,)}{}^k \, , \label{commTS} \\[0pt]
&[\, S^{\,i}{}_j \comma \pr^{\,k} \,] \ = \ \pr^{\,i} \, \d_{\,j}{}^{\,k}  \, ,  \\[0pt]
&[\, \pr_{\,k} \comma S^{\,i}{}_j \,] \ = \ \pr_{\,j} \ \d_{\,k}{}^{\,i}  \, , \\[0pt]
&[\, S^{\,i}{}_j \comma S^{\,k}{}_l \,] \ = \ \d_{\,j}{}^{\,k}S^{\,i}{}_l \, - \, \d_{\,l}{}^{\,i}S^{\,k}{}_j \, .
\end{align}
Notice that the $S^{\,i}{}_j$ operators, the key novelty of the
mixed-symmetry case, close into a $gl(N)$ algebra if $N$ index
families are present. As a result, one can well say that this class
of free theories rests somehow on a non-Abelian structure. These
commutation relations give rise to the rules collected in Appendix
\ref{app:bose}.

Finally, in this paper we use extensively a number of standard tools
related to the symmetric group. These include, in particular, the
Young projectors $Y$, that allow to separate irreducible components
in family-index space and can be built combining contributions from
different Young tableaux $Y_\tau$. In general these Young tableaux
can be identified associating integer labels to the tensor indices
to be projected and allowing all their arrangements within the given
graph such that these integers grow from left to right and from top
to bottom. In some cases, however, this simple procedure can
actually fail to produce an orthogonal decomposition, which can
still be attained by a further Graham-Schmidt orthogonalization.
This difficulty is not present if, for any pair of tableaux, there
is at least a couple of indices belonging to a row of the first that
lie in the same column within the second, and vice versa. Let us
stress that this difficulty is never to be faced in our
constructions, as a result of the particular symmetry properties of
our basic objects.

Outside Section \ref{sec:multiforms}, in this paper Young tableaux
are defined in the \emph{symmetric} basis, so that the projector
corresponding to a tableau $\tau$ containing $n$ boxes takes the
form
\be Y_\tau \, = \, \frac{\lambda(\tau)}{n\,!} \ S\, A \, ,
\label{young_proj_gen} \ee
where $S$ and $A$ are the corresponding products of ``row
symmetrizers'' and ``column antisymmetrizers''. On the other hand,
the results in Section \ref{sec:multiforms} are dealt with more
conveniently in the \emph{antisymmetric} basis, where the roles of
$S$ and $T$ are interchanged. Here $\lambda(\tau)$ denotes the
dimension of the associated representation of the symmetric group,
that can be computed for instance counting the standard ways of
filling the boxes of the corresponding diagram with the numbers
$1,2,\ldots,n$, in increasing order from left to right and from top
to bottom. In general, diagrams and tableaux are specified by
ordered lists of the lengths of their rows, so that, for instance,
the $\{3,2\}$ graph is
\be
\begin{picture}(0,30)(20,0)
\multiframe(0,10.5)(10.5,0){1}(10,10){}
\multiframe(10.5,10.5)(10.5,0){1}(10,10){}
\multiframe(21,10.5)(10.5,0){1}(10,10){}
\multiframe(0,0)(10,0){1}(10,10){}
\multiframe(10.5,0)(10,0){1}(10,10){}
\end{picture}
\ee

Similar techniques are also used extensively in Section
\ref{sec:irreducible} to identify irreducible $gl(D)$ and Lorentz
tensors. We refrain from adding further details, since these and
other related standard facts are discussed extensively in the
literature, and in particular in \cite{branching}.

%%%%%%%%%%%%%%%%%%%%%%%%%%%%%%%%%%%%%%%%%%%%%%%%%%%

\scs{Some useful identities}\label{app:bose}

%%%%%%%%%%%%%%%%%%%%%%%%%%%%%%%%%%%%%%%%%%%%%%%%%%%

Using repeatedly the commutators presented in Appendix \ref{app:MIX}, one can recover the $N$-family counterpart of eq.~(A.1) of \cite{fms}, that collects the key identities for the one-family case:
\begin{align} \label{rules2}
& [\,\pr_{\,k}\,,\,\pr^{\,i_1} \ldots \, \pr^{\,i_p}\, ] \, = \, \sum^{p}_{n\,=\,1} \d_k^{~i_n} \, \Box \prod_{r\,\neq\,n} \pr^{\,i_r} \, , \\
& [\,T_{ij} \,,\, \pr^{\,k_1} \ldots \, \pr^{\,k_p} \, ] \, = \, \sum_{m\,<\,n} \d \tensor[^{\,k_m}_{(\,i\,}]{\d}{_{\,j\,)}^{k_n}} \, \Box \!\!  \prod_{r \,\neq\, m , n} \!\! \pr^{\,k_r} \, + \, \sum^p_{n\,=\,1} \prod_{r \,\neq\, n} \pr \tensor[^{\,k_r}]{\d}{^{\,k_n}} \tensor[_{(\,i\,}]{\pr}{_{\,j\,)}} \, , \\
& [\,\pr_{\,k}\,,\, \h^{i_1j_1} \ldots \, \h^{i_p j_p} \, ] \, = \, \12 \sum^p_{n\,=\,1} \prod_{r \,\neq\, n} \h \tensor[^{i_rj_r}]{\d}{_k^{(\,i_n}^{\quad\! j_n\,)}} \hspace{-22pt} \pr \hspace{17pt} \, , \\
& [\,T_{ij} \,,\, \h^{k_1l_1} \ldots \, \h^{k_pl_p} \,] \, =
\, \frac{D}{2} \sum^{p}_{n\,=\,1} \d
\tensor[_i^{(\,k_n\,}]{\d}{^{\,l_n\,)}_j} \prod_{r \,\neq\, n}
\h^{k_nl_n} + \12 \sum_{m\,<\,n} \Big(\, \d \tensor[_j^{k_m}]{\d}{_i^{(\,k_n \quad\! l_n\,)\,l_m}} \hspace{-30pt} \h \nonumber \\
& + \, \d \tensor[_j^{l_m}]{\d}{_i^{(\,k_n \quad\! l_n\,)k_m}} \hspace{-31pt} \h \hspace{25pt} + \, \d \tensor[_i^{k_m}]{\d}{_j^{(\,k_n \quad\! l_n\,)\,l_m}} \hspace{-30pt} \h \hspace{25pt} + \, \d \tensor[_i^{l_m}]{\d}{_j^{(\,k_n \quad\! l_n\,)\,k_m}} \hspace{-31pt} \h \hspace{25pt} \,\Big) \! \prod_{r \,\neq\, m,n} \! \h^{k_rl_r} \nonumber \\[-2pt]
& + \, \12 \sum^{p}_{n\,=\,1} \prod_{r \,\neq\, n} \h^{k_rl_r} \left(\, \d \tensor[_j^{(\,k_n}]{S}{^{\,l_n\,)}_i} \, + \, \d \tensor[_i^{(\,k_n}]{S}{^{\,l_n\,)}_j} \,\right) \, .
\end{align}
One often applies \eqref{rules2} to expressions that only contain
contracted indices. It is thus convenient to rewrite them explicitly
for cases where the various expressions are contracted against
tensors ${\cal Q}$ possessing identical manifest symmetries for
their family indices
\begin{align}
& [\,\pr_{\,k} \,,\, \pr^{\,i_1} \ldots \, \pr^{\,i_p} \,] \, {\cal Q}_{\,i_1 \ldots \, i_p} \, = \, p \ \Box \, \pr^{\,i_1} \ldots \, \pr^{\,i_{p-1}} \, {\cal Q}_{\,k\, i_1 \ldots \, i_{p-1}} \, , \label{dd^k} \\[10pt]
& [\,\pr_{\,k} \,,\, \h^{i_1 j_1} \ldots \, \h^{i_p j_p} \,]\, {\cal Q}_{\,i_1 j_1,\, \ldots \, ,\,i_p j_p} \, =
\, p \ \h^{i_1 j_1} \ldots \, \h^{i_{p-1} j_{p-1}} \, \pr^{\,l} \, {\cal Q}_{\,kl\,,\, i_1 j_1, \, \ldots \,
,\,i_{p-1} j_{p-1}} \, \label{dh^k} ,\\[10pt]
& [\, T_{kl} \,,\, \pr^{\,i_1} \, \ldots \, \pr^{\,i_p} \,]\, {\cal Q}_{\,i_1 \ldots \, i_p} \, = \, p \, (\,p-1\,) \ \Box \, \pr^{\,i_1} \ldots \, \pr^{\,i_{p-2}} \, {\cal Q}_{\,kl \, i_1 \ldots \, i_{p-2}} \ \nonumber \\
& \phantom{[\, T_{kl} \,,\, \pr^{\,i_1} \, \ldots \, \pr^{\,i_p} \,]\, {\cal Q}_{\,i_1 \ldots \, i_p} \,} + \, p \ \pr^{\,i_1} \ldots \, \pr^{\,i_{p-1}} \, \pr_{\,(\,k}\, {\cal Q}_{\,l\,)\,i_1 \ldots \, i_{p-1}} \, , \\[10pt]
& [\, T_{kl} \,,\, \h^{i_1 j_1} \ldots \, \h^{i_p j_p} \,]\, {\cal Q}_{\,i_1 j_1,\, \ldots \, ,\,i_p j_p} \, = \, p \, D \ \h^{i_1 j_1} \ldots \, \h^{i_{p-1} j_{p-1}} \, {\cal Q}_{\,kl\,,\,i_1j_1,\,\ldots\,,\,i_{p-1}j_{p-1}} \nonumber \\
& + \, \h^{i_1 j_1} \ldots \, \h^{i_{p-1} j_{p-1}} \, \left\{\,
p\,(\,p-1\,) \, {\cal
Q}_{\,i_1\,(\,k\,,\,l\,)\,j_1,\,\ldots\,,\,i_{p-1}j_{p-1}} \, + \, p
\ S^{\,m}{}_{(\,k} \,
{\cal Q}_{\,l\,)\,m\,,\,i_1j_1,\,\ldots\,,\,i_{p-1}j_{p-1}} \,
\right\} \, , \label{Th^k}
\end{align}
while another useful relation is given by
\be \label{T^kh}
\begin{split}
& [\, T_{i_1j_1} \ldots \, T_{i_pj_p} \,,\, \h^{kl} \,] \, {\cal Q}_{\,kl} \, = \, D \, \sum_{n\,=\,1}^p \, \prod_{r\,\neq\,n} T_{i_rj_r} \O_{\,i_nj_n}\, \\
& + \, \sum_{m\,<\,n} \, \prod_{r\,\neq\,m,n} \, T_{i_rj_r} \left(\, T_{i_m\,(\,i_n\,}\O_{\,j_n\,)\,j_m} + \, T_{j_m\,(\,i_n\,}\O_{\,j_n\,)\,i_m} \,\right) \, + \, \sum_{n\,=\,1}^p \, S^{\,k}{}_{(\,i_n\,|}\, \prod_{r\,\neq\,n} \, T_{i_rj_r}\, \O_{\,|\,j_n\,)\,k} \, .
\end{split}
\ee
Of course, identifying all family indices one can recover, as a
special case, the symmetric rules collected in eq.~(A.1) of
\cite{fms}.

The previous results are particularly useful when one tries to
compute the traces and divergences of the tensors $\cF$ and $\cA$
that are needed, for instance, to compute the equations of motion
for the Lagrangians \eqref{lag} and \eqref{laggenb}. Starting from
the Fronsdal-Labastida tensor
\be
\cF \, = \, \Box \, \vf \, - \, \pr^{\,i}\pr_{\,i} \, \vf \, + \, \12 \ \pr^{\,i}\pr^{\,j} \, T_{ij} \,
\vf\, ,
\ee
one can thus obtain
\be \label{gentraceb}
\begin{split}
& \prod_{r\,=\,1}^p T_{i_rj_r} \, \cF \, = \, (\,p+1\,) \ \Box  \prod_{r\,=\,1}^p T_{i_rj_r} \, \vf \, - \, 2 \sum_{n\,=\,1}^p \, \pr_{\,i_n}\pr_{\,j_n} \prod_{r\,\neq\,n}^p T_{i_rj_r} \, \vf \\
& + \, \sum_{n\,=\,1}^p \, \sum_{m\,<\,n} \left(\, \pr_{\,i_n}\pr_{\,(\,i_m}\, T_{j_m\,)\,j_n} + \, \pr_{\,j_n}\pr_{\,(\,i_m}\, T_{j_m\,)\,i_n} \,\right) \!\prod_{r\,\neq\,m\,,\,n}^p\! T_{i_rj_r} \, \vf \\
& - \, \pr^{\,k} \bigg[\ \pr_{\,k} \prod_{r\,=\,1}^p T_{i_rj_r} \, \vf \, - \, \sum_{n\,=\,1}^p \, \pr_{\,(\,i_n}\,T_{\,j_n\,)\,k} \prod_{r\,\neq\,n}^p T_{i_rj_r} \, \vf \ \bigg] + \, \12 \ \pr^{\,k}\pr^{\,l} \, T_{kl} \, \prod_{r\,=\,1}^p T_{i_rj_r} \, \vf
\end{split}
\ee
and
\be
\begin{split}
& \pr_{\,k} \prod_{r\,=\,1}^p T_{i_rj_r} \, \cF \, = \, \sum_{n\,=\,1}^p \, \Box \ \pr_{\,(\,k}\, T_{i_nj_n\,)} \prod_{r\,\neq\,n}^p T_{i_rj_r} \, \vf - \, 2 \sum_{n\,=\,1}^p \, \pr_{\,k\,}\pr_{\,i_n}\pr_{\,j_n} \prod_{r\,\neq\,n}^p T_{i_rj_r} \, \vf \\
& + \, \sum_{n\,=\,1}^p \, \sum_{m\,<\,n} \pr_{\,k} \left(\, \pr_{\,i_n}\pr_{\,(\,i_m}\, T_{j_m\,)\,j_n} + \, \pr_{\,j_n}\pr_{\,(\,i_m}\, T_{j_m\,)\,i_n} \,\right) \!\prod_{r\,\neq\,m\,,\,n}^p\! T_{i_rj_r} \, \vf \\
& - \, \pr^{\,l} \bigg[\ \pr_{\,k\,}\pr_{\,l} \prod_{r\,=\,1}^p T_{i_rj_r} \, \vf \, - \, \sum_{n\,=\,1}^p \, \pr_{\,k\,}\pr_{\,(\,i_n}\,T_{\,j_n\,)\,l} \prod_{r\,\neq\,n}^p T_{i_rj_r} \, \vf \ \bigg] \, + \, \Box \, \pr^{\,l}\, T_{kl} \prod_{r\,=\,1}^p T_{i_rj_r} \, \vf \\
& + \, \12 \ \pr^{\,l}\pr^{\,m} \, \pr_{\,k} \, T_{lm} \prod_{r\,=\,1}^p T_{i_rj_r} \, \vf \, .
\end{split}
\ee
Notice that these two expressions do not contain the $S^{\,i}{}_{j}$
operators, that as a result do not appear in the field equations,
but only emerge in their reduction procedure, and in particular in
the propagators. Let us also stress that eq.~\eqref{gentraceb} is
the relation needed to fix the coefficients
\eqref{coeflab} using the condition of self-adjointness, following
the original derivation of
\cite{labastida}. In particular, in the two-family case the relevant identities are
\begin{align}
& T_{ij} \, \cF \, = \, 2 \, \Box \, T_{ij} \, \vf \, - \, 2 \, \pr_{\,i\,}\pr_{\,j}\, \vf \, + \, \pr^{\,k} \left(\, T_{k\,(\,i}\, \pr_{\,j\,)}\, \vf \, - \, T_{ij} \, \pr_{\,k}\, \vf \,\right)  + \, \12 \, \pr^{\,k}\pr^{\,l} \, T_{ij}\,T_{kl} \, \vf \, , \label{trF} \\[8pt]
& T_{ij} \, T_{kl} \, \cF \, = \, 3 \, \Box \, T_{ij} \, T_{kl} \, \vf \, - \, 3 \left(\, T_{ij} \, \pr_{\,k\,} \pr_{\,l}\, \vf \, + \, T_{kl} \, \pr_{\,i\,}\pr_{\,j}\, \vf \, - \, \frac{1}{3} \ T_{(\,ij}\, \pr_{\,k\,}\pr_{\,l\,)}\, \vf \,\right) \nonumber \\
& - \, \pr^{\,m} \, T_{ij} \, T_{kl} \, \pr_{\,m}\, \vf \, + \,
\pr^{\,m} \left(\, T_{ij} \, T_{m\,(\,k} \, \pr_{\,l\,)}\, \vf \, +
\, T_{kl} \, T_{m\,(\,i} \pr_{\,j\,)}\, \vf \,\right)  + \, \12 \
\pr^{\,m}\pr^{\,n}\, T_{ij} \, T_{kl} \, T_{mn} \, \vf
\label{doubtrF}
\end{align}
and
\begin{align}
& T_{ij} \, \pr_{\,k}\, \cF \, = \, \Box \, T_{(\,ij} \, \pr_{\,k\,)}\, \vf \, - \, 2 \, \pr_{\,i\,}\pr_{\,j\,} \pr_{\,k}\, \vf \, + \, \Box \, \pr^{\,l} \, T_{ij} \, T_{kl} \, \vf \nonumber \\
& \phantom{T_{ij} \, \pr_{\,k}\, \cF \,} + \, \pr^{\,l} \, \left( \, T_{l\,(\,i} \, \pr_{j\,)\,} \pr_{\,k}\, \vf \, - \, T_{ij} \, \pr_{\,k}\, \pr_{\,l}\, \vf \, \right) \, + \, \12 \, \pr^{\,l}\pr^{\,m} \, T_{ij} \, T_{lm} \, \pr_{\,k}\, \vf \, , \label{dtrF} \\[10pt]
& T_{ij} \, T_{kl} \, \pr_{\,m}\, \cF \, = \, 2 \, \Box \, T_{ij} \, T_{kl} \, \pr_{\,m}\, \vf \, + \, \Box \left( \, T_{ij} \, T_{m\,(\,k} \, \pr_{\,l\,)}\, \vf \, + \, T_{kl} \, T_{m\,(\,i} \, \pr_{\,j\,)}\, \vf \, \right) \nonumber \\
& \phantom{T_{ij} \, T_{kl} \, \pr_{\,m}\, \cF \,} - \, 3 \left( \, T_{ij} \, \pr_{\,k\,}\pr_{\,l\,}\pr_{\,m}\, \vf \, + \, T_{kl} \, \pr_{\,i\,}\pr_{\,j\,}\pr_{\,m}\, \vf \, - \, \frac{1}{3} \, T_{(\,ij} \, \pr_{\,k\,}\pr_{\,l\,)\,}\pr_{\,m}\, \vf \, \right) \nonumber \\
& \phantom{T_{ij} \, T_{kl} \, \pr_{\,m}\, \cF \,} - \, \pr^{\,n} \, T_{ij} \, T_{kl} \, \pr_{\,m\,}\pr_{\,n}\, \vf \, + \, \pr^{\,n} \left( \, T_{ij} \, T_{n\,(\,k} \, \pr_{\,l\,)}\, \pr_{\,m}\, \vf \, + \, T_{kl} \, T_{n\,(\,i} \, \pr_{\,j\,)\,}\pr_{\,m}\, \vf \, \right) \nonumber \\
& \phantom{T_{ij} \, T_{kl} \, \pr_{\,m}\, \cF \,} + \, \Box \, \pr^{\,n} \, T_{ij} \, T_{kl} \, T_{mn} \, \vf \, + \, \12 \, \pr^{\,n}\pr^{\,p} \, T_{ij} \, T_{kl} \, T_{np} \,
\pr_{\,m} \, \vf \label{ddoubtrF} \, .
\end{align}

In a similar fashion, starting from the unconstrained gauge
invariant tensor $\cA$,
\be
\cA \, = \, \cF \, - \, \12 \, \pr^{\,i}\pr^{\,j}\pr^{\,k} \, \a_{\,ijk} \, ,
\ee
one can obtain the corresponding expression
\be
\begin{split}
& \prod_{r\,=\,1}^p T_{i_rj_r} \, \cA \, = \, \prod_{r\,=\,1}^p T_{i_rj_r} \, \cF \, - \, 3 \ \Box \sum_{n\,=\,1}^p \, \sum_{m\,\neq\,n} \, \prod_{r\,\neq\,m,n}^p T_{i_rj_r} \, \pr_{\,(\,i_m} \a_{\,j_mi_nj_n\,)} \\
& - 3 \, \sum_{n\,=\,1}^p \, \sum_{\genfrac{}{}{0pt}{}{m\,<\,n}{q\,<\,m}} \, \prod_{r\,\neq\,m,n,q}^p T_{i_rj_r} \left(\, \pr_{\,i_n\,}\pr_{\,(\,i_m\,|\,}\pr_{\,(\,i_q} \, \a_{\,j_q\,)\,|\,j_q\,)\,j_n} \, + \, \pr_{\,j_n\,}\pr_{\,(\,i_m\,|\,}\pr_{\,(\,i_q} \, \a_{\,j_q\,)\,|\,j_q\,)\,i_n} \,\right) \\
& - \, 3 \ \Box \, \pr^{\,k} \sum_{n\,=\,1}^p \, \prod_{r\,\neq\,n}^p T_{i_rj_r} \, \a_{\,i_nj_nk} \, - \, 3 \, \pr^{\,k} \sum_{n\,=\,1}^p \, \sum_{m\,<\,n} \, \prod_{r\,\neq\,m,n}^p T_{i_rj_r} \, \pr_{\,(\,i_n\,|\,}\pr_{\,(\,i_m} \, \a_{\,j_m\,)\,|\,j_n\,)\,k} \\
& - \, \frac{3}{2} \ \pr^{\,k}\pr^{\,l} \sum_{n\,=\,1}^p \, \prod_{r\,\neq\,n}^p T_{i_rj_r} \, \pr_{\,(\,i_n}\, \a_{\,j_n\,)\,kl} \, - \, \frac{1}{2} \ \pr^{\,k}\pr^{\,l}\pr^{\,m} \prod_{r\,=\,1}^p T_{i_rj_r} \, \a_{\,klm} \, ,
\end{split}
\ee
that restricting again the attention to two index families reduce to
\begin{align}
& T_{ij} \, \cA \, = \, T_{ij} \, \cF \, - \, 3 \, \Box \, \pr^{\,k} \, \a_{ijk} \, - \, \frac{3}{2} \, \pr^{\,k}\pr^{\,l} \, \pr_{\,(\,i}\, \a_{j\,)kl} \, - \, \12 \, \pr^{\,k}\pr^{\,l}\pr^{\,m} \, T_{ij} \, \a_{klm} \, , \\[8pt]
& T_{ij} \, T_{kl} \, \cA \, = \, T_{ij} \, T_{kl} \, \cF \, - \, 3 \, \Box \, \pr_{\,(\,i}\, \a_{jkl\,)} \, - \, 3 \, \pr^{\,m} \left(\, \pr_{\,i\,}\pr_{(\,k}\, \a_{\,l\,)\,jm} \, + \, \pr_{\,j\,}\pr_{\,(\,k}\, \a_{\,l\,)\,im}  \,\right) \, \nonumber \\
& \phantom{T_{ij} \, T_{kl} \, \cA \,} - \frac{3}{2} \, \pr^{\,m}\pr^{\,n} \left(\, T_{ij}\, \pr_{\,(\,k}\, \a_{\,l\,)\,mn} + \, T_{kl}\, \pr_{\,(\,i}\, \a_{j\,)\,mn} \ \,\right) \nonumber \\
& \phantom{T_{ij} \, T_{kl} \, \cA \,} - \, 3 \, \Box \, \pr^{\,m}
\left( \, T_{ij} \, \a_{klm}\, + \, T_{kl} \, \a_{ijm} \, \right) -
\, \12 \, \pr^{\,m}\pr^{\,n}\pr^{\,p} \, T_{ij} \, T_{kl} \,
\a_{mnp} \, .
\end{align}
Other useful identities are
\begin{align}
& T_{ij} \, \pr_{\,k} \, \cA \, = \, T_{ij} \, \pr_{\,k}\, \cF \, - \, 3 \, \Box^{\,2} \, \a_{ijk} \, - \, 3 \, \Box \, \pr^{\,l} \, \pr_{\,(\,i}\, \a_{jk\,)\,l} \, - \, \frac{3}{2} \, \Box \, \pr^{\,l}\pr^{\,m} \, T_{ij} \, \a_{klm} \, \nonumber \\
& \phantom{T_{ij} \, \pr_{\,k} \, \cA \,} - \, \frac{3}{2} \, \pr^{\,l}\pr^{\,m} \, \pr_{\,k\,}\pr_{\,(\,i}\, \a_{j\,)\,lm} \, - \, \12 \, \pr^{\,l}\pr^{\,m}\pr^{\,n} \, T_{ij} \, \pr_{\,k}\, \a_{lmn} \, , \\[8pt]
& T_{ij} \, T_{kl} \, \pr_{\,m} \, \cA \, = \, T_{ij} \, T_{kl} \, \pr_{\,m}\, \cF \, - \, 3 \, \Box \left( \, \pr_{\,m\,}\pr_{\,(\,i}\, \a_{jkl\,)} \, + \, \pr_{\,i\,}\pr_{\,(\,k}\, \a_{\,l\,)\,jm} \, + \, \pr_{\,j\,}\pr_{\,(\,k}\, \a_{\,l\,)\,im} \, \right) \, \nonumber \\[2pt]
& - \, 3 \, \pr^{\,n} \left( \, \pr_{\,m\,}\pr_{\,i\,}\pr_{\,(\,k}\, \a_{\,l\,)\,jn} \, + \, \pr_{\,m\,}\pr_{\,j\,}\pr_{\,(\,k}\, \a_{\,l\,)\,in} \, \right) - \, 3 \, \Box^{\,2} \left( \, T_{ij} \, \a_{klm} \, + \, T_{kl} \, \a_{ijm} \, \right) \nonumber \\
& - \, 3 \, \Box \, \pr^n \left( \, T_{ij} \, \pr_{\,m}\, \a_{kln} \, + \, T_{kl} \, \pr_{\,m}\, \a_{ijn} \, \right) - \, 3 \, \Box \, \pr^{\,n} \left( \, T_{ij} \, \pr_{\,(\,k}\, \a_{\,l\,)\,mn} \, + \, T_{kl} \, \pr_{\,(\,i}\, \a_{j\,)\,mn} \, \right) \nonumber \\
& - \, \frac{3}{2} \, \Box \, \pr^{\,n}\pr^{\,p} \, T_{ij} \, T_{kl} \, \a_{mnp} \, - \, \frac{3}{2} \, \pr^{\,n}\pr^{\,p} \left( \, T_{ij} \, \pr_{\,m\,}\pr_{\,(\,k}\, \a_{\,l\,)\,np} \, + \, T_{kl} \, \pr_{\,m\,}\pr_{\,(\,i}\, \a_{j\,)\,np} \, \right) \, \nonumber \\
& - \, \12 \, \pr^{\,n}\pr^{\,p}\pr^{\,q} \, T_{ij} \, T_{kl} \,
\pr_{\,m}\, \a_{npq} \, .
\end{align}
They could be used to compute directly the field equations
\eqref{ep} and \eqref{ephi}, when combined with eqs.~\eqref{trF},
\eqref{doubtrF}, \eqref{dtrF} and \eqref{ddoubtrF}.

\vskip 24pt

%%%%%%%%%%%%%%%%%%%%%%%%%%%%%%%%%%%%%%%%%%%%%%%%%%%%%%%%%%%%%%%%
\scs{Proof of some results used in Section \ref{sec:generalb}}
\label{app:idsb}
%%%%%%%%%%%%%%%%%%%%%%%%%%%%%%%%%%%%%%%%%%%%%%%%%%%%%%%%%%%%%%%%

The construction of Section \ref{sec:generalb} is based on the key
result that all Young projections of multiple traces of $\cA$ with
more than two columns can be related to the constraint tensors
$\cC_{ijkl}$. In fact, these projections can be realized via a sum
of Young tableaux involving at least one symmetrization over four
family indices, since from a product of identical traces one can
only build Young diagrams with even numbers of boxes in each row.
Aside from the two simple cases,
\be
T_{(\,ab}\,T_{cd\,)} \quad \textrm{and} \quad T_{i\,(\,a\,|}\,
T_{j\,|\,b}\, T_{cd\,)}\, , \label{trivialconstr}
\ee
that are manifestly related to the constraints, either directly via
eq.~\eqref{aconstrdt} or via the enlargement of cycles from three to
four family indices, that is automatic for a pair of trace tensors,
one ought to consider the non-trivial case where the four
symmetrized indices are spread over four distinct traces:
\be
T_{i\,(\,a\,|}\,T_{j\,|\,b\,|}\,T_{k\,|\,c}\,T_{d\,)\,l}\ .
\ee
Even these types of terms, however, can be related to the
constraints via eq.~\eqref{aconstrdt}, since
\be
T_{i\,(\,a\,|}\,T_{j\,|\,b\,|}\,T_{k\,|\,c}\,T_{d\,)\,l} \, =
\, T_{(\,ab}\,T_{cd}\,T_{ij}\,T_{kl\,)} \, - \,
T_{(\,ab}\,T_{c\,|\,(\,i\,|}\,T_{|\,d\,)\,|\,j}\,T_{kl\,)} \, - \,
T_{(\,ab}\,T_{cd\,)}\,T_{(\,ij}\,T_{kl\,)}\, ,
\ee
where the first and last terms are manifestly related to the
constraints, while the second term is like the last one in
eq.~\eqref{trivialconstr}. These results clearly imply that, in the
constrained Labastida setting, any similar combination of traces of
$\cF$ tensors vanishes on account of eq.~\eqref{symmcomp}.

Alternatively, one could start from eq.~\eqref{aconstrdt} to
generate expressions of the type
\be
T_{i_1j_1} \ldots T_{i_pj_p} \, T_{(\,ab} \, T_{cd\,)}\, \cA
\label{tpt2a}
\ee
computing further traces. These admit all the Young projections
allowed for the more general expression
\be
T_{i_1j_1} \ldots T_{i_pj_p}\, T_{ab}\, T_{cd}\, \cA\,,
\label{tptta}
\ee
aside from the two-column one. Acting on each individual irreducible
component of \eqref{tpt2a}, however, the permutation group can
generate the entire corresponding irreducible subspaces, which
suffices to extend the statement to
\eqref{tptta}. The conclusion, as above, is that any product of
traces of $\cA$ corresponding to a Young diagram with more than two
columns can be linked to the constraints.

In addition, we need the two key identities \eqref{chain} and
\eqref{id_alpha}, that we list again for convenience:
\begin{align}
&Y_{\{2^p,1^q\}} \!\left[\, 2 \, \pr_{\,k} \!\prod_{r\,=\,1}^p T_{i_rj_r}\! - \!\sum_{n\,=\,1}^p
\pr_{\,(\,i_n}\, \!T_{j_n\,)\,k} \!\prod_{r\,\neq\,n}^p  \!T_{i_rj_r} \right] \!\cA \, = \, (\,p+2\,)
\, Y_{\{2^p,1^q\}} \, \pr_{\,k} \, \cA^{[\,p\,]}{}_{i_1j_1,\,\ldots\,,\,i_pj_p}  \, ,\label{id1b} \\
& \bra \prod_{r\,=\,1}^p T_{i_rj_r} \, \L_{\,k}  \comma  Y_{\{3,2^{p-1}\}} \left[\, \pr_{\,k} \,
\cA^{[\,p\,]}{}_{i_1j_1,\, \ldots\, ,\,i_pj_p} \, - \, \frac{p}{p+2} \ \pr_{\,(\,k} \, \cA^{[\,p\,]}{}_{i_1j_1\,),\,i_2j_2\,,\, \ldots\, ,\,i_pj_p} \,\right] \ket \, = \, 0 \, . \label{id2b}
\end{align}
In the remainder of this Appendix we would like to prove these last
two results.

\subsection*{\sc Proof of eq.~\eqref{id1b}}

A crucial observation for this proof is that one can compute the
$\{2^p,1^q\}$ projection of an expression containing $p$ traces, say
$T_{i_1j_1}
\ldots\, T_{i_pj_p}\,\partial_k$, via the single standard Young tableau
\be \label{youngid1b}
\textrm{
\begin{picture}(30,70)(0,-15)
\multiframe(0,35)(15,0){1}(15,15){{\footnotesize $i_1$}}
\multiframe(15.5,35)(15,0){1}(15,15){{\footnotesize $j_1$}}
\multiframe(0,15.5)(15,0){1}(15,19){\vspace{7pt}$\vdots$}
\multiframe(15.5,15.5)(15,0){1}(15,19){\vspace{7pt}$\vdots$}
\multiframe(0,0)(15,0){1}(15,14.8){{\footnotesize $i_p$}}
\multiframe(15.5,0)(15,0){1}(15,14.8){{\footnotesize $j_p$}}
\multiframe(0,-15.3)(15,0){1}(15,15){{\footnotesize $k$}}
\end{picture}
}
\ee
which corresponds to the choice of standard labeling $i_n=2n-1$,
$j_n=2n$, ($n=1,\ldots,p$), $k=2p+1$. This choice indeed guarantees
that all other standard Young tableaux vanish, due to the symmetry
properties of the $T_{ij}$. The first step in computing the
projection associated to \eqref{youngid1b} is then to enforce the
antisymmetrization in $k$ and the $i_{\,m}$. In order to prove
eq.~\eqref{id1b}, it is sufficient to compare the results obtained
antisymmetrizing the two expressions in eq.~\eqref{id1b},
\begin{align} \label{step1id1b}
& \left[\, 2 \, \pr_{\,k} \!\prod_{r\,=\,1}^p T_{i_rj_r} - \!\sum_{n\,=\,1}^p \pr_{\,(\,i_n}\, T_{j_n\,)\,k} \!\prod_{r\,\neq\,n}^p  T_{i_rj_r} \right] \!\cA \, \longrightarrow \, (\,p+2\,)\, \pr_{\,[\,k}\, T_{i_1\,|\,j_1} \, \ldots \, T_{|\,i_p\,]\,j_p}\, \cA \, , \nonumber \\[5pt]
& \pr_{\,k} \, \cA^{\,[\,p\,]}{}_{i_1j_1,\,\ldots\,,\,i_pj_p} \, \longrightarrow \,
\pr_{\,[\,k} \, \cA^{\,[\,p\,]}{}_{i_1\,|\,j_1,\,\ldots\,,\,|\,i_p\,]\,j_p} \, .
\end{align}
Moreover, one should note that in the expansion
\be
\pr_{\,k} \, T_{i_1j_1} \ldots\, T_{i_pj_p} \, \cA \, = \, \pr_{\,k} \left(\, Y_{\{2^p\}} \, T_{i_1j_1}
\ldots\, T_{i_pj_p}\, + \, Y_{\{4,2^p-1\}} \, T_{i_1j_1} \ldots\, T_{i_pj_p}\, + \, \ldots  \,\right) \cA
\ee
in all available irreducible representations, only the first term,
that defines $\cA^{\,[\,p\,]}$, can survive the antisymmetrization
of all the $i_{\,m}$ indices, so that
\be
\pr_{\,[\,k} \, \cA^{\,[\,p\,]}{}_{i_1\,|\,j_1,\,\ldots\,,\,|\,i_p\,]\,j_p} \, = \, \pr_{\,[\,k}\,
T_{i_1\,|\,j_1} \, \ldots \, T_{|\,i_p\,]\,j_p}\, \cA \, .
\ee
As a result, after the first antisymmetrization the two terms in
\eqref{step1id1b} become proportional, and this property remains
true for the full projection, which finally proves the identity
\eqref{id1b}.

\subsection*{\sc Proof of eq.~\eqref{id2b}}

The presence of the scalar product makes it possible to prove
eq.~\eqref{id2b} computing the projection associated to a single
Young tableaux, as in the previous case. In fact, while the full
$\{3,2^{p-1}\}$ projection results from a sum of different tableaux,
one can choose them in such a way that only one of them contributes
to the scalar product. In particular, one can reduce the full
$Y_{\{3,2^{p-1}\}}$ to the projection associated to
\be \label{youngid2b}
\textrm{
\begin{picture}(30,50)(0,0)
\multiframe(0,35)(15,0){1}(15,15){{\footnotesize $i_1$}}
\multiframe(15.5,35)(15,0){1}(15,15){{\footnotesize $j_1$}}
\multiframe(31,35)(15,0){1}(15,15){{\footnotesize $k$}}
\multiframe(0,15)(15,0){1}(15,19.5){\vspace{7pt}$\vdots$}
\multiframe(15.5,15)(15,0){1}(15,19.5){\vspace{7pt}$\vdots$}
\multiframe(0,0)(15,0){1}(15,14.8){{\footnotesize $i_p$}}
\multiframe(15.5,0)(15,0){1}(15,14.8){{\footnotesize $j_p$}}
\end{picture}
}
\ee
because any symmetrization involving three indices of the set
$\{i_m,j_n\}$ in the left entry of the scalar product extends to a
symmetrization over four indices, on account of the properties of
products of identical tensors. Then, recalling that by definition
$\cA^{\,[\,p\,]}{}_{i_1j_1,\,\ldots\,,\,i_pj_p}$ is already
projected according to
\be \label{youngid2b2}
\textrm{
\begin{picture}(30,50)(0,0)
\multiframe(0,35)(15,0){1}(15,15){{\footnotesize $i_1$}}
\multiframe(15.5,35)(15,0){1}(15,15){{\footnotesize $j_1$}}
\multiframe(0,15)(15,0){1}(15,19.5){\vspace{7pt}$\vdots$}
\multiframe(15.5,15)(15,0){1}(15,19.5){\vspace{7pt}$\vdots$}
\multiframe(0,0)(15,0){1}(15,14.8){{\footnotesize $i_p$}}
\multiframe(15.5,0)(15,0){1}(15,14.8){{\footnotesize $j_p$}}
\end{picture}
}
\ee
one can recognize that the operations needed to build the projection
associated to the tableau \eqref{youngid2b} differ from those
implied by the tableau \eqref{youngid2b2} only in the symmetrization
of the three indices $(i_1,j_1,k)$. More precisely, denoting the
tableau \eqref{youngid2b} by $\t_1$ and the tableau
\eqref{youngid2b2} by $\t_2$ and using the notation of
eq.~\eqref{young_proj_gen} one can recognize that acting on products
of traces
\be
Y_{\t_2} \, = \, 2 \, \frac{\l(\t_2)}{(\,2\,p\,)\,!} \, S_{\,\t_2} \, A_{\,\t_2} \, ,
\ee
where the product $S_{\,\t_2}$ of ``row symmetrizers'' does not
include the operator $S_{(i_1,j_1)}$ because this symmetrization is
already induced by the others. When applied to $T_{i_1j_1} \ldots\,
T_{i_pj_p}\,\partial_k$, the product of the two Young projectors
$Y_{\t_1}$ and $Y_{\t_2}$ then gives
\be \label{youngtableaux12}
Y_{\t_1} \, Y_{\t_2} \, = \, \frac{\l(\t_1)}{(\,2\,p+1\,)\,!} \,
S_{\,(i_1,\,j_1,\,k)} \, S_{\,\t_2} \, A_{\,\t_2} \, Y_{\,\t_2} \, =
\, \frac{\l(\t_1)}{2\,(\,2\,p+1\,)\,\l(\t_2)} \,
S_{\,(i_1,\,j_1,\,k)} \, (\,Y_{\t_2}\,)^{\,2} \, .
\ee
The ratio that appears in eq.~\eqref{youngtableaux12} is
\be
\frac{\l(\t_1)}{(\,2\,p+1\,)\, \l(\t_2)} \, = \, \frac{p}{p+2}
\ee
so that
\be
Y_{\t_1} \, \pr_{\,k} \, \cA^{[\,p\,]}{}_{i_1j_1\,,\,\ldots\,
,\,i_pj_p} \, = \, \frac{p}{p+2} \, \pr_{\,(\,k} \,
\cA^{[\,p\,]}{}_{i_1j_1\,),\,i_2j_2\,,\, \ldots\, ,\,i_pj_p} \, ,
\ee
which finally proves eq.~\eqref{id2b}, since $Y_{\{3,2^{p-1}\}}$ can
be replaced with $Y_{\,\t_1}$ within the scalar product.

\end{appendix}

\newpage

%%%%%%%%%%%%%%%%%%%%%%%%%%%%%%%%%%%%%%%%%%%%%%%%%%%%%%%%%%%%%%%%%%%%%

\end{document}